\documentclass[prb, twocolumn, superscriptaddress, notitlepage]{revtex4-2}
\usepackage{amsmath,amsfonts, amssymb, amsthm, dsfont}
\usepackage{yfonts}
\usepackage{bm}
\usepackage{mathrsfs}
\usepackage{graphicx}
\usepackage{verbatim}
\usepackage{hyperref}
\usepackage{multirow}
\usepackage[normalem]{ulem}
\usepackage{tikz}
	\usetikzlibrary{calc}
	\usetikzlibrary{shapes.multipart}
	\usetikzlibrary{decorations}
	\usetikzlibrary{snakes}

\newcommand{\di}{\mathrm{d}}

\renewcommand{\vec}[1]{{\mathbf #1}}
\newcommand{\ket}[1]{|#1\rangle}

\renewcommand{\vr}{{\vec{R}}}

\newcommand{\vx}{{\vec{x}}}
\renewcommand{\ol}[1]{\overline{#1}}
\newcommand{\comments}[1]{}
\newcommand{\mb}[1]{\mathbf{#1}}

\renewcommand{\Ref}[1]{ Ref.~[\onlinecite{#1}]}
\newcommand{\Refs}[1]{ Refs.~[\onlinecite{#1}]}

\def\U{\mathrm{U}(1)}
\def\H{\mathcal{H}}
\def\Z{\mathbb{Z}}

\newcommand{\stkout}[1]{\ifmmode\text{\sout{\ensuremath{#1}}}\else\sout{#1}\fi}

\makeatletter
\def\l@subsubsection#1#2{}
\makeatother

\usetikzlibrary{decorations.pathreplacing,decorations.markings}
\tikzset{middlearrow/.style={
        decoration={markings,
            mark= at position 0.55 with {\arrow{#1}} ,
        },
        postaction={decorate}
    }
}
\begin{document}

\title{Rotation Symmetry-Protected Topological Phases of Fermions}

\author{Meng Cheng}
\email{m.cheng@yale.edu}
\affiliation{Department of Physics, Yale University, New Haven, CT 06511-8499, USA}

\author{Chenjie Wang}
\email{cjwang@hku.hk}
\affiliation{Department of Physics and HKU-UCAS Joint Institute of Theoretical and Computational Physics, The University of Hong Kong, Pokfulam Road, Hong Kong, China}
\affiliation{Department of Physics, City University of Hong Kong,
Tat Chee Avenue, Kowloon, Hong Kong, China}

\date{\today}

\begin{abstract}
%Symmetry-protected topological phases with crystalline and internal symmetries were recently proposed to be equivalent in boson systems. 

We study classification of interacting fermionic symmetry-protected topological (SPT) phases with both rotation symmetry and Abelian internal symmetries in one, two, and three dimensions. By working out this classification, on the one hand, we demonstrate the recently proposed correspondence principle between crystalline topological phases and those with internal symmetries through explicit block-state constructions.  We find that for the precise correspondence to hold it is necessary to change the central extension structure of the symmetry group by the $\mathbb{Z}_2$ fermion parity.  On the other hand, we uncover new classes of intrinsically fermionic SPT phases that are only enabled by interactions, both in 2D and 3D. Several 3D Majorana-type fermionic SPT phases are identified. Moreover, several new instances of Lieb-Schultz-Mattis-type theorems for Majorana-type fermionic SPT phases are obtained and we discuss their interpretations from the perspective of bulk-boundary correspondence.
\end{abstract}

\maketitle

\tableofcontents

\section{Introduction}
\subsection{Background and motivation}
Symmetries can greatly enrich gapped phases of quantum matter. In condensed matter systems, crystalline symmetries of lattice systems are among the most common symmetries, besides a few internal symmetries such as  charge or spin conservation and time reversal. Recently, a rich variety of crystalline symmetry-protected topological (SPT) phases have been discovered and classified~\cite{FuPRL2011,  SlagerNP2012, TeoPRL2013, ShiozakiPRB2014, ShiozakiPRB2017, SongPRX2017, KruthoffPRX2017, BradlynNature2017, FangArxiv2017, PoNC2018, SongNC2018, KhalafPRX2018, HanArxiv2018}, in particular for band insulators of non-interacting electrons, culminating in exhaustive lists of possible topological materials~\cite{ZhangCatalogue2018, Vergniory2018, Tang2018}. Similar phases for interacting bosonic/spin systems have also been constructed, and systematic classifications have been achieved in some cases~\cite{hermele2016, HuangPRB2017, RasmussenArxiv2018}. 

Investigations of bulk-boundary correspondence in crystalline SPT phases have also been fruitful. For SPT phases with internal symmetries, it is known that the boundary must have 't Hooft anomalies and can not have symmetric short-range entangled (SRE) boundary states\cite{tHooftbook,KapustinPRL2014,WittenRMP2016}. As a result, SPT boundaries can have gapless excitations, spontaneously break the protecting symmetries, or  develop symmetric gapped states with topological order when the spatial dimension of the boundary  is greater than 1~\cite{vishwanath2013}. The same principle applies to crystalline SPT phases as well, as long as the boundary preserves the protecting crystalline symmetries. However, the  fact that symmetries involve spatial coordinate transformations do bring in new twists to the bulk-boundary correspondence. For instance, in many cases, boundaries of crystalline SPT phases admit tensor product structure both for the Hilbert space and the boundary symmetry action and can be viewed as a well-defined lattice systems on their own. In these cases, 't Hooft anomalies lead to various generalizations of Lieb-Schultz-Mattis-Oshikawa-Hastings theorems~\cite{LSM, OshikawaLSM, HastingsLSM, ChengPRX2016, HuangPRB2017, JianPRB2018, PoPRL2017, ChengArxiv2018, YangPRB2018, LuLSMSPT}. Moreover,  it is realized that if one allows non-uniformity on the boundary, one can trivially gap out almost everywhere except certain lower-dimensional regions, i.e. corners or hinges. This phenomena was dubbed ``higher-order'' boundary states~\cite{BenalcazarScience, SchindlerSciAdv2018, BenalcazarPRB2017, SongPRL2017, YuxuanWang2018, JosiasPRL2017, Matsugatani2018, KhalafPRB2018, YouHO2018, DubinkinHO2018}.

In this work we consider interacting fermionic SPT (FSPT) phases protected by spatial rotation and internal symmetries. Many previous works studied either free fermions or bosonic systems with rotation symmetry. The physics of strongly interacting fermionic phases remains a major open question. We focus on rotation, a basic point group operation, to develop systematic understanding of crystalline fermionic SPT phases.  More concretely, we study fermionic systems with the symmetry group $C_M \times G$, where $G$ is an Abelian internal symmetry group, and $C_M$ denotes $M$-fold rotations. 

Another motivation for this work is to gain insight into the classification of interacting FSPT phases with internal symmetry only. Much progress has been made on this problem, for example a complete picture of how interactions affect the periodic table for topological band insulators and superconductors has been obtained~\cite{ChongWangPRB2014, freed2016}. Outside the periodic table, theories of FSPT phases with $\Z_2^f\times G$ symmetry with a general finite $G$ have also been developed~\cite{GuPRB2014, Kapustin2015b, KapustinThorngren, WangPRX2018}. However, they often involve complicated constructions of exactly-solvable models~\cite{GuPRB2014, WangPRX2018}, or employ sophisticated algebraic topology techniques~\cite{Kapustin2015b, freed2016}, and it is not straightforward to extract physical properties for such FSPT phases.
Recently, it has become clear that the topological classifications for gapped phases with spatial symmetries is closely related to those with internal symmetries, as long as the abstract group structures match~\cite{ChengPRX2016, ThorngrenPRX2018}.  It is perhaps clearer for orientation-preserving symmetries, since one can introduce lattice defects serving as fluxes of the symmetries to probe the topological properties. Such correspondence is formalized as a ``crystalline equivalence principle'' in \Ref{ThorngrenPRX2018}.  In fermionic systems, the global symmetry group is a central extension of the physical, ``bosonic'' symmetry group by the $\Z_2^f$ fermion parity. We will show that the equivalence between classifications with crystalline and internal symmetries requires a change in the group extension structure.  

The goal of this paper is to apply the dimensional reduction approach \cite{SongPRX2017} to interacting FSPT phases with $C_M\times G$ symmetry, in particular 3D FSPT phases which are much less explored, to study their classification and physical properties. We also aim to explore the crystalline equivalence principle for FSPT phases and find that it comes with a new feature compared to the bosonic version (see Sec.~\ref{sec:correspondence} for a summary). In addition, we study Lieb-Schultz-Mattis-type theorems associated with Majorana-type FSPT phases and discuss their interpretations from the perspective of bulk-boundary correspondence. In the rest of the introduction, we describe the general idea of the dimensional reduction approach in Sec.~\ref{sec:general-principle} and summarize our main results in Sec.~\ref{sec:mainresults}. We discuss organization of the paper in Sec.~\ref{sec:organization}

\subsection{Classification scheme}
\label{sec:general-principle}
We study the classification of interacting fermionic SPTs with both rotation and internal Abelian symmetry. We will follow the dimensional reduction approach introduced in \Ref{SongPRX2017}. In a crystallgraphic symmetry group, the only allowed rotations are $C_M$ with $M=2,3,4,6$. In this work the only spatial symmetry under consideration is rotation, so there is no limitation on the values of $M$. We will take any value $M>1$. In addition to rotation, we also consider an internal symmetry group $G$, such that the whole symmetry group is $C_M \times G$ (with fermions, we need to specify how rotations are extended by the fermion parity $\Z_2^f$, which we will define more precisely in Sec.~\ref{sec:symmetry}).

\begin{figure*}[t]
	\centering
	\includegraphics{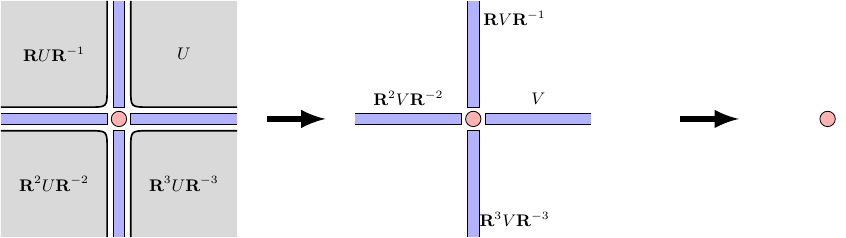}
	\caption{Dimensional reduction procedure for $C_4$ rotation SPT phase.}
	\label{fig:disentangle}
\end{figure*}

Let us lay out the general principles of dimensional reduction for rotation symmetry in $d$ spatial dimensions. 
Even though we always consider lattice systems, let us assume the space is $\mathbb{R}^d$ for a moment to describe the general principles. We first divide the space into open disjoint regions related to each other by the point group symmetries, labeled by $\mathcal{M}_j$ (see Fig.~\ref{fig:disentangle}).  For example, in 2D with polar coordinates, 
\begin{equation}
\mathcal{M}_j=\left\{ (\rho, \theta)\Big| \rho>0, \ \frac{2\pi (j-1)}{M}<\theta < \frac{2\pi j}{M}\right\}
\end{equation}
where $j=1,\cdots, M$. Note that this leaves out the origin  $\rho=0$, and $M$ half lines $\rho>0,\ \theta=\frac{2\pi j}{M}$. We will denote them by $\ol{\mathcal{M}}$, the complement of $\bigcup_j \mathcal{M}_j$. For physical reasons, it is convenient to ``thicken'' $\ol{\mathcal{M}}$, e.g. we take $\ol{\mathcal{M}}$ to be the union of a small disk around the rotation center, plus narrow strips centered at the rays $\theta=\frac{2\pi j}{M}$.  Generalizations to higher dimensions are straightforward.

The dimensional reduction procedure takes three steps to study rotation SPT phases (Fig.~\ref{fig:disentangle}). First, consider one of the regions, $\mathcal{M}_j$. The state on $\mathcal{M}_j$ may already be an SPT state protected by the onsite symmetry $G$.  In addition, we will also consider the case that $\mathcal{M}_j$ hosts an invertible topological phase, e.g., 1D Majorana chain and 2D $p+ip$ superconductors, which do not need any symmetry for protection. (It is generally believed that there is no invertible topological phase in 3D\cite{freed2021reflection,Kapustin2015b}. We will take this assumption throughout this work). Due to rotation symmetry, other regions must be in the same SPT/invertible phase. An important question that we need to check is whether this whole state is compatible with the rotational symmetry, when all regions $\{\mathcal{M}_j\}$ are combined. We answer this question constructively: When the regions $\{\mathcal{M}_j\}$ are disconnected from each other, they have nontrivial boundary states. We can then glue the boundary states together from neighboring regions, in a rotation symmetric way. Notice that this is always possible because neighboring boundaries have opposite orientations. The only subtlety here arises near the rotation center, where one may find some remaining symmetry-protected low-energy modes. In some cases, one can remove the degeneracies by introducing lower-dimensional SPT states in $\ol{\mathcal{M}}$ (one such example is discussed in Appendix \ref{app:CplusZZ2D}). If not, we conclude that the bulk SPT phase is not compatible with the rotation symmetry, and should be excluded from the classification. However, as we will discuss, in two dimensions where the rotation center is 0D, we can also interpret such an obstruction to having rotationally invariant SPT phases as instances of LSM-type theorems for the internal symmetry SPT phase.  

Second, if the state in $\mathcal{M}_j$ is trivial, it can be transformed into a product state by a finite-depth local unitary that preserves all the internal symmetries. Let $U$ be the local unitary applied to region $\mathcal{M}_1$. Because of the rotation invariance, we can apply $\vr^{j}U\vr^{-j}$ to disentangle $\mathcal{M}_{j+1}$, where $\vr$ is the rotation operator. When combined, the local unitary $\prod_j \vr^j U\vr^{-j}$ respects both $G$ and $C_M$.  Now entanglement remains only in the $\ol{\mathcal{M}}$ region. If we focus on one of the hyper half-planes away from the rotation center, we can ask whether there is a nontrivial SRE phase in this $(d-1)$-dimensional system. With rotations, all the $(d-1)$-dimensional hyper half-planes must have the same SRE states on them. Importantly, their $(d-2)$-dimensional boundaries must meet near the rotation center, so we should impose the condition that $M$ copies of these boundary states can be gapped out preserving the internal and the $C_M$ symmetry. That is, it is required that  we are able to glue the $(d-1)$-dimensional half-plane symmetrically. We will see that in many cases, certain $(d-1)$ SPT phases cannot be glued in a symmetric way.

Third, if the $(d-1)$-dimensional half-planes are topologically trivial, we can disentangle the states on them using symmetric local unitary, similar to above. Then, we further ask whether the $(d-2)$-dimensional rotation center is in an SRE phase or not. At the center, rotation becomes an internal symmetry. Accordingly, the $(d-2)$-dimensional SRE phases should have $\mathbb{Z}_M\times G$ internal symmetry, where we use $\mathbb{Z}_M$ to reflect the fact that rotation acts as if an internal symmetry.

When describing the dimensional reduction approach above, we have used some \emph{constructive} viewpoints. In fact, the whole procedure can be ``inverted'' and viewed constructively: $d$-dimensional rotation SPT phases with $C_M\times G$ symmetry can be built out of $d$-dimensional and lower dimensional $G$ symmetric \emph{block states} by gluing them together in certain rotation symmetric fashion.  Therefore, depending on which building blocks are used, we define the following three groups:
\begin{itemize}
	\item $\mathscr{G}_0$, the group of $d$ dimensional SPT phases protected by $G$ and compatible with the $C_M$ symmetry. When building these $d$ dimensional ``block states'', lower dimensional states may also be used to successfully glue the blocks. 
	\item $\mathscr{G}_{-1}$, the group of $(d-1)$ dimensional block states, built from $(d-1)$-dimensional SRE states with the internal symmetry group $G$, by gluing $M$ copies of them  together while preserving both $G$ and rotation.
	\item $\mathscr{G}_{-2}$, the group of $(d-2)$ dimensional block states, which are basically a $(d-2)$ dimensional SRE states with $\Z_M\times G$ internal symmetry, located at the center of rotation. 
\end{itemize}
A special attention should be paid to the states in $\mathscr{G}_{-2}$. One should notice that there are additional equivalence relations between these phases, so $\mathscr{G}_{-2}$ is generally \emph{different} from the actual classification for $(d-2)$D SRE phases with $\mathbb{Z}_M\times G$ symmetry. The additional equivalence relations lead to the \emph{trivialization} of certain $(d-2)$ dimensional block states, which are nontrivial as SPT states with strictly internal $\Z_M \times G$ symmetry.

Generally speaking, SPT phases form an Abelian group under the stacking operation. Thus the remaining task in the classification problem is to determine the group structure of the SPT phases. We have already identify three subgroups: $\mathscr{G}_0$, $\mathscr{G}_{-1}$ and $\mathscr{G}_{-2}$. Their own group structures can be derived from the known classification of (internal symmetry) SPT phases in the corresponding dimensions, modulo further trivializations mentioned above.  Let us denote the group of all $C_M\times G$ SPT phases by $\mathscr{G}$, and all SPT states consisting of $(d-1)$ and $(d-2)$ block states by $\mathscr{G}_{\leq -1}$ (in this notation $\mathscr{G}\equiv \mathscr{G}_{\leq 0}$).  There is no general formula to determine $\mathscr{G}$ from the known subgroups, and the problem has to be solved on a case-by-case basis. However,  we do know that these groups should fit into the following two short exact sequences: 
\begin{equation}
	\begin{gathered}
	1\rightarrow \mathscr{G}_{-2}\rightarrow \mathscr{G}_{\leq -1}\rightarrow \mathscr{G}_{-1}\rightarrow 1\\
	1\rightarrow \mathscr{G}_{\leq -1}\rightarrow \mathscr{G}\rightarrow \mathscr{G}_{0}\rightarrow 1.
	\end{gathered}
	\label{eq:extension}
\end{equation}
Both group extensions are central. See Appendix \ref{sec:groupext} for a brief review of group extension. In general these two sequences do not necessarily split. Physically, the first short exact sequences essentially says that if we stack two $(d-1)$ dimensional block states, the result must be another $(d-1)$-dimensional block state, possibly with a $(d-2)$ dimensional block state. The second short exact sequence is similar. It is worth pointing out that in the bosonic case, with $C_M\times G$ symmetry the extensions are all trivial. However, as we will see later, for fermions this is not the case.

To summarize, our task is to: (i) obtain the groups $\mathscr{G}_0$, $\mathscr{G}_{-1}$ and $\mathscr{G}_{-2}$ and (ii) work out the group extensions in Eq.~\eqref{eq:extension}. We remark that this dimension reduction or ``gluing'' approach works for both bosonic and fermionic systems. We have applied it to bosonic SPT phases with $C_M\times G$ symmetry in Appendix \ref{sec:bosonic}, and find that the group extensions in Eq.~\eqref{eq:extension} are always trivial. That is, for bosonic SPT phases, we always have $\mathscr{G} = \mathscr{G}_{0}\times \mathscr{G}_{-1}\times\mathscr{G}_{-2}$. Nevertheless, we will see that for FSPT phases, the extensions in \eqref{eq:extension} may be nontrivial. We also remark that given $\mathscr{G}_{0}$,  $\mathscr{G}_{-1}$, and $\mathscr{G}_{-2}$,  Eq.~\eqref{eq:extension} alone does not determine $\mathscr{G}$. Mathematically, there could be many $\mathscr{G}$'s satisfying Eq.~\eqref{eq:extension}. To determine $\mathscr{G}$, we need to study physical properties of FSPT phases which vary in dimensions (see, e.g., Sec.~\ref{sec:inversion2} for a 1D example and Sec.~\ref{sec:stackgroup} for a 3D example).

Note that we will include the Majorana chain into $\mathscr{G}_0$ for 1D fermionic SPTs, as Majorana chains do extend other SPT phases. However, we will not include $p+ip$ superconductors into $\mathscr{G}_0$ for 2D fermionic SPTs. This is because $p+ip$ superconductors are chiral and of infinite order. They do not extend SPT phases which are non-chiral and always have finite order. Combining the 2D block SPT states and $p+ip$ superconductors gives a $\mathscr{G}_0\times \mathbb{Z}$ classification. (A single layer of $p+ip$ superconductor may not be compatible with the rotation symmetry, so there is a subtlety on what is the root state of $\mathbb{Z}$. See Appendix \ref{append:p+ip} for a discussion.)

\subsection{Main results}
\label{sec:mainresults}

Based on the dimensional reduction approach explained above, we derive a systematic classification for FSPT phases protected by $C_M\times G$ symmetry with $G$ being a unitary finite Abelian group. While the counterparts of 1D and 2D FSPT phases with internal symmetries were known previously, our work provides an alternative viewpoint and makes a connection between SPT phases with internal and spatial symmetries. For 3D FSPT phases, many of our results are new, whose counterparts with internal symmetries are not well understood. Examples of our classification are summarized in Tables \ref{tab12D}, \ref{tab1} and \ref{tab2}. 

While the basic construction is parallel to the bosonic case, fermionic systems exhibit several notable new features and subtleties. We summarize a few important results as follows:
\begin{enumerate}
	\item We construct intrinsically fermionic crystalline SPT phases which can only exist with strong interactions. We show that this is the case for all nontrivial rotational FSPT phases with Abelian internal symmetries in 3D, and discuss one example in 2D where the internal symmetry is the BDI class in the periodic table. We find obstructions in the dimensional reduction construction that prevent gluing together lower-dimensional block states while preserving rotation symmetries. No such obstructions were present in bosonic systems previously studied~\cite{HuangPRB2017, RasmussenArxiv2018}.
	\item We formulate the precise correspondence between classifications with rotation symmetry and those with only internal symmetry, which can be used to map out the classification of 3D interacting fermionic SPT phases protected by internal Abelian unitary symmetries. The correspondence is stated in Sec.~\ref{sec:correspondence}.
	\item We identify several new instances of Lieb-Schultz-Mattis (LSM) type theorems for 2D FSPT phases. Roughly speaking, certain FSPT phases are only compatible with rotation symmetry projectively represented. Otherwise there has to be ``anomalous'' degrees of freedom at the rotation center. These LSM theorems also indicate that there exist 3D ``trivial'' bulk states, whose boundary states are symmetric SRE, but can not be realized in strictly 2D systems (unless under the LSM-type conditions\footnote{Certain types of anomalies involving crystalline symmetries can be implemented by allowing symmetries to act \emph{projectively}, instead of \emph{linearly}, such that the anomalous system can be realized in the same dimension without a bulk. Not every anomaly can be realized in this way. The conditions making this to happen are refereed to as ``LSM-type conditions'' (see e.g. Ref.~\cite{ChengPRX2016}).}). 
\end{enumerate}

\subsection{Organization of the paper}
\label{sec:organization}

The rest of the paper is organized as follows. In Sec.~\ref{sec:generalities}, we discuss a few generalities including symmetry groups in fermionic systems, a subtlety of the trivial state in crystalline topological phases, and the crystalline equivalence principle of FSPT phases that we obtain. In Secs.~\ref{sec:1D}, \ref{sec:2D} and \ref{sec:3D}, we apply the general classification scheme in Sec.~\ref{sec:general-principle} to various examples of interacting crystalline FSPT phases in 1D, 2D and 3D, respectively. We discuss several new instances of Lieb-Schultz-Mattis type theorems for 2D FSPT phases in Sec.~\ref{sec:sptlsm}. We conclude in Sec.~\ref{sec:conclusion}. Appendix \ref{sec:bosonic} includes a study on bosonic rotation SPT phases through the dimension reduction approach. We review some 1D and 2D FSPT phases with internal symmetries in Appendix \ref{append:review}, and discuss the compatibility of $p+ip$ superconductors with rotation symmetry in Appendix \ref{append:p+ip}. More 2D and 3D examples are studied in Appendices \ref{app:2d} and \ref{sec:more3D}, respectively. 

\section{Generalities}
\label{sec:generalities}
\subsection{Symmetries in fermionic systems}
\label{sec:symmetry}
We make some general remarks here regarding symmetries in fermionic systems. Our remarks apply to any symmetries, but we will focus on the rotation. For fermionic superconductors, it is important to distinguish two cases: $\vr^M=\mathds{1}$ or $\vr^M=P_f$. (For insulators, one can simply redefine $\vr$ by a $\U$ rotation, and the two cases become equivalent.)  We will also refer to them as $C_M^\pm$ regarding the action on single-particle states. To be precise, the fermion creation/annihilation operators transform linearly under rotation:
\begin{equation}
\vr	
		c_\vx
	\vr^{-1}=
	U_\vr(\vx)
		c_{\vr(\vx)}.
	\label{}
\end{equation}
Here $U_\vr$ is a unitary transformation, and we suppress the spin/orbital/$\cdots$ indices. $\vr^M=(\pm\mathds{1})^{N_f}$ correspond to 
\begin{equation}
	\prod_{j=1}^M U_\vr\big(\vr^j(\vx)\big)=\pm \mathds{1}.
	\label{}
\end{equation}
Note that for odd $M$, we could simply redefine $\vr$ by $\vr P_f$ so the two choices are equivalent. Mathematically, these choices correspond to possible central extensions of $C_M$ by the fermion parity symmetry (i.e. $\H^2[\Z_M, \Z_2]=\Z_{(M,2)}$).

For example, if the rotation only operates on the spatial degrees of freedom, then we expect to have $\vr^M=\mathds{1}$. On the other hand, for spin-$1/2$ electrons, naturally rotation affects both the orbital and spin degrees of freedom, so should satisfy $\vr^M=P_f$ because $2\pi$ spin rotation results $-1$. In this case, we can also combine $\vr$ with a $\frac{2\pi}{M}$ spin rotation in the opposite direction to get $\vr^M=\mathds{1}$. Even for spinless fermions, if an odd-parity pairing order parameter is present the rotation symmetry has to satisfy $\vr^M=P_f$. Therefore in this paper we will consider both $C_M^\pm$ symmetries. 

We also include an Abelian internal symmetry group $G$.  In principle one also has to specify how $G$ is extended by the fermion parity, but for simplicity we will just consider the trivial extension for $G$ in this work. In fact, this form of symmetry already allows to compare to the most general FSPT phases with Abelian internal symmetries.

%In fact, for Abelian $G$, it is always possible to redefine the generators for $C_M\times G$ such that the group extension only occurs in the $C_M$ part.

\subsection{Trivial states}
\label{sec:triviality}

We now discuss what it means for a state to be trivial in systems with point group symmetries. Usually, trivial states are defined to be those that can be adiabatically connected  to a product state (i.e. an atomic insulator in the context of band insulators).
However, for SPT phases protected by point group symmetries, we need to refine the notion of triviality. 

First of all, we allow for a more general notion of trivial states. This was discussed in \Ref{SongPRX2017} for mirror reflection symmetry.  Consider a state of the following form: in any of the $\mathcal{M}_i$ we place a lower-dimensional short-range entangled phase (which may be a nontrivial invertible phase), denoted by ${A}$, and use the point group to fill the other regions. This state can be made into a true product state by the following transformation.
We fill the rest of $\cup_i\mathcal{M}_i$ with product states. We then adiabatically generate pairs of $\ol{A}$ and ${A}$ from the product state. We can then pair annihilate all $A$ and $\ol{A}$'s adiabatically.  Essentially we move the ${A}$'s to infinity using this procedure. We will consider states of this form trivial.

Another subtlety is that there can be topologically distinct classes of product states, in particular in one and two dimensions~\cite{HuangPRB2017}. That is, two product states can not be adiabatically connected preserving the symmetry. This situation occurs when there are degrees of freedom in the rotation center. For simplicity, we will assume that the microscopic degrees of freedom do not live exactly on the rotation center to avoid this subtlety.  

\subsection{Relation to FSPT phases with internal symmetries}
\label{sec:correspondence}
A close relation between bosonic topological phases protected by crystalline symmetries and those by internal symmetries was recently identified, dubbed as ``crystalline equivalence principle'' by Else and Thorngren in Ref.~\onlinecite{ThorngrenPRX2018}. The equivalence principle states that the classification of crystalline topological phases (both SPTs and SETs) of symmetry $G$ is the same as that of topological phases with internal symmetry $G$. For this equivalence to work, an orientation-reversing spatial symmetry should be mapped to anti-unitary internal symmetry. In Appendix \ref{sec:bosonic} we show explicitly that the classifications of $d$D bosonic SPT phases protected by $C_M\times G$ symmetry are identical to $\H^{d+1}[C_M\times G, \U]$, for $d=2$ and $3$.

The crystalline equivalence principle is expected to hold for fermionic topological phases too. We now state the precise form of the correspondence for rotation FSPT phases:

\noindent\fbox{
	\parbox{0.94\columnwidth}{The $C_n \times G$ SPT classification for $\vr^M=(\pm 1)^{N_f}$ is equivalent to $\Z_M\times G$ SPT classification with $\mb{g}^M=(\mp 1)^{(M-1)N_f}$, where $\mb{g}$ is the generator of $\Z_M$.}
}

That is, for even $M$, the two possible extensions of $C_M$ by the fermion parity get swapped, when mapped to internal $\Z_M$ symmetry.  Intuitively, the difference should be attributed to the topological spin of fermions, i.e. a $2\pi$ rotation results in $-1$ phase factor.  A similar twist of signs is known to occur for the correspondence between reflection symmetry and time-reversal symmetry. This correspondence is obtained from all the examples that we work out. We do not have a mathematically rigorous proof without working out individual examples.

\section{Inversion FSPT phases in 1D}
\label{sec:1D}

As a warm-up exercise, we study 1D FSPTs with rotation symmetry. In 1D, the only sensible rotation symmetry is inversion $I$ (notice that the inversion is orientation-reversing, unlike rotations in higher dimensions). In fermionic systems, there exist two possibilities, $I^2=\mathds{1}$ and $I^2= P_f$. Below we derive the classification of inversion FSPTs following the general classification principles outlined in Sec.~\ref{sec:general-principle}. While most results in this section are not new~\cite{ShiozakiPRB2017b, ShapourianPRL2017}, the derivation touches on conceptual subtleties that will be important for higher-dimensional systems, so we include them here for pedagogical purpose.

According to the correspondence principle in Sec. \ref{sec:correspondence}, the classification of inversion FSPTs in the two cases should be the same as the classification of 1D time-reversal FSPT phases with $T^2=P_f$ and $T^2=\mathds 1$, respectively. The latter classifications are known to be $\mathbb{Z}_2$ for $T^2=P_f$, and $\mathbb{Z}_8$ for $T^2=\mathds 1$~\cite{Fidkowski2011}. Our results are completely consistent with the crystalline correspondence principle.

\subsection{$I^2=\mathds{1}$}
\label{sec:inversion1}
We first consider $I^2=\mathds{1}$. According to the general classification scheme, we need to consider (i) possible 0D-block states and (ii) 1D invertible topological phases that are compatible with $I^2=\mathds 1$. 

For the 0D block, the total symmetry group reduces to $\Z_2^f\times\Z_2$, where the latter $\mathbb{Z}_2$ represents inversion acting on the 0D block. There are four 0D-block states, corresponding to the four irreducible representations of $\mathbb{Z}_2^f\times \mathbb{Z}_2$. The two root states are:
\begin{enumerate}
	\item The fermion parity of the 0D block is odd. 
	\item The inversion eigenvalue of the 0D block is $-1$. 
\end{enumerate}
However, the second root state is actually trivial. To see that, we consider spinless fermions on a chain, with a bond-centered inversion $I$ defined as $Ic_nI^{-1}=c_{1-n}$. It is easy to design a gapped Hamiltonian such that the ground state is $\prod_n c_n^\dag |0\rangle$. This is obviously a trivial state, as there is no entanglement between any two fermions.  Then, the 0D-block state with only two sites is 
\begin{equation}
\ket{\psi}_{\rm 0D}=c_0^\dag c_1^\dag \ket{0}
\end{equation}
Under inversion symmetry, the state $|\psi\rangle_{\rm 0D}$ has an eigenvalue $-1$: $I\ket{\psi}_{\rm 0D}=c_1^\dag c_0^\dag\ket{0}=-\ket{\psi}_{\rm 0D}$. Accordingly, the 0D-block state with inversion eigenvalue $-1$ does not correspond to any nontrivial FSPT state.  

Now we further show that the first root state with odd fermion parity in the 0D block is indeed non-trivial. We will define the following many-body topological invariant: consider an open chain with boundary conditions preserving the inversion symmetry. It is always possible to lift any degeneracy (i.e. from accidental zero modes at the ends) and have a unique, inversion-symmetric ground state. The fermion parity of the ground state is a many-body topological quantum number invariant under fermionic finite-depth local unitary circuit, and distinguishes two phases.  
Hence, 0D-block states lead to a $\mathbb{Z}_2$ classification.

The only ``invertible topological phase'' in 1D is the Kitaev Majorana chain. We argue that the Majorana chain is not compatible with $I^2=\mathds{1}$. %It is already evident from the well-known realization in spinless fermion chain with $p$-wave pairing. More generally,
To see that, we first imagine cutting the Majorana chain in the middle (i.e. the inversion center), which leaves two edge Majorana zero modes $\gamma_{l}$ and $\gamma_r$. Under inversion symmetry,  the two Majorana zero modes transform into one another:
\begin{equation}
	I:\ \gamma_l\leftrightarrow \gamma_r.
	\label{eq:inv-1}
\end{equation}
So far, the full symmetry is preserved. Next, we try to glue the two half chains, by removing the zero modes $\gamma_l$ and $\gamma_r$. However,  the only coupling term $i\gamma_l\gamma_r$ is odd under $I$, so we cannot glue the chains. In fact, the zero modes can never be removed in an inversion symmetric way, even when additional 0D-block states are decorated. This follows from the observation that the two-dimensional Hilbert space spanned by $\ket{0}$ and $a^\dag\ket{0}$, with $a\equiv (\gamma_l + i \gamma_r)/2$,  forms a projective representation of $\mathbb{Z}_2^f\times\mathbb{Z}_2$. Indeed, in this Hilbert space, we have $\gamma_l= \sigma^x$, $\gamma_r=\sigma^y$, $P_f=\sigma^z$,  and 
\begin{equation}
I = \left(
\begin{matrix}
0 & e^{i\pi/4} \\
e^{-i\pi/4} & 0
\end{matrix}
\right),
\end{equation}
where $\sigma^i$ are Pauli matrices. This representation fulfils the transformation \eqref{eq:inv-1} and the condition $I^2=\mathds 1$. It is easy to see that $P_f I = - I P_f$, which is a sufficient condition showing that the Hilbert space is a projective representation of $\mathbb{Z}_2^f\times\mathbb{Z}_2$. Hence, the two-fold degeneracy cannot be lifted, even if additional 0D-block states (i.e., linear representations) are attached. Accordingly, 1D Majorana chain is not compatible with $I^2=\mathds 1$.

Combining the above results, we conclude that the classification of 1D inversion FSPTs is $\mathbb{Z}_2$, the same as the class DIII superconductors, i.e., $T^2=P_f$ fermion systems. 

\subsection{$I^2=P_f$}
\label{sec:inversion2}

\begin{figure}
\centering
\includegraphics{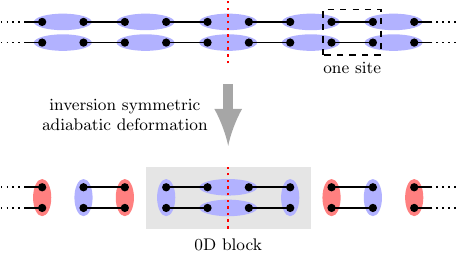}
\caption{Inversion symmetric adiabatic deformation of a double Majorana chain (top) to a state (bottom) in which inter-site entanglement only occurs in the 0D block (gray rectangle). Each dot represents a Majorana fermion $\gamma$, and every two dots connected by a solid line represent a physical complex fermion. Each lattice site of the double chain contains two complex fermions. Elliptically shaded two dots are Majorana fermions that are paired up, where ``blue'' represents pairing through $-i\gamma\gamma'$ and ``red'' represents pairing through $i\gamma\gamma'$.}
\label{fig:inversion}
\end{figure} 

Next we consider $I^2=P_f$. We need to consider (i) possible 0D-block states and (ii) 1D invertible topological phases that are compatible with $I^2=P_f$. 

For the 0D block, the symmetry group is the unitary $\Z_4^f$, with $I$ being the generator. There are four 0D-block states, corresponding to the four irreducible representations of $\Z_4^f$, i.e., with inversion eigenvalue being $1, i, -1$, and $-i$, respectively. Similar trivialization procedures to those in the $I^2=\mathds{1}$ case are attempted, but all the four 0D-block states remain distinct. It suggests that they  represent different FSPT states.  Next, we consider if the 1D Majorana chain is compatible with $I^2=P_f$. We follow the same cutting and gluing setting in Sec.~\ref{sec:inversion1}. Now, the middle Majorana zero modes $\gamma_l$ and $\gamma_r$ satisfy the following inversion transformation
\begin{equation}
	I\gamma_l I^{-1}=\gamma_r, \quad I\gamma_rI^{-1}=-\gamma_l
	\label{}
\end{equation}
to comply the requirement $I^2=P_f$. The coupling term $-i\gamma_l\gamma_r$ is symmetric under $I$, so we can successfully glue the two half Majorana chains in the middle. Hence, 1D Majorana chain is compatible with $I^2=P_f$.

Combining all together, we have identified $8$ distinct phases. The group structure of the eight FSPTs under stacking, i.e, how the 1D Majorana chain extends the 0D-block states, remains to be identified. To do that, we consider stacking two identical Majorana chains (top of Fig.~\ref{fig:inversion}).  We will show that, without closing the energy gap and breaking the inversion symmetry, the double chain can be adiabatically deformed to a state in which inter-site entanglement only exists between the two sites near the inversion center (bottom of Fig.~\ref{fig:inversion}). To do that, we first consider four Majoranas $\gamma_1, \gamma_2, \gamma_1', \gamma_2'$ and show that there is a smooth deformation between the following states:
\begin{equation}
\begin{tikzpicture}[scale=0.7, baseline={([yshift=-.5ex]current  bounding  box.center)}]
\fill[blue!30] (0.5,0) ellipse (0.7 and 0.2);
\fill[blue!30] (0.5,-0.7) ellipse (0.7 and 0.2);
\fill (0,0) circle(0.1);
\fill (1,0) circle(0.1);
\fill (0,-0.7) circle(0.1);
\fill (1,-0.7) circle(0.1);
\node at (-0.5,0){$\gamma_1$};
\node at (-0.5,-0.7){$\gamma_1'$};
\node at (1.5,0){$\gamma_2$};
\node at (1.5,-0.7){$\gamma_2'$};
\draw [line width=0.7pt, latex-latex] (2,-0.35)--(3,-0.35);
\begin{scope}[xshift=4cm]
\fill[red!50] (0,-0.35) ellipse (0.22 and 0.55);
\fill[blue!30] (1,-0.35) ellipse (0.22 and 0.55);
\fill (0,0) circle(0.1);
\fill (1,0) circle(0.1);
\fill (0,-0.7) circle(0.1);
\fill (1,-0.7) circle(0.1);
\node at (-0.5,0){$\gamma_1$};
\node at (-0.5,-0.7){$\gamma_1'$};
\node at (1.5,0){$\gamma_2$};
\node at (1.5,-0.7){$\gamma_2'$};
\end{scope}
\end{tikzpicture}
\label{eq:inv-2}
\end{equation}
where the ellipses represent that the two Majoranas are paired up. Indeed, consider the following Hamiltonian
\begin{equation}
H(\theta) =\cos \theta \left(-i\gamma_1\gamma_2 -i\gamma_1'\gamma_2'\right) + \sin\theta \left(i\gamma_1\gamma_1' -i\gamma_2\gamma_2'\right).
\label{eq:inv-3}
\end{equation}
When $\theta=0$, the ground state is the one on the left in \eqref{eq:inv-2}; when $\theta=\pi/2$, the ground state is the one on the right of \eqref{eq:inv-2}. Note that the sign of $i\gamma_1\gamma_1'$ is positive in \eqref{eq:inv-3}, represented by a ``red'' color in \eqref{eq:inv-2} for the ground state. (The key here is that the signs in front of $i\gamma_1\gamma_1'$ and $i\gamma_2\gamma_2'$ are opposite; it does not matter which one is positive and which is negative.) It is not hard to find that energy eigenvalues of $H(\theta)$ are 
\begin{equation}
E=\pm \sqrt{2(1-s)},
\end{equation}
where $s=\pm1$ is the eigenvalue of the conserved quantity $\gamma_1\gamma_2\gamma_1'\gamma_2'$. The whole spectrum is independent of $\theta$, and the ground state has energy $-2$. Accordingly, the two states in \eqref{eq:inv-2} are indeed adiabatically connected. Next, we apply this smooth deformation to the whole double chain in an inversion symmetric fashion, and obtain the state at the bottom of Fig.~\ref{fig:inversion}. In this state, the only inter-site entanglement occurs between the two sites near the inversion center. These two sites are viewed as the 0D block.

It remains to calculate the inversion eigenvalue of the 0D-block state. This 0D block can be viewed as a single short Majorana chain with periodic boundary condition. Let us label the Majorana fermions under the following convention
\begin{equation}
\begin{tikzpicture}[scale=0.7, baseline={([yshift=-.5ex]current  bounding  box.center)}]
\fill [gray!20] (2.5, -1.2) rectangle(6.5, 0.7);
%\node at (4.5, -1.4)[scale=0.9]{0D block};
\fill[blue!30] (4.5,0) ellipse (0.7 and 0.2);
\fill[blue!30] (4.5,-0.5) ellipse (0.7 and 0.2);
\foreach \x in {3}
	{
	\fill[blue!30] (\x,-0.25) ellipse (0.22 and 0.45);
	}
\foreach \x in {6}
	{
	\fill[blue!30] (\x,-0.25) ellipse (0.22 and 0.45);
	}
\foreach \x in {3,4, ..., 6}
	{
	\fill (\x,0) circle(0.1);
	\fill (\x,-0.5) circle(0.1);
	}
\foreach \x in { 3, 5}
	{
	\draw [thick](\x,0)--(\x+1,0);
	\draw [thick](\x,-0.5)--(\x+1,-0.5);
	}

\node at (4, 0.4)[scale=0.9]{$\gamma_1$};
\node at (3, 0.4)[scale=0.9]{$\gamma_2$};
\node at (3, -0.9)[scale=0.9]{$\gamma_3$};
\node at (4, -0.9)[scale=0.9]{$\gamma_4$};
\node at (5, 0.4)[scale=0.9]{$\gamma_1'$};
\node at (6, 0.4)[scale=0.9]{$\gamma_2'$};
\node at (6, -0.9)[scale=0.9]{$\gamma_3'$};
\node at (5, -0.9)[scale=0.9]{$\gamma_4'$};
\end{tikzpicture}
\end{equation}
The Majoranas are related to the complex fermions  in the following way
\begin{align}
c_1 & = (\gamma_2 + i\gamma_1)/2, \quad c_2 = (\gamma_3+i\gamma_4)/2 \nonumber\\
c_1' & = (\gamma_2' + i\gamma_1')/2, \quad c_2' = (\gamma_3'+i\gamma_4')/2
\end{align} 
Under inversion symmetry, we have $\gamma_i \rightarrow \gamma_i'$ and $\gamma_i' \rightarrow -\gamma_i$. With the gapping Hamiltonian
\begin{equation}
H = -i\gamma_1\gamma_1' -i\gamma_4\gamma_4' - i\gamma_2\gamma_3-i\gamma_2'\gamma_3',
\end{equation}
it is not hard to find the ground state of the 0D block:
\begin{align}
|\psi\rangle_{\rm 0D} = (c_1^\dag - c_2^\dag & -i c_1'^\dag+ic_2'^\dag  - c_1^\dag c_1'^\dag c_2'^\dag + c_2^\dag c_1'^\dag c_2'^\dag  \nonumber\\
& +ic_1^\dag c_2^\dag c_1'^\dag - ic_1^\dag c_2^\dag c_2'^\dag)|0\rangle
\label{eq:inv-0d-state}
\end{align}
One can easily check that the 0D-block state satisfies
\begin{equation}
I|\psi\rangle_{\rm 0D} = i |\psi\rangle_{\rm 0D}
\end{equation}
Accordingly, two Majorana chains stack into the root state of 0D-block states. 

Hence, the group structure of inversion FSPTs with $I^2=P_f$ is $\mathbb{Z}_8$, agreeing with the classification of 1D $T^2=\mathds 1$ superconductors. 

\begin{table}
\begin{tabular}{|c|c|c|c|}
 \hline  \  Dimension \ & $\quad$ Symmetry $\quad$  &  $\quad \mathscr{G} \quad $ & \quad  Comments \quad  \\
%\hline $C_M^+$ &  $\Z_1$\\
\hline  \multirow{2}{*}{1D} &  $I^2=1$ &  $\Z_2$ &\\
 &  $I^2=P_f$ &  $\Z_8$ & \\
\hline  \multirow{6}{*}{2D} & $C_{M}^-$ &  $\Z_M$ & $M$ odd\\
 & $C_M^-$ &  $\Z_{4M}$ & $M = 2 (\mathrm{mod} \ 4)$ \\
 & $C_M^-$ &  $\Z_{2M}\times \Z_2$ & $M = 0 (\mathrm{mod} \ 4)$ \\
 & $C_M^+$ &  $\quad \Z_{M/2} \quad$ & $M$ even\\
  & $C_M^+$ &  $\quad \Z_{M} \quad$ & $M$ odd\\
\hline
\end{tabular}
\caption{Examples of classifications of 1D and 2D FSPT phases with inversion or rotation symmetry. }
\label{tab12D}
\end{table}

\section{Rotation FSPT phases in 2D}
\label{sec:2D}

We now study 2D rotation FSPTs, both with and without additional Abelian onsite unitary symmetries. We will discuss the case of FSPT phases with $C_M^\pm$ symmetry only in the main text, and present an example of intrinsically interacting fermionic SPT phase with $C_M^\pm\times \Z_2^\mathsf{T}$ symmetry.  The classifications are summarized in Table \ref{tab12D}. We leave the classifications with an additional Abelian internal symmetries to Appendix \ref{app:2d}.

\subsection{$C_M^-$}
\label{sec:2D-CM-minus-1}

\begin{figure}
\centering
\includegraphics{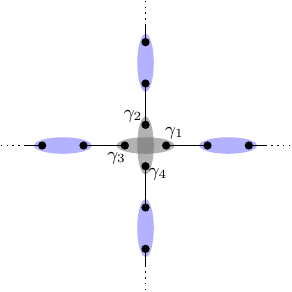}
\caption{Building 1D-block state by gluing half Majorana chains across the origin in a $C_M^-$ symmetric way, for even $M$. We take $M=4$ for illustration. }
\label{fig:rotation-connect}
\end{figure}

\begin{figure*}
\centering
\includegraphics{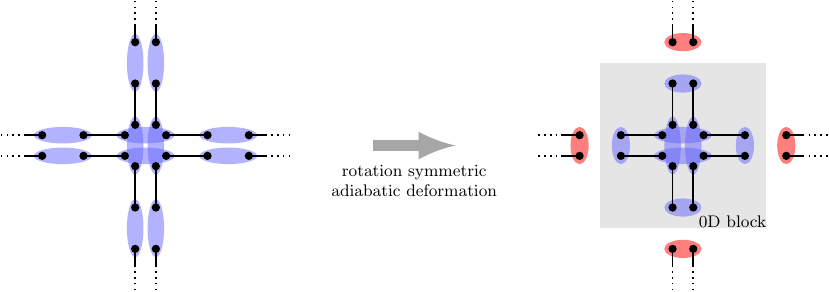}
\caption{Rotation symmetric adiabatic deformation of two stacked 1D-block states. After adiabatic deformation, inter-site entanglement only occurs in the 0D block (gray square). Physical meanings of the graphs are the same as in Fig.~\ref{fig:inversion}. }
\label{fig:1Dblock-stack}
\end{figure*}

Let us begin with FSPTs with $C_M^-$ symmetry only. Accordingly to the general classification scheme in Sec.~\ref{sec:general-principle}, we need to consider (i) possible 0D-block states, (ii) possible 1D-block states and (iii) 2D invertible topological phases that are compatible with $C_M^-$ symmetry.

For 0D-block states, the onsite symmetry group is isomorphic to $\Z_{2M}^f$ with the generator being the rotation $\vr$. There are $2M$ 0D-block states, with the rotation eigenvalue being $1, e^{i\pi/M}, \dots, e^{i(2M-1)\pi/M}$, respectively. These states form a group $\mathbb{Z}_{2M}$.  However, not every 0D-block state represents a distinct FSPT state, i.e, there exists ``trivialization''. Consider a system with a fermion $c_{j,\alpha}$ on each site, where $(j,\alpha)$ is the site index with $j=1,\dots,M$ and $\alpha$ is an additional label. Under rotation $R$, the fermions transform as follows 
\begin{align}
c_{j,\alpha} \rightarrow c_{j+1,\alpha}, \ 1\leq j<M; \quad
c_{M,\alpha} \rightarrow -c_{1,\alpha}
\end{align}
The ``$-$'' sign for transformation of $c_{M,\alpha}$ complies with $R^M=P_f$. It is easy to design a rotation symmetric gapped Hamiltonian such that the ground state is a simple atomic insulator $\prod_{n,\alpha} c^\dag_{n,\alpha}\ket{0}$. Let $\alpha=0$ represent the lattice sites closest to the origin. Then, the 0D-block state is
\begin{equation}
\ket{\psi}_{\rm 0D} = c_{1,0}^\dag c_{2,0}^\dag \dots c_{M,0}^\dag \ket{0}
\label{eqn:psi0D}
\end{equation}
It is easy to see that, 
\begin{equation}
	\vr\ket{\psi}_{\rm 0D}= c_{2,0}^\dag c_{3,0}^\dag \dots (-c_{1,0}^\dag) \ket{0} =(-1)^{M} \ket{\psi}_{\rm 0D}
	\label{eq19}
\end{equation}
Accordingly, when $M$ is odd, this 0D-block state has rotation eigenvalue $-1$. Hence, the 0D-block states reduce to a $\mathbb{Z}_M$ classification. Equation \eqref{eq19} does not lead to trivialization for even $M$. We find no other trivialization processes either. Therefore, the classification of 0D-block states remains $\mathbb{Z}_{2M}$ for even $M$. 

Next, we consider 1D-block states. Consider $M$ semi-infinite 1D lines, arranged in a rotation symmetric way round the origin. One each semi-infinite line, we may have a Majorana chain. Whether it forms a 1D-block state depends on whether the Majorana zero modes at the origin can be gapped out in a rotation symmetric way. When $M$ is odd, there are odd number of Majorana zero modes at the origin. It is obvious that we cannot gap them out. When $M$ is even, one can show that the Majorana zero modes can be gapped out while preserving $\vr$. One just needs to glue all pairs of the half-chains that are opposite to one another (see Fig.~\ref{fig:rotation-connect}). More specifically, denote the Majorana zero modes at the origin by $\gamma_1, \gamma_2, \dots, \gamma_M$. Under rotation $\vr$,
\begin{equation}
\gamma_j \rightarrow \gamma_{j+1}, \ 1\leq j<M; \quad
\gamma_M \rightarrow -\gamma_1
\label{eq:rot_CM-1}
\end{equation}
Then, the zero modes can be gapped out by the following Hamiltonian
\begin{equation}
H = -i\sum_{j=1}^{M/2}\gamma_{j}\gamma_{j+M/2}
\end{equation}
Moreover, $H$ is symmetric under the transformation \eqref{eq:rot_CM-1}. Hence, we obtain a rotation symmetric 1D-block state. 

Does the 1D-block state extend the 0D-block states? To obtain the group structure of FPSTs, we consider stacking two 1D-block states (the left panel of Fig.~\ref{fig:1Dblock-stack}). On each axis, we have a double Majorana chain. We apply a similar adiabatic deformation as in Fig.~\ref{fig:inversion}, using a Hamiltonian like \eqref{eq:inv-3}. It is not hard to see that the stacked 1D-block state can be deformed to the state on the right side of Fig.~\ref{fig:1Dblock-stack}. An important feature of that state is that inter-site entanglement only occurs in the neighborhood of the origin. We choose these sites as the 0D block. Then, we need to calculate the rotation eigenvalue of this 0D-block state. The calculation is very similar to that for 1D inversion FSPTs. In fact, the 0D-block state here is $M/2$ copies of the 0D-block state in Fig.~\ref{fig:inversion} [given in Eq.~\eqref{eq:inv-0d-state}]. With this understanding, we find
\begin{equation}
\vr|\psi\rangle_{\rm 0D} = e^{i\frac{\pi}{2}(M-1)} |\psi\rangle_{\rm 0D} \nonumber
\end{equation}
where $|\psi\rangle_{\rm 0D}$ denotes the 0D-block state in Fig.~\ref{fig:1Dblock-stack}. That is, the 0D-block state is nontrivial. However, the rotation eigenvalue can be modified, if we stack a 0D-block state to the original 1D-block state before stacking. If a 0D-block state with a rotation eigenvalue $e^{i\pi p /M}$ is attached to each 1D-block state, the rotation eigenvalue $r$ of the 0D-block state in Fig.~\ref{fig:1Dblock-stack} becomes 
\begin{equation}
r=e^{i\frac{\pi}{2}(M-1)+ i \frac{2\pi p}{M}} \nonumber
\end{equation}
There are two cases: (a) when $M=0\ ({\rm mod}\ 4)$, we can take $p= \frac{1}{4}(1-M)M$ such that $r=1$; (b) when $M=2 \ ({\rm mod} \ 4)$, there exists no integer $p$. Therefore, when $M$ is a multiple of 4, there is an appropriate 1D-block state which itself forms a $\mathbb{Z}_2$ structure under stacking. When $M$ is an odd multiple of 2, 1D-block states extend the 0D-block states, and all together they form a $\mathbb{Z}_{4M}$ group. 

Finally, we need to consider 2D invertible topological phases that are compatible with $C_M^-$ symmetry. 2D fermionic invertible topological phases are generated by the $p_x\pm ip_y$ states. Conventionally, they are not considered as FSPTs since they are topologically nontrivial even in the absence of any symmetries. We show in Appendix \ref{append:p+ip} that $p_x\pm ip_y$ are compatible with $C_M^-$ symmetry. However, since these states form the group $\mathbb{Z}$ which is of infinite order, they can never extend the 0D- and 1D-block state.  The fact that $p_x\pm ip_y$ superconductors are only compatible with $C_M^-$ symmetry will be important in our later discussions.

Combining these results together, FSPTs with $C_M^-$ symmetry are classified by the following groups under stacking
\begin{equation}
	\mathscr{G}= \left\{
\begin{array}{ll}
\mathbb{Z}_M, & \text{$M$ is odd}\\[3pt]
\mathbb{Z}_{4M}, & M=2 \ ({\rm mod} \ 4)\\[3pt]
\mathbb{Z}_{2M}\times \mathbb Z_2, & M=0 \ ({\rm mod} \ 4)
\end{array}
\right.
\end{equation}
This agrees with the classification of 2D FSPTs with onsite unitary symmetry $\mathbb{Z}_2^f\times \mathbb{Z}_M$.

\subsection{$C_M^+$}
\label{sec:2dcm+}
Let us first consider 0D block state, corresponding to linear representations of $\Z_2^f\times \Z_M$ group. Naively the rotation eigenvalues are $e^{\frac{i 2\pi l}{M}}$ where $l=0,1,\dots, M-1$. However, we should again consider a state like Eq. \eqref{eqn:psi0D} which may trivialize some of the rotation eigenvalues. Indeed, we have
\begin{equation}
	\vr\ket{\psi}_\text{0D}=(-1)^{M-1} \ket{\psi}_\text{0D}.
	\label{}
\end{equation}
Thus for even $M$, the rotation eigenvalue $-1$ in fact corresponds to a trivial phase. Thus we obtain $\Z_{M/2}$ classification. For odd $M$, the classification remains $\Z_M$.

\begin{figure*}
\centering
\includegraphics{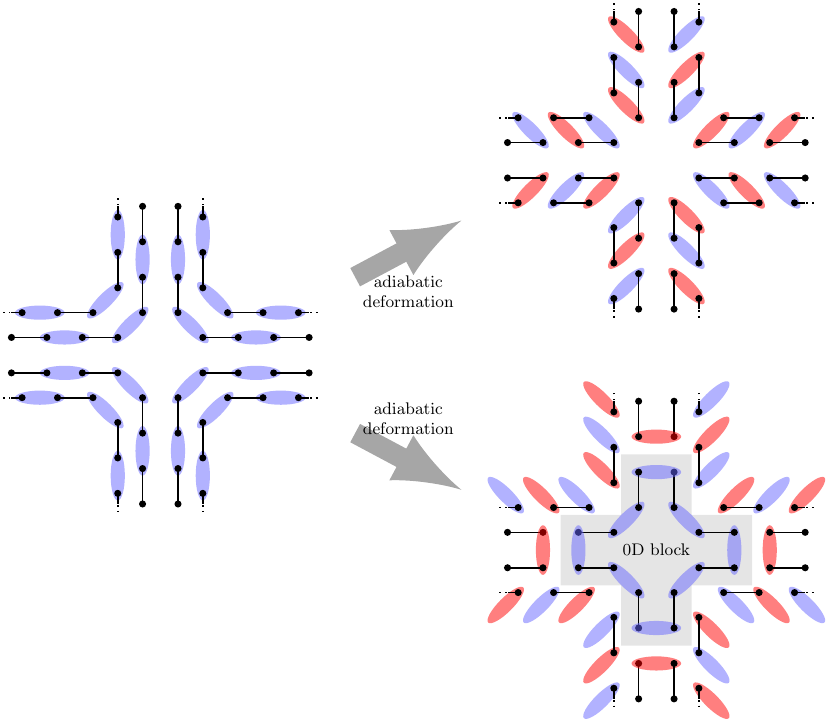}
\caption{Two smooth deformations of the state on the left. The state on the top right can be further deformed, in a rotation symmetric way, to the trivial product state. The 0D-block state on the bottom right is a short Majorana chain, arranged rotation symmetrically around the origin, which is known to have an odd fermion parity. Physical meanings of the graphs are the same as in Fig.~\ref{fig:inversion}. }
\label{fig:2DMajoranaChains}
\end{figure*}

The 0D-block state with odd fermion parity can always be trivialized. It is obvious for odd $M$. The atomic insulator state in \eqref{eqn:psi0D} has odd fermion parity. For even $M$, it is somewhat less obvious.  Consider a state of $M$  Majorana chains, arranged in a rotation-symmetric configuration, as illustrated in Fig. \ref{fig:2DMajoranaChains} for $M=4$. We will assume that the system is infinite. On the one hand, this state can be adiabatically deformed to a trivial state, by disentangling two neighboring chains. On the other hand, there exists another adiabatic deformation, namely choosing an alternative ``dimerization'' pattern when disentangling neighboring chains, which removes the entanglement between fermions in a 0D block and those sitting outside. The remaining state in the 0D block of Fig.~\ref{fig:1Dblock-stack} is nothing but a single Majorana chain with \emph{periodic} boundary condition. (Instead, if we had $C_M^-$ symmetry, the boundary condition would be \emph{anti-periodic}.) It is a well-known fact that the ground state of a Majorana chain with periodic boundary condition has odd fermion parity. Combining the two adiabatic deformations, it proves that the 0D block state with odd fermion parity is indeed trivial. We note that in a finite system, one can start from a product state, and adiabatically deform it into a state with odd fermion parity in the origin, and a Majorana chain sitting on the boundary.

Now we turn to 1D block states.  Again consider $M$ semi-infinite Majorana chains meeting at the rotation center. To construct a SPT phase, one must be able to gap out the $M$ Majorana zero modes in a rotationally invariant way. We can prove that this is impossible. For odd $M$ this is obvious, so we will assume $M$ is even. Denote the Majorana zero modes by $\gamma_j, j=1,2,\dots, M$. Rotation acts on them by $\vr:\gamma_j\rightarrow \gamma_{j+1}$. Consider the fermion parity near the rotation center. Without loss of generality we have
\begin{equation}
	P_f = i^{M/2}\prod_{j=1}^M \gamma_j.
	\label{}
\end{equation}
It is easy to show that
\begin{equation}
	\vr P_f\vr^{-1}=i^{M/2}\gamma_2\cdots \gamma_M\gamma_1 = (-1)^{M-1} P_f=-P_f.
	\label{eq:proj2d}
\end{equation}
The anticommutation between $\vr$ and $P_f$ forbids a non-degenerate ground state.

Lastly we consider 2D block states. In this case, we find that $p_x\pm ip_y$ superconductors (or any state with an odd Chern number) are \emph{not} compatible with the $C_M^+$ symmetry. Therefore only those with even Chern numbers are allowed. Interestingly, there is a way around this obstruction: if the system has a (single) Majorana mode at the rotation center, then one can realize a $p_x\pm ip_y$ superconductor with $C_M^+$ symmetry. We will elaborate more on this in Sec. \ref{sec:sptlsm}.

To conclude, we have found that the classification is 
\begin{equation}
	\mathscr{G}= \begin{cases}
		\Z_{M/2} & M\text{ is even}\\
		\Z_M & M\text{ is odd}
	\end{cases}
	\label{}
\end{equation}
All these states are characterized by an ``angular momentum'', i.e. rotation eigenvalues. In a sense they are all ``bosonic'' SPT phases.

\subsection{$C_M\times \Z_2^\mathsf{T}$}
In this section we consider time-reversal symmetry in the BDI class, i.e. $\mathsf{T}^2=\mathds{1}$. We will not attempt to give a full classification, but rather focus on an example of interacting intrinsically fermionic SPT phase protected by $C_M$ and $\Z_2^\mathsf{T}$. 

We will consider a 1D block state. The blocks are 1D class BDI topological superconductor, each consisting of $\nu$ Majorana chains~\cite{Fidkowski2011}. Since the interacting classification for BDI superconductors is $\Z_8$, interactions can gap out the end states of 8 Majorana modes. Therefore, we may construct a time-reversal invariant SPT state when $M\nu$ is a multiple of $8$. It remains to check if the gapping interactions can be $C_M$ symmetric. If so, the states obtained this way are only enabled by strong interactions at the rotation center.

Here we study $M=4$, $\nu=2$ in detail. We show that there indeed exist $C_M$ invariant interactions which gap out the Majorana zero modes in the rotation center. The edge mode of the 1D BDI superconductor is a complex fermion. We have four edge modes in the rotation center, $c_j$ with $j=1,2,3,4$. They transform under the time-reversal symmetry as $c_j\rightarrow c_j^\dag$.  Under $C_4^\pm$ they transform as
\begin{align}
	& c_j\rightarrow c_{j+1},\  j=1,2,3\nonumber \\ 
	& c_4\rightarrow sc_1.
	\label{}
\end{align}
where $s=\pm 1$ corresponding to $C_4^\pm$.

Let us denote $n_j=c_j^\dag c_j$. First let us add the following interaction to the Hamiltonian:
\begin{equation}
	\begin{split}
		H_1= U\left[(n_1-\frac{1}{2})(n_3-\frac{1}{2})+(n_2-\frac{1}{2})(n_4-\frac{1}{2})\right].
	\end{split}
	\label{}
\end{equation}
with  $U>0$. With this interaction, one and only one of the two fermions $1$ and $3$ is occupied. It is the same for fermions $2$ and $4$. Then, there is a two-fold ground-state degeneracy from $1$ and $3$, which can be viewed as a spin-$1/2$ degree of freedom. So is it for $2$ and $4$. We will denote the two spins by $\bm{\tau}_{13}$ and $\bm{\tau}_{24}$. More precisely,
\begin{equation}
	\tau_{ab}^\mu = (c_{a}^\dag,c_b^\dag) \sigma^\mu 
	\left(
	\begin{matrix}
	c_{a}\\
	c_b
	\end{matrix}
	\right)	
	\label{}
\end{equation}
where $(a,b)=(1,3)$ or $(2,4)$, and $\sigma^\mu$ with $\mu=x,y,z$ are Pauli matrices. Under $C_4$, they transform as
\begin{equation}
	\begin{gathered}
	\tau_{13}^\mu\rightarrow \tau_{24}^\mu\\
	\tau_{24}^{x}\rightarrow s\tau_{13}^{x}, \ \tau_{24}^{y}\rightarrow -s\tau_{13}^{y}, \ \tau_{24}^z\rightarrow -\tau_{13}^z, \quad 
	\end{gathered}
	\label{}
\end{equation}
Under the time-reversal symmetry, they transform as
\begin{equation}
	\tau_{ab}^\mu\rightarrow -\tau_{ab}^\mu.
	\label{}
\end{equation}
Now we further add an additional term $H_2$ to lift the degeneracy and obtain a unique ground state. For $s=1$, we add
\begin{equation}
	H_2=J(\tau_{13}^x\tau_{24}^x + \tau_{13}^y\tau_{24}^z - \tau_{13}^z\tau_{24}^y)
	\label{}
\end{equation}
and for $s=-1$, we add
\begin{equation}
	H_2=J(\tau_{13}^y\tau_{24}^y + \tau_{13}^x\tau_{24}^z - \tau_{13}^z\tau_{24}^x)
	\label{}
\end{equation}
where $J$ is positive. It is easy to check that the Hamiltonian $H_2$ preserves all symmetries.

Notice that our Hamiltonian preserves the $\U$ symmetry. 
Therefore we can also view the system as a topological insulator in class AIII, with $\U\times\Z_2^\mathsf{T}$ symmetry (i.e. time-reversal acts as particle-hole transformation). An example of such 1D topological insulator is the famous Su-Schriffer-Heeger chain.

We can also check that the 1D block construction goes through for $M=8, \nu=1$ for both $C_M^\pm$.
%\begin{equation}
%	-\sum_{i}(c_{Ai}^\dag c_{Bi} + c_{Bi}^\dag c_{A,i+1} + \text{h.c.}).
%\end{equation}
%Here $c_A\rightarrow c_A^\dag, c_B\rightarrow -c_B^\dag$

\section{Rotation FSPT phases in 3D}
\label{sec:3D}

We study 3D FSPT phases with $C_M^\pm \times G$ in this section, where $G$ is a finite Abelian unitary group. It is known that with only finite unitary symmetry group, there are no free FSPT phases in 3D~\cite{Ludwig_review,ChenArxiv2018}. We will see that this is true with rotation symmetry as well, as expected from the crystalline equivalence principle. We will construct a series of FSPT phases which are all stabilized by interactions.

For 3D systems, we should consider (i) 3D internal FSPT phases. (ii) 2D block states and (iii) 1D block states on the rotation axis. 

Let us first consider 3D block states. These are internal FSPT phases with $\Z_2^f\times G$ symmetry. Recently complete classifications of such phases have been proposed in \Refs{KapustinThorngren, WangPRX2018}, extending an earlier partial classification based on group super-cohomology~\cite{GuPRB2014}. We conjecture that all 3D internal FSPT phases with Abelian unitary symmetries are consistent with the $C_M^-$ symmetry. More specifically, as we will argue later (see Sec. \ref{sec:sptlsm}), the ``beyond-supercohomology'' FSPT phases, or Majorana-decorated SPT phases, are only compatible with $C_M^-$ symmetry. We conjecture that the group supercohomology phases can be compatible with both $C_M^\pm$ symmetries.

Let us comment on the general strategy to study 2D block states. It turns out that for our purpose, all relevant 2D block states have free fermion realizations, so we can easily obtain low-energy theories of 1D gapless edges. 
They are either Majorana or Dirac fermions. Let us set up some notations that will be used throughout the section. The chiral Majorana edge mode of a $p_x+ip_y$ superconductor is denoted by $\gamma$, and the Majorana edge mode of a $p_x-ip_y$ superconductor is denoted by $\ol{\gamma}$ which has opposite chirality to $\gamma$. The chiral Dirac edge mode of a Chern insulator with Chern number $C=1$ is denoted by $\psi$, while $\ol{\psi}$ denotes the chiral Dirac edge modes for $C=-1$.

In order to build a fully gapped bulk phase, we demand that these edges can be gapped out without breaking the internal symmetries or the $C_M$ symmetries. Once we focus on the edges, we can imagine ``folding'' all the half planes so that they can be treated as a multi-layer system, and the rotation acts as a cyclic permutation of layers, i.e. an internal symmetry. Then the requirement is that the multi-layer system is a trivial SPT phase under all the symmetries.

The 1D block states are 1D FSPT phases with only internal symmetries, as $C_M^\pm$ acts as internal symmetry on the rotation axis. Another issue one needs to deal with is the possible group extension of $\mathscr{G}_{-1}$ by $\mathscr{G}_{-2}$. Some 2D block states may stack into 1D block states, leading to nontrivial group extension of $\mathscr{G}_{-1}$.

In the main text we consider the classifications of FSPT phases with $C_M^\pm$ symmetry only, and those with $C_M^-\times \Z_N$ symmetry, to highlight the main technicalities and the subtleties that may arise. To simplify our discussions, we will assume that the orders of groups are all powers of $2$, i.e., 
\begin{equation}
	M=2^m, \ N=2^n.
	\label{}
\end{equation}
with $m\ge 1$ and $n\ge0$. We have also considered other Abelian internal symmetries and the details can be found in Appendix \ref{app:more3D}. We summarize the classification s for $C_M^\pm\times \Z_N$ in Table \ref{tab1} and \ref{tab2}, which we believe are complete.

\begin{comment}
\begin{tabular}{|c|cc|c|c|c|}
\hline
& \multicolumn{2}{|c|}{$\mathscr{G}_{-2}$ } & \multicolumn{2}{|c|}{$\mathscr{G}_{-1}$ } & $\mathscr{G}$ \\
\cline{4-5}
\hline
\multirow{2}{*}{$m\geq n$} & \multicolumn{2}{c|}{$\Z_N^\text{B}$} & %
    $n=1$ & $n\geq 2$ \\
	\cline{4-5}
	& & & $\begin{cases} \Z_M^\text{B} & m=1 \\ \Z_M & m=2\\ \Z_8^\text{NA} & m\geq 3 \end{cases}$ & $\begin{cases} \Z_2^{\text{NA}}\times\Z_N & m=n \\ \Z_2^{\text{NA}}\times\Z_{2N} & m>n \end{cases}$ \\
% & \multicolumn{2}{c|}{Value} & \multicolumn{2}{c|}{Value} & \\
%\cline{2-5}
 %&  &  & $n=1$ & $n\geq 2$ & \\
\hline
\multirow{2}{*}{$m< n$} & \multicolumn{2}{c|}{$\Z_{2M}$} & %
    $n=2$ & $n> 2$ \\
	\cline{4-5}
	& & & $\Z_2^\text{B}$ & $\Z_2^\text{NA}\times\Z_M$\\
\hline
% etc. ...
\end{tabular}
\end{comment}

\begin{table*}
\begin{tabular}{|c|c|c|c|}
\hline & $\mathscr{G}_{-2}$ & $\mathscr{G}_{-1}$ & $\mathscr{G}$\\
\hline 
%$n=0$ & & $\Z_1$ & $\Z_1$ \\
$n=1$ & \multirow{5}{*}{$\quad \Z_{(2M,N)} \quad $}  & $\Z_{(M,8)}$ & $\Z_2\times \Z_{(M,8)}$ \\
$m=1,n=2$ & &  $\Z_2$ & $\Z_4\times \Z_2$ \\
$m=1,n\ge3$ & & $\Z_{2}\times \Z_2^{\rm NA}$ & $\Z_{2}\times \Z_8^{\rm NA}$\\
$\quad m\ge2, n\ge2\quad$ &  & $\quad \Z_{(M, 2N)}\times \Z_2^{\rm NA}\quad $ & $\quad \Z_{(2M,N)}\times \Z_{(M, 2N)}\times \Z_2^{\rm NA}\quad $\\
\hline
\end{tabular}

\begin{tabular}{|c|c|}
\hline  $\quad$ Symmetry $\quad$  &  $\mathscr{G}$  \\
\hline $C_2^-\times \Z_2$ &  $\Z_2\times \Z_2$\\
\hline $C_4^-\times \Z_2$ &  $\Z_2\times \Z_4$\\
\hline $C_{8}^-\times \Z_2$ &  $\Z_2\times \Z_8^{\rm NA}$\\
\hline $C_{16}^-\times \Z_2$ &  $\Z_2\times \Z_8^{\rm NA}$\\
\hline $C_2^-\times \Z_4$ &  $\Z_2\times \Z_4$\\
\hline $C_4^-\times \Z_4$ &  $\Z_4\times \Z_4\times\Z_2^{\rm NA}$\\
\hline $C_8^-\times \Z_4$ &  $\Z_4\times \Z_8 \times \Z_2^{\rm NA}$\\
\hline $C_{16}^-\times \Z_4$ &   $\Z_4\times \Z_8 \times \Z_2^{\rm NA}$\\
\hline $C_2^-\times \Z_8$ &  $\Z_2\times \Z_8^{\rm NA}$\\
\hline $C_4^-\times \Z_8$ &  $\Z_4\times \Z_8\times \Z_2^{\rm NA}$\\
\hline $C_8^-\times \Z_8$ &  $\quad \Z_8\times \Z_8\times \Z_2^{\rm NA}\quad $\\
\hline
\end{tabular}
\caption{\emph{Upper}: Classification of 3D FSPT phases with $C_M^-\times \Z_N$ symmetry, where $M=2^m, N=2^n$ with $m, n\ge 1$. The superscript ``NA''  means that the generator is a non-Abelian (Majorana) FSPT phase. This classification should be the same as that with the internal symmetry $\Z_2^f\times \Z_M \times \Z_N$.   \emph{Lower}: a few examples.}
\label{tab1}
\end{table*}

\begin{table}
\begin{tabular}{|c|c|}
\hline  $\quad$ Symmetry $\quad$  &  $\quad \mathscr{G} \quad $ \\
%\hline $C_M^+$ &  $\Z_1$\\
\hline $C_2^+\times \Z_2$ &  $\Z_2$\\
\hline $C_{4}^+\times \Z_2$ &  $\Z_4$\\
\hline $C_{8}^+\times \Z_2$ &  $\Z_4$\\
\hline $C_2^+\times \Z_4$ &  $\Z_2$\\
\hline $C_4^+\times \Z_4$ &  $\Z_2\times \Z_4$\\
\hline $C_8^+\times \Z_4$ &  $\quad \Z_2\times \Z_8 \quad$\\
\hline
\end{tabular}
\caption{Classification of 3D FSPT phases with $C_M^+\times \Z_{N}$ symmetry, where $M=2^m, N=2^n$ with $m, n\ge 1$. The general formula is $\mathscr{G}=\Z_{(M,N)/2}\times \Z_{(M, 2N)}$ (see Appendix \ref{app:more3D2} for detailed discussions). Note that none of them are Majorana FSPT phases. }
\label{tab2}
\end{table}

\subsection{$C_M^-$}
When only the rotation symmetry is present, there is no need to consider 3D block states.  For 1D block states, the rotation symmetry becomes an internal symmetry $\Z_{2M}^f$. This case is covered by Sec. \ref{sec:3Dm1d} below. It turns out that there is no nontrivial 1D block states. Furthermore, we do not have to consider 2D block states, since they would have to be a class D topological superconductor which is classified by $\Z$. There is no way that their \emph{chiral} edge modes near the rotation axis can be gapped out, and thus  can not be used to build 2D block states. In conclusion, we find that there are no nontrivial FSPT phases in this case.

\subsection{$C_{M}^- \times \Z_{N}$}
\label{sec:CMmZN}
We now study FSPT phases with $C_M^-\times\Z_N$ symmetry. All nontrivial FSPT phases found here are enabled by strong interactions. The classification is summarized in Table. \ref{tab1}.

%\begin{table}
%\begin{tabular}{|c|cc|c|c|}
%\hline
%& \multicolumn{2}{|c|}{$\mathscr{G}_{-2}$ } & \multicolumn{2}{|c|}{$\mathscr{G}_{-1}$ }  \\
%\cline{4-5}
%\hline
%\multirow{2}{*}{$m\geq n$} & \multicolumn{2}{c|}{$\Z_N^\text{B}$} & %
%    $n=1$ & $n\geq 2$ \\
%	\cline{4-5}%
%	& & & $\begin{cases} \Z_M^\text{B} & m=1 \\ \Z_M & m=2\\ \Z_8^\text{NA} & m\geq 3 \end{cases}$ & $\begin{cases} \Z_2^{\text{NA}}\times\Z_N & m=n \\ \Z_2^{\text{NA}}\times\Z_{2N} & m>n \end{cases}$ \\
% & \multicolumn{2}{c|}{Value} & \multicolumn{2}{c|}{Value} & \\
%\cline{2-5}
 %&  &  & $n=1$ & $n\geq 2$ & \\
%\hline
%\multirow{2}{*}{$m< n$} & \multicolumn{2}{c|}{$\Z_{2M}$} & %
%    $n=2$ & $n> 2$ \\
%	\cline{4-5}
%	& & & $\Z_2^\text{B}$ & $\Z_2^\text{NA}\times\Z_M$\\
%\hline
% etc. ...
%\end{tabular}
%\caption{Classification for $C_M^-\times \Z_N$, where $M=2^m, N=2^n$. Superscript B means the generator is a bosonic phase, and NA means the generator is a non-Abelian (Majorana) SPT phase.}
%\label{tab1}
%\end{table}

\subsubsection{1D block states}
\label{sec:3Dm1d}
In a 1D block state, the internal symmetry on the rotation axis is $\Z_{2M}^f\times \Z_N$. Hence, we need to deal with 1D FSPT phases with $\Z_{2M}^f\times \Z_N$ internal symmetry, which we review in Appendix \ref{append:review}. The ``bosonic'' part of the symmetry group is $G_b=\Z_M\times\Z_N$. It is extended nontrivially by $\Z_2^f$. The group extension can be described by a 2-cocycle $\nu\in \mathcal{H}^2[G_b, \mathbb{Z}_2]$. We label the group elements of $G_b$ by $a=(a_1,a_2)$ where $a_1\in \Z/M\Z, a_2\in \Z/N\Z$, and group multiplication is denoted additively. Then, an  explicit representation of $\nu$ is
\begin{equation}
	\nu(a, b) = \frac{a_1+b_1-[a_1+b_1]_{M}}{M}.
	\label{}
\end{equation}
where $[...]_M$ denotes ``modulo $M$''.

As reviewed in Appendix \ref{append:review}, we have three types of 1D FSPT phases: (i) the Majorana chain, (ii) FSPT phases described by $\mu\in \mathcal H^1[G_b, \mathbb{Z}_2]$ and (iii) bosonic SPT phases described by an element in $\mathcal{H}^2[G_b, U(1)]$. It is known from Ref.~\onlinecite{Fidkowski2011} that the Majorana chain is not compatible with a nontrivial $\nu$, which is exactly our case. So, we do not need to consider the Majorana chain. Next,  we pick a $\mu\in \mathcal{H}^1[G_b, \Z_2]$ for the second type of FSPT phases (i.e., $\mu$ is a homomorphism $G_b\rightarrow \mathbb{Z}_2$). However, not all $\mu$'s are valid. Mathematically, $\mu$ leads to an FSPT phase only if the so-called the obstruction class $[O]\in \H^3[G_b, \U]$ is a trivial 3-cocycle (see Appendix \ref{append:review}). More explicitly, $O$ is given by 
\begin{equation}
	\begin{split}
	O(a,b,c) = &\frac{1}{2}\mu(a)\nu(b,c) \\
	        = & \frac{\mu( e_1 )}{2M} a_1(b_1+c_1-[b_1+c_1]_M) \\ 
			&+  \frac{\mu( e_2 )}{2M} a_2(b_1+c_1-[b_1+c_1]_M)
	\end{split}
	\label{}
\end{equation}
where  $e_1=(1,0) $ and $ e_2=(0,1)$. To determine whether $[O]$ is trivial or not, we compute the invariants for group cohomology classes defined in \Ref{WangPRB2015}:
\begin{equation}
	\begin{gathered}
	\Theta_1 = \pi \mu(e_1), \quad \Theta_2=0,\\
	\Theta_{12}=\frac{\pi N}{(M,N)}\mu(e_2)
	\end{gathered}
	\label{}
\end{equation}
where $(M, N)$ denotes the greatest common divisor of $M$ and $N$. It was shown in Ref.~\onlinecite{WangPRB2015} that $[O]$ is trivial if and only if all the above invariants vanish. Therefore, we must have $\mu(e_1)=0$ and
\begin{equation}
	\mu(e_2)=
	\begin{cases}
		0, 1 & \text{if } m<n\\
		0 & \text{if } m\geq n
	\end{cases}
	\label{}
\end{equation}
For the nontrivial case $\mu(e_2)=1$ when $m<n$, we also have to consider bosonic SPT phases. They are classified by $\H^2[\Z_M\times\Z_N, \U]=\Z_{(M,N)}=\Z_M$. It can be shown that the FSPT phases and bosonic SPTs together form a $\mathbb{Z}_{2M}$ group under stacking. Hence, the overall classification is given by $\Z_{2M}$ with a root fermionic SPT when $m<n$, and $\Z_{(M,N)}=\Z_N$ for $m\ge n$ with all SPTs being bosonic. Putting the two cases together, the classification of codimension-2 SPT phases is given by $\mathscr{G}_{-2} = \mathbb{Z}_{(2M, N)}$. These 1D block states cannot be trivialized.

We note that none of the 1D SPT phases with $1\leq m<n$ can be realized in non-interacting systems. The reason is that for free fermions, $\Z_{2M}^f$ with $M$ even is automatically enhanced to a $\U$ symmetry. We can then diagonalize the single-particle Hamiltonian according to eigenvalues of the $\Z_N$ symmetry, and in each subspace with a given eigenvalue, the single-particle Hamiltonian falls in class A, which has no non-trivial states in 1D. An interacting commuting projector model for $\Z_4^f \times \Z_4$ 1D SPT phase is recently presented in \Ref{TantivasadakarnPRB2018}.

\subsubsection{2D block states for $n=1$}
\label{sec:2db1}

We now turn to 2D block states. The situations for $n=1$ and $n>1$ are different. We discuss the $n=1$ case in this subsection, and leave the latter to the next subsection. On the half planes are 2D FSPT phases with $\Z_2$ symmetry, whose classification is reviewed in Appendix \ref{append:review}. The key question is whether we can glue the 2D blocks at the rotation axis by gapping out the edge modes without breaking the symmetries.

To see whether the gapless edge modes can be gapped out preserving all symmetries, we fold the 2D blocks to a multi-layer system. Then, the rotation symmetry becomes the cyclic permutation of layers, i.e., an internal $\mathbb{Z}_{2M}^f$ symmetry. The gappability of the edge modes is equivalent to that the multi-layer state is topologically trivial. This can be checked by computing the topological response through inserting symmetry defects with fluxes corresponding to $\mb{g}$ and $\vr$, the generators of $\mathbb{Z}_2$ and $C_M^-$ respectively. The SPT phase is trivial if and only if the topological response of these defects are ``trivial''. The precise meaning of ``triviality'' will be clear below. More specifically,  according to Ref.~\onlinecite{WangPRB2017}, the SPT phase with $\mathbb{Z}_{2M}^f\times\mathbb{Z}_2$ symmetry is trivial as long as the topological twists of the $\mb{g}$ and $\vr$ are trivial, which we calculate now.

To do that, we first describe how the edge modes transform under the symmetries. 2D FSPT phases with internal $\Z_2$ symmetry are classified by $\Z_8$. The edge of a nontrivial SPT phase contains $\nu$ pairs of counter-propagating Majornana modes, $\gamma_a$ and $\ol{\gamma}_a$, where $a=1,\dots,\nu$ and $\nu=1,2,\dots,7$. Since we have $M$ copies of 2D blocks, there are $M\nu$ pairs of Majorana modes, $\gamma_{j,a}$ and $\ol{\gamma}_{j,a}$, with $j=1,\dots, M$. Under rotation $\vr$, they transform as
\begin{align}
	&\gamma_{j,a}\rightarrow \gamma_{j+1,a},  \quad 1\leq j\le M-1\nonumber\\
	&\gamma_{M,a}\rightarrow -\gamma_{1,a},
	\label{eq:r-sym}
\end{align}
The transformations of $\{\ol{\gamma}_{j,a}\}$ are the same. At the same time, under the $\mathbb{Z}_2$ symmetry $\mb{g}$, they transform as
\begin{equation}
\gamma_{j,a} \rightarrow -\gamma_{j,a}, \quad \ol{\gamma}_{j,a} \rightarrow \ol{\gamma}_{j,a}
\label{eq:z2sym}
\end{equation}
That is, $\gamma_{j,a}$ is neutral while $\ol{\gamma}_{j,a}$ is charged under $\mathbb{Z}_2$ symmetry.

To proceed, we transform the Majorana fields to an eigenbasis of the rotation symmetry. We define the following Dirac fermion modes
\begin{equation}
	\begin{split}
\psi_{l,a}&=\sum_{j=1}^{M} \omega^{-lj} \gamma_{j,a }, \quad
\bar{\psi}_{l,a}=\sum_{j=1}^{M} \omega^{-lj} \bar{\gamma}_{j,a}, 
	\end{split}
	\label{cfermion}
\end{equation}
where $\omega= e^{\frac{i\pi}{M}}$ and $l$ is an odd integer. Without loss of generality, we take $l=1,3,\dots, M-1$ and there are $M/2$ distinct values of $l$. It is straightforward to check these fermions satisfy the anti-commutation relations of Dirac fermions. Under rotation  $\vr$, 
\begin{align}
	\psi_l\rightarrow & \sum_{j=1}^{M-1} \omega^{-lj}\gamma_{j+1} - \omega^{-lM}\gamma_1 \nonumber\\
	&=\sum_{j=2}^{M}\omega^{-l(j-1)}\gamma_j + \gamma_1\nonumber\\
	&=\omega^{l}\sum_{j=1}^{M}\omega^{-lj}\gamma_j\nonumber\\
	&=\omega^{l}\psi_l,
	\label{cf-transformation}
\end{align}
where we have omitted the index $a$ for clarity. It is the same for the symmetry transformation of $\ol{\psi}_{l,a}$. Accordingly, in this basis,  all Dirac fermion modes transform diagonally under the rotation $\vr$. It is obvious that the Dirac fermions also transform diagonally under the $\mathbb{Z}_2$ symmetry $\mb{g}$.

To check whether the SPT phase is trivial or not, we calculate the topological spins corresponding to $\mb{g}$ and $\vr$ fluxes. A basic fact that will be used (and repeatedly used later) is the following: in a Chern insulator with Chern number $C$, a $2\pi\phi$ flux, where $\phi$ is a rational number, has a topological twist factor $e^{{\pi i}C\phi^2}$. In this section (as well as continuations in Appendix \ref{app:more3D}), we denote topological twist factor as $e^{2\pi i h}$ where $h$ is the topological spin. According to \eqref{eq:z2sym}, the $\mb{g}$ flux corresponds to the collection of $\pi$ fluxes from the ${\psi}_{l,a}$ modes, each corresponding to a $C=1$ Chern insulator. Hence, we have
\begin{equation}
	h_\mb{g}= \frac{\nu M}{16} 
	\label{}
\end{equation}
The triviality of the multi-layer SPT phase requires that $2h_\mb{g}\equiv 0$ modulo integers. The factor of 2 follows from the fact that $\mb{g}$ is an order-2 group element (c.f. Refs.~\onlinecite{ChengPRB2018,WangPRB2017}). Hence, $\nu M$ should be a multiple of $8$. This is just saying that $M$ copies of the 2D SPT states must be trivial. To calculate the topological spin of the $\vr$ flux, we notice that $\psi_{l,a}$ and $\ol{\psi}_{l,a}$ transform in the same way under $\vr$. However, they contribute oppositely to the topological spin of $\vr$. Accordingly, we have
\begin{equation}
h_{\vr} = 0
\end{equation}
Hence, the overall requirement for the multi-layer SPT to be trivial is $M\nu \equiv 0 \ {(\rm mod \ 8)}$.

%To compute $\theta_{\mb{g}\vr}$, we notice that $\mb{g}\vr$ acts as
%\begin{equation}
%	\psi_{la}\rightarrow -\omega_M^l\psi_{la}, \bar{\psi}_{la}%\rightarrow \omega_M^l\bar{\psi}_{la}.
%	\label{}
%\end{equation}
%It follows that
%\begin{equation}
%	\begin{split}
%		h_{\mb{g}\vr}&= \frac{\nu}{2}\left(\sum_l\left(\frac{l+M}%{2M}\right)^2 - \sum_l\left(\frac{l}{2M}\right)^2\right)\\
%		&=\frac{\nu}{4}\sum_l \left( \frac{l}{M}+\frac{1}{2} \right)\\
		%&=\frac{\nu}{4} \left( \frac{1}{M} M\frac{M-1}{4} + \frac{M-1}{4} \right)\\
%	&=\frac{M\nu}{8}.
%	\end{split}
%	\label{}
%\end{equation}

%Since $(\mb{gr})^M=P_f$, the invariant is $\theta_{\mb{gr}}^{2M}=e^{\frac{\pi i M^2\nu}{4}}$. 

To summarize, as long as the 2D SPT blocks satisfy the ``fusion'' requirement $M\nu \equiv 0 \ {(\rm mod \ 8)}$, they can be glued together to form a 3D FSPT phase. The group structure follows immediately from that of the 2D phases:
\begin{enumerate}
	\item For $m=1$, the root state corresponds to $\nu=4$ on the half plane, forming a $\Z_2$ subgroup. This is in fact a bosonic SPT.
	\item For $m=2$, the root state corresponds to $\nu=2$ on the half plane, forming a $\Z_4$ subgroup. This phase corresponds to the interacting FSPT phase found in \Ref{ChengPRX2018} with $\Z_2^f\times\Z_2\times\Z_4$ symmetry.
	\item For $m\geq 3$, the root state corresponds to $\nu=1$ on the half plane, forming a $\Z_8$ subgroup. This phase corresponds to the interacting ``beyond-supercohomology'' FSPT phase found in \Refs{WangPRX2018,Zhou2021} with $\Z_2^f\times\Z_2\times\Z_8$ symmetry.
\end{enumerate}
In a closed form, the classification is $\mathscr{G}_{-1}=\mathbb{Z}_{\gcd(M, 8)}$. 

Finally, we need to check the overall group structure of SPTs under stacking operation, after taking both 1D and 2D block states into consideration. This is discussed in Sec.~\ref{sec:stackgroup}. The result is that there is no group extension, so the overall classification is given by
\begin{align}
\mathscr{G} &= \mathscr{G}_{-1} \times \mathscr{G}_{-2} = \Z_{2}\times \Z_{\gcd(M,8)}
\end{align}
The results are summarized in Table \ref{tab1}.

\subsubsection{2D block states for $n\ge 2$}
\label{sec:2db2Z}
%\noindent\textbf{2D reduction for $n>1$}
When $n\ge2$, 2D FSPT phases with $\Z_N$ internal symmetry are classified by $\Z_2\times\Z_{2N}$. Here the $\Z_2$ subgroup is generated by the so-called ``non-Abelian'' root SPT phase, with Majorana edge modes. The $\Z_{2N}$ subgroup is generated by an ``Abelian'' SPT phase with Dirac edge modes (however, two copies of the root state is equivalent to a bosonic state). Both root phases admit free fermion realizations, which are reviewed in Appendix \ref{append:review}. 

First, we consider the 2D blocks to be in the $\Z_2$ Majorana root state.  As long as $m\geq 1$, the edge modes at the rotation axis all together can be gapped out. One just needs to check whether they can be gapped out without breaking symmetries.  According to Ref.~\onlinecite{WangPRB2017}, the edge theory of each 2D FSPT block consists of $\frac{N^2}{4}$ chiral Majorana edge modes $\gamma_{j a}$, with $j=1, \dots, M$ and $a=1, \dots, N^2/4$, and $\frac{N^2}{8}$ chiral Dirac edge modes $\ol{\varPsi}_{j b}$ for $b=1, \dots, N^2/8$. Note that $\gamma_{ja}$ and $\ol{\varPsi}_{jb}$ have opposite chiralities. Again, let us denote $\mb{g}$ as the generator of $\mathbb{Z}_N$ symmetry and $\vr$ as the rotation. Under $\mb{g}$, the fields transform as 
\begin{align}
& \gamma_{j,1} \rightarrow -\gamma_{j,1}, \nonumber\\
& \gamma_{j,a} \rightarrow \gamma_{j,a}, \quad a\ge 2 \nonumber \\
& \ol{\varPsi}_{j,b}\rightarrow e^{i2\pi/N} \ol{\varPsi}_{j,b}
\label{eq:ng2-g}
\end{align}
Under rotation $\vr$, we have
\begin{align}
& \gamma_{j,a} \rightarrow \gamma_{j+1,a},\ \ol{\varPsi}_{j,b}\rightarrow \ol{\varPsi}_{j+1,b}, \quad j\le M-1 \nonumber\\
& \gamma_{M,a} \rightarrow -\gamma_{1,a},\  \ol{\varPsi}_{M,b}\rightarrow -\ol{\varPsi}_{1,b} 
\end{align}
We define the following complex fermions to diagonalize the rotation symmetry transformation:
\begin{align}
	\psi_{la}& =\frac{1}{\sqrt{M}}\sum_{j=1}^{M} \omega^{-lj} \gamma_{ja} \nonumber\\
	\ol{\psi}_{pb} & = \frac{1}{\sqrt{M}}\sum_{j=1}^M \omega^{-pj}\ol{\varPsi}_{jb}
	\label{}
\end{align}
where $\omega = e^{i\pi/M}$, $l$ is an odd integer taking values in the range $1,3,\dots, M-1$, and $p$ is also an odd integer taking values in the range $ 1,3,\dots, 2M-1$.  Under the rotation $\vr$, these complex  fermions transform as
\begin{equation}
\psi_{la} \rightarrow \omega^l \psi_{la}, \quad \ol{\psi}_{pa} \rightarrow \omega^p  \ol{\psi}_{pa}
\end{equation}

%The $\Z_N$ symmetry acts as
%\begin{equation}
%	\psi_{l,1}\rightarrow -\psi_{l,1}, \bar{\psi}_{pb}\rightarrow  \omega_N\bar{\psi}_{pb}.
%	\label{}
%\end{equation}
%Here $\omega_N= e^{\frac{2\pi i}{N}}$.

With the above preparation, we now check if the 2D blocks can be glued at the rotation axis while preserving the symmetries. It is equivalent to check if the edge of the multi-layer system, obtained by folding the 2D blocks,  can be gapped out without breaking $\vr$ and $\mb{g}$, i.e., if the multi-layer system is topologically trivial. The symmetries become internal $\mathbb{Z}_{2M}^f\times\mathbb{Z}_N$ symmetry in the multi-layer system. According to Ref.~\onlinecite{ChengPRB2018, WangPRB2017}, to see if the SPT is trivial or not, it is enough to check the topological spins of the $\vr$, $\mb{g}$ and $\mb{g}\vr$ fluxes. With similar calculations as in Sec.~\ref{sec:2db1}, it is easy to check that 
\begin{equation}
h_{\mb{g}} = h_{\vr} =0
\end{equation}
To calculate the topological spin of the $\mb{g}\vr$ flux, we first write down how the complex fermions transform under the combined $\mb{g}\vr$ symmetry: 
\begin{equation}
	\begin{split}
		\psi_{l1}&\rightarrow -\omega^{l} \psi_{l1}\\
		\psi_{la}&\rightarrow \omega^{l} \psi_{la}, \quad a\ge2\\
		\bar{\psi}_{pb}&\rightarrow e^{i2\pi/N}\omega^p \bar{\psi}_{pb}
	\end{split}
	\label{}
\end{equation}
We can then obtain
\begin{equation}
	\begin{split}
		h_{\mb{g}\vr}&=\frac{1}{2}\bigg[\sum_l\left( \frac{M+l}{2M} \right)^2 + \sum_l\left(\frac{N^2}{4}-1\right) \left(\frac{l}{2M}\right)^2\bigg]\\
		&\quad - \frac{N^2}{16}\sum_p \left( \frac{1}{N}+\frac{p}{2M} \right)^2\\
	&=-\frac{M}{64}(N^2+4N-4)\\
	&=-\frac{M}{16}\left[\left( \frac{N}{2} \right)^2+N-1\right]
	\end{split}
	\label{}
\end{equation}
Notice that $\left( \frac{N}{2} \right)^2+N-1$ is an odd integer.  For the SPT phase to be trivial, we should have  
\begin{equation}
h_{\mb{g}\vr}\times [2M,N]=0, \ ({\rm mod} \ 1) \nonumber
\end{equation}
where $[\dots]$ stands for ``least common multiple''. The factor $[2M,N]$ is the order of the group element $\mb{g}\vr$. This factor is needed for the purpose of  finding ``trivial'' topological spins (c.f. Refs.~\onlinecite{ChengPRB2018} or \onlinecite{WangPRB2017}). Then, we have the following two cases:
\begin{enumerate}
	\item When $m\geq n\geq 2$, $[2M,N]=2M$. Then, we always have $2Mh_{\mb{g}\vr}$ to be an integer. 
\item When $m < n$, then $[2M,N]=N$. Then, we should have
\begin{equation}
	Nh_{\mb{g}\vr} = -\frac{MN}{16} = 2^{m+n-4} \nonumber
	\label{}
\end{equation}
where the equations are defined modulo integers. This requirement is not satisfied, i.e. the multi-layer SPT is nontrivial, only if $m=1, n=2$. All other cases with $n>m\ge 1$ the multi-layer SPT is trivial.  
\end{enumerate}
Therefore, the 2D blocks can be glued without breaking the symmetries for most cases, only except for $C_2^-\times \Z_4$. A similar obstruction arises in the construction of ``beyond-supercohomology'' phases with $\Z_2^f\times\Z_2\times\Z_4$ symmetry~\cite{WangPRX2018}.

Next, we consider the other choice of 2D block states, by putting an Abelian FSPT phase on the half plane. The Abelian FSPT phases with $\mathbb{Z}_N$ symmetry are classified by $\mathbb{Z}_{2N}$. Let us put each 2D block in such an FSPT phase with index $\nu$, with $\nu=1,\dots, 2N-1$. The edge of each block consists of $\nu$ pairs of counter-propagating Dirac fermions $\varPsi_{a}$ and $\ol{\varPsi}_a$, with $a=1, \dots, \nu$. Since we have $M$ copies of 2D blocks, there are in total $M\nu$ pairs of Dirac fermions at the rotation axis. Under rotation $\vr$ and the $\mathbb{Z}_N$ generator $\mb{g}$, the fermions transform as follows
\begin{align} 
	\vr: & \ \varPsi_{j,a}\rightarrow \varPsi_{j+1,a}, \ \ol{\varPsi}_{j,a}\rightarrow \ol{\varPsi}_{j+1,a},  \quad  1\leq j<M \nonumber\\
	& \ \varPsi_{M,a}\rightarrow -\varPsi_{1,a}, \  \ol{\varPsi}_{M,a}\rightarrow -\ol{\varPsi}_{1,a}\nonumber\\
	\mathbf{g}: & \ \varPsi_{j,a}\rightarrow e^{\frac{2\pi i}{N}}\varPsi_{j,a}, \ \bar{\varPsi}_{j,a}\rightarrow \bar{\varPsi}_{j,a}.
	\label{}
\end{align}
Again, we Fourier transform them to $\psi_{la}$ and $\bar{\psi}_{la}$:
\begin{equation}
\psi_{la} = \frac{1}{\sqrt{M}}\sum_{j=1}^M\omega^{-lj}\varPsi_{ja}, \ \ol{\psi}_{la} = \frac{1}{\sqrt{M}}\sum_{j=1}^M\omega^{-lj}\ol{\varPsi}_{ja}
\end{equation}
Note that here $l$ ranges from $\pm 1, \pm 3, \dots, \pm (M-1)$. Under rotation $\vr$, we have
\begin{equation}
	\begin{gathered}
		\vr: \psi_{la}\rightarrow \omega^l\psi_{la}, \bar{\psi}_{la}\rightarrow \omega^l \bar{\psi}_{la}.
	\end{gathered}
	\label{}
\end{equation}
The $\mb{g}$ transformation takes the same form as $\varPsi_{ja}$ and $\ol{\varPsi}_{ja}$. Under the combined $\mb{g}\vr$ transformation, we have
\begin{equation}
	\mb{g}\vr: \psi_{la}\rightarrow e^{i2\pi/N}\omega^l\psi_{la},\ \bar{\psi}_{la}\rightarrow \omega^l \bar{\psi}_{la}.
	\label{}
\end{equation}

With the above preparation, we now check if the 2D blocks can be glued without breaking the symmetries. As before, it is enough to check the topological spins of the $\mb{g}$, $\vr$ and $\mb{g}\vr$ fluxes in a multi-layer system after folding the 2D blocks. After similar calculations as before, we find
\begin{align}
h_{\vr} & =0 \nonumber\\
%h_{\mb{g}} & =  \frac{\nu M}{2N^2}\nonumber\\
h_{\mb{g}} & = h_{{\mb{g}\vr}} = \frac{\nu M}{2N^2}
\end{align}
Again, considering the orders of the elements $\mb{g}$ and $\mb{gR}$ are $N$ and $[2M, N]$ respectively, the multi-layer SPT is trivial if
\begin{equation}
N \times \frac{\nu M}{2N^2} = \frac{\nu M}{2N} = 0  \ ({\rm mod} \text{\ 1}) \nonumber
\end{equation}
That is, we should have $\nu M =0 \ {(\rm mod}\ 2N)$.  More specifically, we have the following two cases:
\begin{enumerate}
	\item When $m> n\geq 2$, then $\nu M$ is always a multiple of $2N$. Thus, the 2D block state is ``obstruction-free'' and the classification is $\Z_{2N}$.
	\item When $m\leq n$, for $\nu M$ to be a multiple of $2N$, we require $\nu $ to be a multiple  of $ \frac{2N}{M}$. Then, the Abelian 2D block states are classified by $\Z_M$.
\end{enumerate}
Putting the two cases together, we have the Abelian 2D block states are classified by $\mathbb{Z}_{(M, 2N)}$, where $(M, 2N)$ is the greatest common divisor of $M$ and $2N$.

%\begin{equation}
%	\begin{split}
%		h_{\mb{g}\vr}&=\frac{1}{2}\sum_l \left[\left( \frac{1}{N}+\frac{l}{2M} \right)^2 - \left( \frac{l}{2M} \right)^2\right]\\
%	&=\frac{1}{2}\sum_l \left(\frac{1}{N^2}+\frac{l}{MN}\right)\\
%	&=\frac{M}{2N^2}.
%	\end{split}
%	\label{}
%\end{equation}

To summarize, the 2D block states are classified by
\begin{equation}
\mathscr{G}_{-1} = \left\{
\begin{array}{ll}
\mathbb{Z}_2, & \ \text{$C_2^-\times \mathbb{Z}_4$}\\[5pt]
 \mathbb{Z}_{(M, 2N)}\times \mathbb{Z}_2^{\rm NA}, & \ \text{others}
\end{array}
\right.
\end{equation}
For $C_2^-\times \mathbb{Z}_4$, the 2D block states are  Abelian FSPT phases. For other symmetries,  the generator of $\mathbb{Z}_2^{\rm NA}$ corresponds the non-Abelian Majorana FPST phases and $\mathbb{Z}_{\gcd(M, 2N)}$ corresponds to Abelian FSPT phases. 

Finally, we check the overall group structure of the 1D and 2D block states under stacking operation.  This is done in Sec.~\ref{sec:stackgroup}. The result is that $\mathscr{G}_{-2}$ extends $\mathscr{G}_{-1}$ only when $M=2$, and the extension occurs only for $\Z_{2}^{\rm NA}$. That is, when $M=2$, the root $\mathbb{Z}_2^{\rm NA}$ blocks state stacks into the 1D root block state. Since 1D block states are classified by $\mathbb{Z}_{(2M, N)} = \mathbb{Z}_4$, it leads to an overall $\mathbb{Z}_8$ classification. Accordingly, the overall classification is given by
\begin{equation}
\label{eq:group-cm-zn}
\mathscr{G} = \left\{
\begin{array}{ll}
\mathbb{Z}_2\times\mathbb{Z}_4, & \ m=1,n=2\\[5pt]
 \mathbb{Z}_{2}\times \mathbb{Z}_{8}, & \ m=1, n\ge 3 \\[5pt]
\mathbb{Z}_{(M, 2N)}\times \mathbb{Z}_{(2M, N)}\times \mathbb{Z}_2^{\rm NA}, & \ m\ge 2, n\ge 2 
\end{array}
\right.
\end{equation}
where we have used the explicit form of $\mathscr{G}_{-1}$ and $\mathscr{G}_{-2}$. 

One comment is that both $C_2^-\times\mathbb{Z}_P$ and $C_P^-\times\mathbb{Z}_2$ correspond to internal symmetry $\mathbb{Z}_2^f\times\mathbb{Z}_2\times\mathbb{Z}_P$, and thereby their FSPT phases should give the same classification. Indeed, the above result agrees with that in Sec.~\ref{sec:2db1}. Also, for $m,n\ge 2$, the classification is symmetric in $M$ and $N$, agreeing with the fact that $C_M^-\times \Z_N$ corresponds to $\mathbb{Z}_2^f\times\mathbb{Z}_M \times \Z_N$ internal symmetry, where $\Z_M$ and $\Z_N$ are in symmetric positions.

\subsubsection{Group structure of FSPT phases}
\label{sec:stackgroup}

\begin{figure*}
\centering
\includegraphics[width=5in]{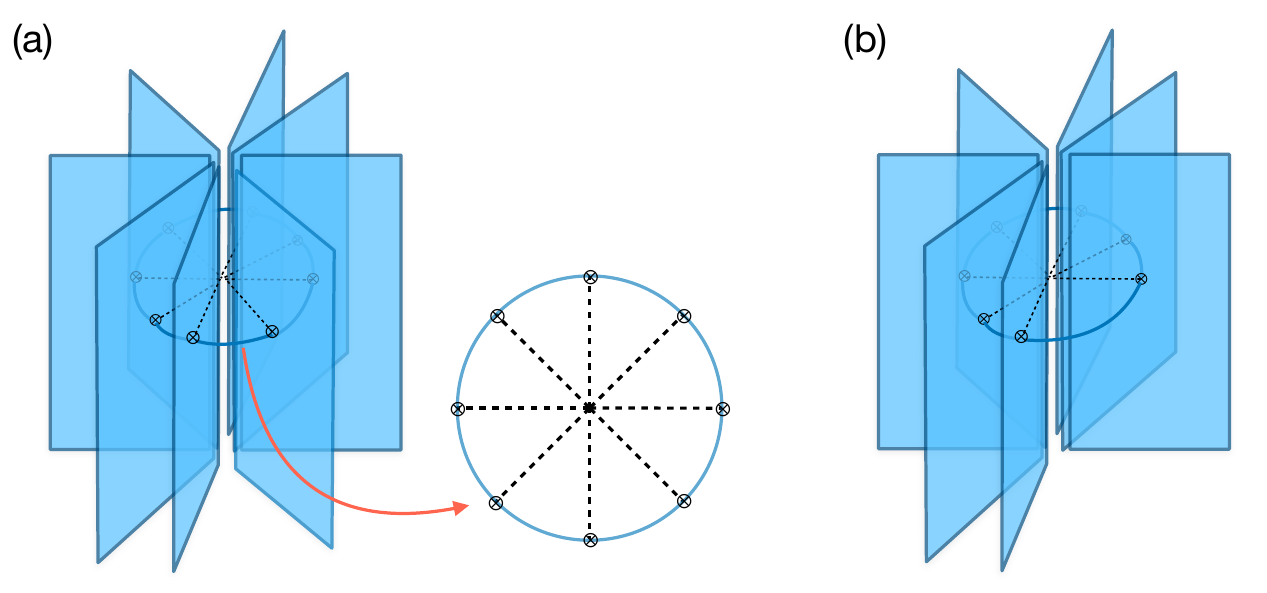}
\caption{Symmetry defects in 2D-block states with $C_M^-\times \Z_N$ symmetry. We have set $M=8$ for illustration. (a) $\Z_N$ symmetry defect loop (blue solid circle) is a collection of 2D defect points. The disk bounded by the loop is 2D FSPT state (1D-block state). The dashed lines are branch cuts. (b) A link of disclination (rotation defect) and $\Z_N$ defect loop.}
\label{fig:flux}
\end{figure*}

To obtain the aforementioned group $\mathscr{G}$, we make use of the following two properties. First, if a 2D-block state is bosonic, it will not stack into a 1D-block state. This can be seen from our discussion on bosonic rotation SPT phases in Appendix \ref{sec:bosonic}. Second, if we insert a loop-like symmetry defect into a 3D FSPT state, the membrane bounded by the loop supports a 2D FSPT state. This point is demonstrated in \Ref{ChengPRX2018}. It can be used to infer the group structure of $\mathscr{G}$.

The first property is enough to determine $\mathscr{G}$ for $n=1$. In this case, the root state in $\mathscr{G}_{-1}$ is either bosonic itself or stacking into a bosonic SPT. So, $\mathscr{G}_{-1}$ will not be extended by $\mathscr{G}_{-2}$, making $\mathscr{G}=\mathscr{G}_{-1}\times \mathscr{G}_{-2}$. It is similar for the ``Abelian'' SPT states for $n\ge 2$. Therefore, the only possible nontrivial extension of $\mathscr{G}_{-1}$ by $\mathscr{G}_{-2}$ comes from ``non-Abelian'' SPT phases. 

To check if two non-Abelian 2D-block states will stack into a 1D-block state, we consider symmetry defects. Figure \ref{fig:flux}(a) shows a unit defect of $\Z_N$. It consists of a set of defect points on the 2D planes, which collectively form a loop-like symmetry defect in 3D. More precisely, in each 2D plane, we create a pair of defect points, connected by a branch cut. We then bring one end of the branch cut onto the rotation axis, and glue all branch cuts together on the rotation axis. For 2D ``non-Abelian'' SPT states, the unit $\Z_N$ defect carries a Majorana zero mode. So, a branch cut can be thought of as a Majorana chain. When we glue the branch cut together, they form a rotation symmetric 1D-block state on the 2D membrane bounded by the defect loop. Under stacking, these membrane FSPTs should follow the same group as 2D $C_M^-$ FSPT phases. According to Sec.~\ref{sec:2D-CM-minus-1},  nontrivial extension occurs only when $M=2$. That is, for $M=2$, 1D-block states on the membrane stack into 0D-block states on the membrane. It implies that the corresponding 3D FSPT state must be a 1D-block state. Similar analysis can be done on the rotation defect (i.e., disclination), shown in Fig.~\ref{fig:flux}(b), which does not change the stacking structure. Therefore,  $\Z_2^{\rm NA}$ is extended by $\mathscr{G}_{-2}$ only for $m=1$ and $n\ge 3$. This gives the results in Eq.~\eqref{eq:group-cm-zn}.

Two comments are in order. First, the group $\mathscr{G}$ can also be obtained by studying the surface.  Suppose the bulk is a root 2D block state in $\mathscr{G}_{-1}$. On the rotation axis, edges of the $M$ blocks are gapped out without breaking any symmetry. On the surface, the $M$ edges, arranged rotation symmetrically, remain gapless. If we stack $|\mathscr{G}_{-1}|$ copies of the root state, the edge modes on the surface can be gapped out, leaving the only subtlety at the intersection point of the surface and rotation axis. This point may carry a symmetry-protected zero mode. Whether it hosts a zero mode or not corresponds exactly to whether the stack of $|\mathscr{G}_{-1}|$ copies of the root state is a 1D-block state or not. However, it is technically not very easy to do so. Second, symmetry defects themselves have interesting properties. Here, we mention one: For a ``non-Abelian'' 3D FSPT state, if a unit $\Z_N$ defect loop is linked to a disclination, both will carry an odd number of Majorana zero modes [see Fig.~\ref{fig:flux}(b)]. It is obvious that the $\Z_N$ defect loop carries an odd number of Majorana zero modes. Since the overall number of Majorana zero modes must be even, the disclination must also host an odd number of Majorana zero modes.  Note that an odd number of Majorana zero modes are robust even after they mobilize on the loop. This property has recently been studied in internal FSPT states in \Ref{Zhou2021}.

\subsection{$C_M^+$}
\label{sec:3dCplus}
Like in the $C_M^-$ case, we only need to consider 1D block states for $C_M^+$ symmetry. For 1D block states, we have an internal $\Z_M\times \Z_2^f$ symmetry on the rotation axis. As reviewed in Appendix \ref{append:review}, classification of 1D FSPT phases with internal symmetry is given by a triple $(\mu,\omega, \gamma)$, where $[\mu]\in \H^1[\Z_M, \Z_2]=\Z_2$, $[\omega]\in\H^2[\Z_M,\U]=\Z_1$ and $\gamma=0,1$ specifies whether it is a Majorana chain or not. The four 1D block states form a $\mathbb{Z}_2\times\mathbb{Z}_2$ classification\cite{Kapustin2015b}. We show that all these 1D block states can be trivialized. Hence, there are no nontrivial SPTs for the $C_M^+$ symmetry.

We first consider $\gamma=1$, i.e. a Majorana chain on the axis. Let us consider a surface perpendicular to the rotation axis, which preserves the symmetry. We see that there is a Majorana zero mode at the rotation center on the surface. This Majorana zero mode can be eliminated by the following construction: as we show in Appendix \ref{append:p+ip}, a fully gapped $p_x+ip_y$ superconductor in 2D is only consistent with $C_M^-$ symmetries. If we enforce a $C_M^+$ symmetry, we can build a $p_x+ip_y$ superconductor with a Majorana zero mode at the rotation center. We can then stack this 2D state to the 3D surface, and couple the two Majorana zero modes to gap them out. We thus find a fully gapped, short-ranged entangled surface on the surface, meaning that the bulk is also trivial. 

\begin{figure}[t]
	\centering
	\includegraphics[width=\columnwidth]{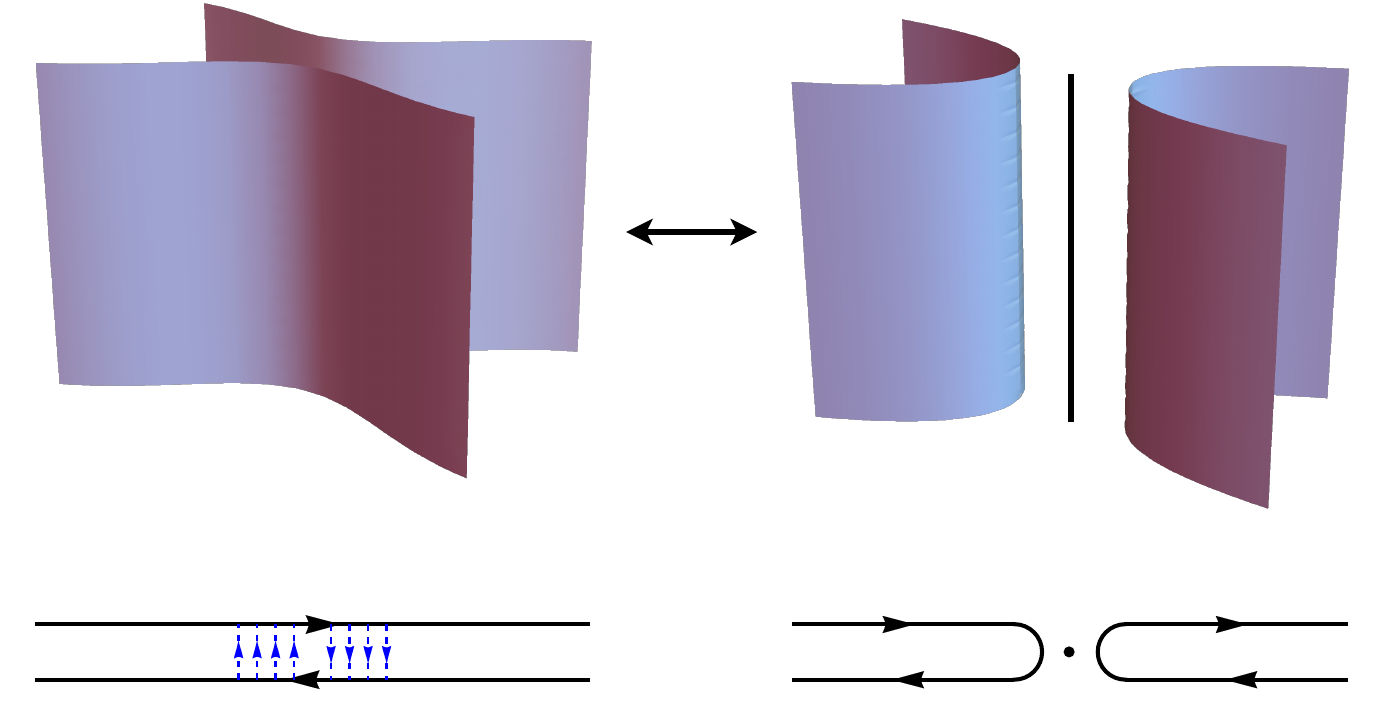}
	\caption{Illustration of how to trivialize a Majorana chain on $C_2$ rotation axis. On the left we show two planes of chiral topological superconductors parallel to the inversion axis, with opposite chiralities. Turning on the inversion-symmetric inter-plane couplings near the inversion axis, one can deform the system to the right picture where the two planes are reconnected and a Majorana chain is left in the middle. The bottom of the figure shows a top view of the system, where chiral Majorana edge modes are reconnected.}
	\label{fig:piptrivialize}
\end{figure}

One can in fact directly trivialize the bulk, using the construction in Sec. \ref{sec:triviality}~\footnote{This argument is due to D. Freed}. We will illustrate this construction for $M=2$. First, we consider two layers of 2D superconductors parallel to the inversion axis, one $p_x+ip_y$ and one $p_x-ip_y$, such that the two layers are mapped to each other under rotation. This is illustrated in the left panel of Fig. \ref{fig:piptrivialize}. Now we turn on inter-layer couplings in the region close to the inversion axis, to ``reconnect'' the two planes. The regions with inter-layer coupling turned on are then disentangled. However, if we require that the inter-layer coupling preserves the $C_2$ symmetry, the disentangling can not be complete; there is actually a Majorana chain left in the middle, shown in the right panel of Fig. \ref{fig:piptrivialize}. This can be seen from the surface, by exactly the same argument in Appendix \ref{append:p+ip}. 

Now we start from the state with a Majorana chain on the inversion axis. We can create two cylinderal ``bubbles'' of chiral $p_x- ip_y$ superconductors, bring them close to the inversion axis, and use the deformation process described in the previous paragraph to eliminate the Majorana chain while reconnecting the two cylinders into one larger cylinder enclosing the axis. Then we can push this topological superconductor close to the surface. As shown above, a gapped surface necessarily harbors a $p_x+ ip_y$ superconductor, which can be trivialized together with the one created from the bulk. Now the whole state is trivialized.
While we just described a particular construction, we conjecture that this is what happens in general: given a finite ($C_M$-symmetric) region, any adiabatic deformation that disentangles the Majorana chain on the axis necessarily creates a $p_x\pm ip_y$ superconductor on the boundary.

We note that this is an interesting kind of bulk-boundary correspondence: while the bulk is indeed trivial, its surface is nevertheless a nontrivial invertible topological phase. In fact, the invertible phase realized on the surface is ``anomalous'', in the sense that one can not find the same phase with the given symmetry properties in strictly 2D systems. In this case, the boundary realizes a $p_x\pm ip_y$ superconductor with $C_M^+$ symmetry. This is impossible unless the Hilbert space of the 2D system contains an odd number of Majorana modes in any $C_M$-symmetric region, which is of course what the bulk provides. The ``anomalous'' invertible phase can only be realized on the boundary of a trivial bulk state but still with nontrivial entanglement. If we remove the bulk entanglement, i.e. transforming the bulk into a product state, we necessarily remove the boundary state as well.

Next we consider a nontrivial $[\mu]$, which means that on the edge, the generator of $\Z_M$ (namely, $\vr$) does not commute with the fermion parity. We can realize such a phase by two decoupled Majorana chains, whose Majorana edge modes are denoted by $\gamma_1$ and $\gamma_2$, and let $\vr$ maps to $(-1)^{N_1}$. One might worry that such a representation of $\Z_M$ is not faithful. This can be easily resolved by attaching a completely trivial state where $\vr$ acts faithfully, without affecting any of the discussions we will have. The $\vr$ transformation acts on the boundary as
\begin{equation}
	\vr: \gamma_1\rightarrow -\gamma_1, \ \gamma_2\rightarrow \gamma_2,
	\label{anomlous-spt2}
\end{equation}
under which the local fermion parity $i\gamma_1\gamma_2$ changes. That is, $\vr P_f = -P_f \vr$, forming a projective representation of $\mathbb{Z}_2^f\times \mathbb Z_{M}$. 

This 1D block state is trivial, which we show in a similar way as above. Again, consider a surface perpendicular to the rotation axis. We see that there is a pair of Majorana modes on the surface, transforming under rotation according to the projective representation \eqref{anomlous-spt2}. Now we stack a purely 2D short-range entangled state onto the surface to eliminate the Majorana modes. The surface state we stack is the 1D block state of 2D $C_M^+$ symmetric state, discussed in Sec.~\ref{sec:2dcm+}. It is built out of $M$ copies of semi-infinite Majorana chains, arranged in a rotation symmetric fashion. The Majorana modes at the ends of the chains form a projective representation of $C_M^+$ [see Eq.~\eqref{eq:proj2d}]. Then, when this 2D state is attached to the surface, these Majorana modes together with the original pair of Majorana modes can be gapped out without breaking the symmetry. This is because that $C_M^+$ has only one projective representation, which is of order 2.  Hence, we have shown that the 1D block state admits a fully gapped, short-range entangled surface, meaning that the bulk is trivial. One may also try to directly trivialize the bulk by creating ``bubbles'' in the bulk, which now should be loops of Majorana chains. The argument is very similar to that for the $p_x+ip_y$ superconductors, which we will leave for the readers. 

Similarly to $C_M^+$ symmetric $p_x+ip_y$ superconductors, the attached 2D state is ``anomalous''. It is not compatible with $C_M^+$ symmetry if we require it to be fully gapped and to live in strictly 2D. It can only live on the surface of a trivial 3D bulk, if fully gapped.

%\subsection{Summary and Examples}
%We have not completely determined the group structures of SPT phases, besides the short exact sequences. This can in principle be done since we have explicitly constructed all states, but will be left for future works.
%\begin{enumerate}
%	\item $C_2^-\times\Z_4$.  We find $\mathscr{G}_{-2}=\Z_4$ with the root phase being an interacting fermionic SPT phase, and $\mathscr{G}_{-1}=\Z_2$.
%	\item $C_4^-\times \Z_2$. We find $\mathscr{G}_{-2}=\Z_2$ and $\mathscr{G}_{-1}=\Z_4$.
%	\item $C_4^-\times \Z_4$. We find $\mathscr{G}_{-2}=\Z_4$ and $\mathscr{G}_{-1}=\Z_2\times\Z_4$.
%	\item $C_2^-\times \Z_8$. We find $\mathscr{G}_{-2}=\Z_4$ and $\mathscr{G}_{-1}=\Z_2\times\Z_2$.
%	\item $C_8^-\times \Z_2$. We find $\mathscr{G}_{-2}=\Z_2$ and $\mathscr{G}_{-1}=\Z_8$.
%\end{enumerate}

\section{LSM anomaly for FSPT phases}
\label{sec:sptlsm}

In the derivation of the classification, we have found several cases, all with $C_M^+$ symmetry, that a FSPT phase can be realized only in a system where degrees of freedom in a rotationally-invariant region transform ``anomalously'' under the symmetry. Conversely, in such a system, a SRE ground state has to be the associated FSPT phase. These are new examples of Lieb-Schultz-Mattis theorems for SPT phases. Previously similar theorems were derived for systems with magnetic translation symmetries~\cite{LuLSMSPT, YangPRB2018}. 

The basic example is a 2D system of fermions with $C_M^+$ symmetry, and a Majorana zero mode at the rotation center. When $M$ is even, the ground state has to be a topological superconductor with odd Chern number. This is closely related to the ``no-go'' that a $p_x+ip_y$ superconductor is only compatible with $C_M^-$ symmetry; with $C_M^+$ symmetry there is necessarily an unpaired Majorana zero mode at the rotation center, which is demonstrated in Appendix \ref{append:p+ip}. For odd $M$, we can obtain a gapped ground state by a 1D block construction with $M$ number of Majorana chains meeting at the rotation center. 

Let us briefly outline a proof of this LSM theorem, generalizing the technique used in \Ref{PoPRL2017}. Let us consider the $C_2$ subgroup of the $C_M$, and denote the inversion by $\mb{I}$. Imagine inserting two fermion parity fluxes to the system, and place them in $C_2$-invariant positions. Under $\mb{I}$, the Hamiltonian is not invariant since the branch line between the two fluxes change location. Denote by $\Sigma$ the region encoded by the union of the branch lines before and after applying $\mb{I}$. The inversion symmetry can be restored by combining $\mb{I}$ with a fermion parity symmetry transformation restricted to the $\Sigma$ region. However, this new inversion anti-commutes with the global fermion parity, because there are odd number of Majorana modes inside the region. Therefore, we conclude that there must be at least two ground states in the presence of the fermion parity fluxes, with different fermion parities.  Obviously, this kind of non-local degeneracy can only arise in topological superconductors with odd Chern numbers.

Building on this theorem, we can easily obtain several others when additional symmetries are present:  
\begin{itemize}
	\item With a global $\Z_2$ symmetry, consider a system with a fermion mode $c$ at the rotation center which transforms as $c\rightarrow c^\dag$ under the $\Z_2$ symmetry. We can prove that a symmetric ground state must be a $\Z_2$ 2D FSPT phase.
	\item With a global $\Z_2^\mathsf{T}$ symmetry and $\mathsf{T}^2=P_f$, we conjecture that a system with a Majorana Kramers doublet at the rotation center must have a class DIII TSC as the ground state. This was recently discussed in \Ref{SongArxiv2018}
\end{itemize}

Similar phenomena can happen for 3D systems. Consider a class DIII topological superconductor. They are labeled by an integer $\nu$ mod $16$. We will argue that the odd $\nu$ ones are only compatible with $C_M^-$ symmetry. Consider creating a time-reversal domain wall in the bulk (i.e. by adding time-reversal breaking mass terms). Since the time-reversal symmetry is broken, on either side of the domain wall one can continuously deform the state into a trivial one. However, there must appear a 2D class D topological superconductor with odd Chern number at the domain wall, a defining feature of the bulk state. Now suppose the domain wall lies in a plane perpendicular to the rotation axis. The setup exactly preserves the rotation symmetry, so it is only compatible with $C_M^-$. We can conclude that the original bulk state shares the same property since everything we have done preserves rotational invariance.
If the symmetry is $C_M^+$, we are forced to introduce a Majorana zero mode at the intersection of the domain wall and the rotation axis. This implies that the rotation axis must host a helical Majorana fermion, i.e. the edge state of a 2D class DIII topological superconductor.

The same arguments apply to ``beyond supercohomology'' FSPT phases~\cite{WangPRX2018, KapustinThorngren}. These phases can be thought as decorating 2D Majorana FSPT states on domain walls, and we know that these 2D Majorana FSPT phases are only compatible with $C_M^-$ symmetry. By considering a domain wall perpendicular to the rotation axis, we conclude that the same is true for the 3D FSPT phase.

Our discussion in Sec. \ref{sec:3dCplus}  provides a bulk interpretation for these SPT-LSM theorems, in terms of a trivial but neverthess entangled bulk. While we focus on rotations, similar interpretations hold for other SPT-LSM theorems. For instance, \Ref{LuLSMSPT} proved that for a 2D fermionic system with an odd number of Majorana modes together with a $\pi$ flux per unit cell, SRE ground states preserving the magnetic translation symmetry must have odd Chern number. The LSM anomaly in this theorem can be understood as follows: in systems that do not obey the conditions of the LSM theorem, it is impossible to realize $p_x\pm ip_y$ superconductors with magnetic translation symmetry. This is best understood if one gauges the fermion parity to obtain an Ising topological order. There are three types of quasiparticles $I, \sigma,\psi$, where $\sigma$ is the fermion parity flux, an Ising anyon, and $\psi$ is the fermion. Now the magnetic translation symmetry in the ungauged fermionic system becomes an usual translation symmetry in the gauged system, i.e. a symmetry-enriched Ising topological order. However, this interpretation requires that the $\psi$ quasiparticle transform projectively. From the general classification of symmetry-enriched topological phases~\cite{SET, Tarantino_SET}, we know that the fermion $\psi$ in the Ising topological phase must carry the same symmetry representation as the vacuum, i.e. linear representation, as both of them appear in $\sigma\times\sigma$ fusion channels. Therefore, no projective representation is allowed on $\psi$, including magnetic translation symmetry. The only way out is that the system is realized on the surface of a 3D bulk, in this case a stack of Kitaev chains.

We can further generalize the argument to conclude that with a global unitary symmetry $G$, if fermions carry nontrivial projective representations of $G$ then it is impossible to realize topological superconductors with odd Chern numbers.  This ``no-go'' covers both the $C_M^+$ rotation (there is a twist in the projective representation as one interprets the symmetry as an internal one) case, as well as the SPT-LSM theorem with magnetic translation symmetry discussed earlier. We also conjecture that such topological superconductors with inconsistent symmetries can be realized on surfaces of 3D trivial but entangled bulk.

\section{Conclusion}
\label{sec:conclusion}

In conclusion, we have applied the dimensional reduction approach to study 1D, 2D and 3D interacting fermionic SPT phases with a symmetry group $C_M\times G$, where $C_M$ consists of rotations and $G$ contains internal Abelian symmetries. We obtain the classification of fermionic SPT phases for various symmetry groups. In addition, 2D and 3D fermionic crystalline SPT phases that can only exist with strong interaction are constructed. By comparing our results with known classifications of internal SPT phases, we formulate a precise crystalline equivalence principle for fermionic systems. We also identify several new instances of Lieb-Schultz-Mattis-type theorems for 2D FSPT phase. The main results have been summarized in Sec.~\ref{sec:mainresults}.

In this work we only study FSPT phases with rotation symmetry and Abelian internal symmetries. Moreover, the total symmetry group is a direct product of the two. For future studies, it is interesting but challenging to study the classification of more general crystalline symmetry groups, in particular those in which the crystalline symmetries are extended by the internal symmetries. The crystalline equivalence principle of these symmetry groups are particularly interesting to look at. Also, a general proof of the crystalline equivalence principle for FPST phases is highly demanded.

\emph{Related works.} Several relevant works on crystalline SPT phases appeared on arXiv around the same time that a preprint of this work was posted (in October 2018; see the note below). \Refs{SongArxiv2018, ShiozakiArxiv2018, ElseArxiv2018} presented general frameworks for the classification of crystalline SPT phases. In particular, our results partially overlap with \Ref{ShiozakiArxiv2018}. \Ref{RasmussenArxiv2018b} also constructed interacting intrinsically fermionic SPT phases with crystalline symmetries.

\emph{Note added.} The first version of this paper was posted on arXiv in October 2018 (arXiv:1810.12308v1), more than three years before its submission for peer review. Compared to that version, the main improvement is the derivation of classification group $\mathscr{G}$ of 3D FSPT phases (see  Sec.~\ref{sec:stackgroup} and related discussions in Appendix \ref{sec:more3D}). In addition, a new example is 
included in Appendix \ref{app:e3}.

\begin{acknowledgments}
M.C. acknowledges discussions with Dominic Else,  Adrian Po,  Ashvin Vishwanath and Dominic Williamson. C.W. and M.C. are grateful to Yang Qi and Zheng-Cheng Gu for enlightening conversations. We are also grateful to Jian-Hao Zhang for pointing out a mistake in an earlier version of the manuscript. M.C. is supported by startup funds from Yale University. C.W. is supported in part by the Research Grant Council of Hong Kong (GRF 17300220).  C.W and M.C. acknowledge Aspen Center of Physics (supported by NSF grant PHYS-1607611) for hospitality where part of the work were completed.
\end{acknowledgments}

\appendix

\section{Group extension}
\label{sec:groupext}
Let $G$ be a finite group, and $N\subset G$ is a normal subgroup. Then we can form the quotient group $Q=G/N$. We say that $G$ is an extension of $Q$ by $N$.  Equivalently, the three groups $N, G, Q$ fit in the following short-exact sequence:
\begin{equation}
    1\rightarrow N\rightarrow G\rightarrow Q\rightarrow 1.
\end{equation}
What the short exact sequence means is that there is a surjective map $\pi: G\rightarrow Q$, whose kernel is exactly $N$. It is then easy to see that $N$ must be normal and $Q=G/N$.

The extension is said to be \emph{central} if $N$ is in the center of $G$, which also implies that $N$ is an Abelian group. All extensions considered in this paper are central, and in fact in all cases $N, G$ and $Q$ are Abelian. The extension problem refers to the determination of $G$ from $N$ and $Q$. For central extension, the additional information needed to determine $G$ is a group 2-cocycle $\omega\in Z^2[Q,N]$.  More explicitly, $\omega$ is a function from $Q\times Q$ to $N$ satisfying the 2-cocycle condition:
\begin{equation}
    \omega(q_1,q_2)\omega(q_1q_2,q_3)=\omega(q_1,q_2q_3)\omega(q_2,q_3),
\end{equation}
for all $q_1,q_2,q_3\in Q$. Once $\omega$ is given, $G$ can be explicitly constructed as follows: we represent $G$ set-wise as the Cartesian product $Q\times N=\{(q,n)|q\in Q, n\in N\}$, equipped with the following group multiplication law:
\begin{equation}
    (q_1,n_1)\times (q_2,n_2)=(q_1q_2, n_1n_2\omega(q_1,q_2)).
\end{equation}
The 2-cocycle condition of $\omega$ ensures that the multiplication is associative. The map $\pi$ is then given by $\pi((q,n))=q$. 
One can prove that the group extension only depends on the cohomology class $[\omega]$, so one-to-one correspondent to the group cohomology $\H^2[Q,N]$. If the cohomology class $[\omega]$ is trivial, then we have $G=Q\times N$. In this case we say the short exact sequence splits, or simply the extension is trivial.

To give a concrete example, consider the following short-exact sequence:
\begin{equation}
    1\rightarrow \Z_2\rightarrow G\rightarrow \Z_2\rightarrow 1.
\end{equation}
Since $\H^2[\Z_2,\Z_2]=\Z_2$, there are two possibilities. If we write $Q=\Z_2=\{1,q\}$ and $N=\Z_2=\{1,n\}$, the two possibilities correspond to $\omega(q,q)=1$ or $\omega(q,q)=n$ (and $\omega(1,1)=\omega(q,1)=\omega(1,q)=1$). It is easy to see that if $\omega(q,q)=1$, then the extension is trivial, i.e. $G=N\times Q=\Z_2\times\Z_2$. If $\omega(q,q)=n$, then we have
\begin{equation}
    (q,1)\times (q,1)=(1,n).
\end{equation}
In this case, the extension $G$ is actually isomorphic to $\Z_4$, with $(q,1)$ as the generator.

\section{Bosonic $C_n\times G$ SPT phases}
\label{sec:bosonic}
In this Appendix, we consider bosonic systems with symmetry group $C_n \times G$. In the dimensional reduction, we will restrict ourselves to block states that can be captured by group cohomology classification. 
%We will derive the classification in the two approaches and prove their equivalence explicitly.

Let us apply the dimensional reduction in two and three dimensions:
\begin{itemize}
	\item To determine $\mathscr{G}_{-2}$, we classify the corresponding $(d-2)$D SPT states protected by $\Z_n\times G$. The cohomology group can be computed using K\"unneth formula. 
For $d=2$ we have
\begin{equation}
	\H^{1}[G\times\Z_n, \U]=
			\Z_n\oplus   \H^{1}[G, \U].
\end{equation}
The first $\Z_n$ represents phases protected by rotation alone. The last factor $\H^{1}[G, \U]$ obviously means placing a 0D $G$-charge at the rotation center. According to the discussion in Sec. \ref{sec:triviality}, if we can split a $G$-charge into $n$ parts, then the state is trivial. Thus we actually have the quotient $\H^1[G, \U]/n\H^1[G, \U]$ to take into account the trivialization.

For $d=3$ we have
\begin{equation}
	\H^2[G\times\Z_n,\U]=\H^{1}[G, \Z_n]\oplus \H^{2}[G, \U].
	\label{}
\end{equation}
Similar to $d=2$, the $\H^2[G, \U]$ corresponds to placing 1D SPT states protected by $G$ symmetry at the rotation axis. If a 1D SPT can be split into $n$ parts, it is trivial. This trivialization gives a classification of $\H^2[G, \U]/n\H^2[G, \U]$.

Now we discuss the $\H^1[G, \Z_n]$ factor. Physically, it corresponds to nontrivial commutation relations between $G$ and $\Z_n$ transformations on the boundary of the 1D SPT. Suppose we choose a cocycle $[b]\in \H^1[G, \Z_n]$, which is basically a homomorphism from $G$ to $\Z_n$, we represent it as a function $b(\mb{g})$ with $b^n(\mb{g})=1$. Suppose that the localized $\mb{g}$ symmetry transformation on the boundary is $U_\mb{g}$ for $\mb{g}\in G$, and $U_\vr$ for the generator of the $\Z_n$ group (which is a $2\pi/n$ rotation restricted on the rotation axis) then
\begin{equation}
	U_\mb{g} U_\vr= b(\mb{g})U_\vr U_\mb{g}.
	\label{}
\end{equation}

We will also need to check the trivialization condition. For simplicity, we assume $G$ is Abelian and unitary. To see that state is nontrivial, we define the following physical invariant: for $\mb{g}\in G$, we insert a $\mb{g}$ flux loop wrapping around the rotation axis

\item $\mathscr{G}_{-1}$ are formed by order-$n$ elements of $\H^d[G, \U]$, if the boundaries can be trivially gapped without breaking $G$ and $C_n$. We will show below that the construction always works for any order-$n$ element.

\item $\mathscr{G}_0$ is basically the internal SPT phases protected by $G$, as long as they are compatible with the $C_n$ symmetry. This is always the case so $\mathscr{G}_0=\H^{d+1}[G, \U]$, as shown explicitly in \Ref{ThorngrenPRX2018}.
\end{itemize}

Now we show that in both cases, the classification agrees completely with those of internal $G\times \Z_n$ SPT phases~\cite{ThorngrenPRX2018, RasmussenArxiv2018}.
 We can use Kunneth formula to compute $\H^{d+1}[G\times\Z_n,\U]$. For $d=2$, 
	\begin{equation}
			\H^{3}[G\times \Z_n, \U] = \Z_n\oplus \H^{2}[G, \Z_n] \oplus \H^{3}[G, \U]
				\label{}
	\end{equation}
	We can easily identify the $\Z_n$ factor as the pure $C_n$ SPT phases in $\mathscr{G}_{-2}$, and $\H^3[G, \U]$ as those protected just by $G$. To match the $\H^2[G, \Z_n]$ part with the dimension reduction approach, we need to use the following relation between cohomology groups~\cite{WenPRB2015}:
\begin{equation}
	\begin{split}
	\H^d[G, M]&=\H^d[G, \mathbb{Z}]\otimes_\Z M \oplus \H^{d+1}[G, \Z]\boxtimes_\Z M\\
	&= \H^{d-1}[G, \U]\otimes_\Z M \oplus \H^{d}[G, \U]\boxtimes_\Z M
	\end{split}
	\label{eqn:kunneth2}
\end{equation}
Here $\otimes_\Z$ denotes the tensor product with respect to the module $\Z$, and $M_1\boxtimes_\Z M_2$ denotes the torsion of the two modules $\mathrm{Tor}_{\Z}(M_1,M_2)$. For $M=\Z_n$, they can be understood in more elementary terms:
\begin{equation}
	A\otimes_\Z \Z_n=A/nA, \quad A\boxtimes_\Z \Z_n=\{a\in A|a^n=0\}.
	\label{}
\end{equation}
Clearly $\H^{2}[G, \U]\boxtimes_\Z \Z_n$ is identified with $\mathscr{G}_{-1}$ and $\H^{1}[G, \U]\otimes_\Z \Z_n$ is the remaining part of $\mathscr{G}_{-2}$.

For $d=3$ we similarly obtain
	\begin{equation}
			\H^{4}[G\times \Z_n, \U]=
			\H^{1}[G, \Z_n] \oplus \H^3[G, \Z_n] \oplus {\H^{4}[G, \U]}
			\label{}
	\end{equation}
	We can also use Eq. \eqref{eqn:kunneth2} to match $\H^3[G, \Z_n]$ with part of $\mathscr{G}_{-2}$ and $\mathscr{G}_{-1}$.

We now show that $(d-1)$-dimensional block states can always be constructed, as long as the bosonic SPT block has order $n$. Consider a bosonic SPT phase corresponding to a cocycle $[\omega]\in \H^{d+1}[G, \U]$, and $[\omega]$ has order $n$, i.e. $\omega^n=\delta \nu$. We can then redefine $\omega\rightarrow \omega (\delta \nu)^{1/n}$ (the ambiguity in the $n$-th root has no effect) to make $\omega^n=1$. We will assume this gauge in the following.

We consider the group-cohomology model of such a SPT phase. The boundary can be modeled as a $(d-1)$D lattice, with group elements on each site $\ket{g_i}$ where $g\in G$. The symmetry transformation reads
\begin{equation}
	U_g\ket{\{g_i\}}= S(g)\ket{\{gg_i\}}.
	\label{}
\end{equation}
Here $S(g)$ is a phase factor that can be expressed in terms of group cocycles. We will not need its specific form, just noticing that $S^n(g)=1$ in our gauge.

Now consider $n$ copies of the boundary. We denote the basis as $\ket{g_i^\alpha}$ where $\alpha=1,2,\dots,n$ is the ``layer'' index. 
\begin{equation}
	\ket{\Psi}=\prod_i \frac{1}{\sqrt{|G|}}\sum_{g_i\in G} \ket{g_i^\alpha=g_i, \,\forall \alpha}
	\label{}
\end{equation}
Namely, $\ket{\Psi}$ is a product state, and on each site $i$. It is straightforward to show that the ground state is invariant under arbitrary permutation of layers from $S_n$.

\section{Review of 1D and 2D FSPTs with onsite symmetries}
\label{append:review}
In this section we review the classifications of FSPT phases in 1D~\cite{Fidkowski2011, KapustinPRB2018, TurzilloYou} and 2D~\cite{ChengPRB2018, WangPRB2017}.

\subsection{1D FSPT phases}
We follow the algebraic description in \Ref{Fidkowski2011} and \Ref{KapustinPRB2018}. Here we include Majorana chains in the definition of FSPT phases. We denote the ``bosonic'', or physical symmetry group by $G_b$. The total symmetry group is a central extension of $G_b$ by $\Z_2^f$, characterized by a nontrivial $2$-cocycle $[\nu]\in \H^2[G_b, \Z_2]$. All fermionic SPT phases are labeled by a triple $(\mu, \omega, \gamma)$ where $[\mu]\in \H^1[G_b, \Z_2]$ and satisfies 
\begin{equation}
	\delta\omega = \frac{1}{2} \nu\cup \mu,
	\label{eq:B1}
\end{equation}
where ``$\cup$'' is the cup product. In writing this formula we represent $\Z_2$ additively as $\Z_2=\{0,1\}$. Since $\mu$ and $\nu$ are $\Z_2$-valued cocycles, the right-hand side of \eqref{eq:B1} is an integer. $\omega$ is a $\mathbb{R}/\Z$-valued $2$-cochain on $G$. Physically, $\mu(\mb{g})$ for $\mb{g}\in G_b$ specifies whether the local unitary implementing $\mb{g}$ on an end of the system commute or anti-commute with the fermion parity.

In other words, $\frac{1}{2}\nu\cup \mu$ defines an obstruction class in $\H^3[G_b, \U]$. $\gamma=0,1$ indicates whether the state has unpaired Majorana zero modes on the edge, and $\gamma=0$ if $[\nu]$ is cohomologically nontrivial. 

\subsection{2D FSPT phases}
A complete classification of 2D FSPT phases with $\Z_2^f\times G$ symmetry has been obtained in \Ref{ChengPRB2018}.  It turns out that these are all we need for the block state construction. 

Let us first present the general algebraic description. 2D FSPT phases are classified by triples $(\rho, \nu, \omega)$. Here $[\rho]\in \H^1[G, \Z_2], [\nu]\in \H^2[G, \Z_2]$ and $\omega$ is $\mathbb{R}/\Z$-valued 3-cochain. They need to satisfy
\begin{equation}
	\delta \omega = \frac{1}{2} \nu\cup \nu.
	\label{}
\end{equation}
It turns out that for Abelian unitary $G$, the obstruction class $\frac{1}{2}\nu\cup\nu$ always vanishes.

For our purpose, we actually need concrete models for edge states of 2D FSPT phases, for $G=\Z_N$ and $\Z_{N_1}\times\Z_{N_2}$. Thus we will now focus on these two groups. We will list the classifications and explicit free fermion constructions of root phases, following the discussions in \Ref{WangPRB2017}.

First we consider $G=\Z_N$. Without loss of generality we will assume $N=2^n$. It was found in \Ref{ChengPRB2018} that the classification is $\Z_{4N}$ for $n=1$, and $\Z_2\times \Z_{2N}$ for $n>1$. 

The $n=1$ case is well known, so we only present constructions for $n>1$. The root phase for the $\Z_2$ subgroup in the classification has the following construction: consider two-component fermions, say spin up and down. Let the spin-up fermions form $N^2/4$ copies of $p_x+ip_y$ superconductors, and spin-down fermions form a Chern insulator with $C=-N^2/8$. We view the Chern insulator as $N^2/8$ copies of the Chern number $C=-1$ phase. The internal symmetry is generated by
\begin{equation}
	U_\mb{g}=(-1)^{N^f_{\uparrow1}} e^{\frac{2\pi i}{N}N^f_\downarrow},
	\label{}
\end{equation}
where $N_{\uparrow1}^f$ is the particle number operator of the first copy of the $N^2/4$ spin-up fermions, and $N_\downarrow^f$ is the total particle number of all spin-down fermions. It is clear that a $\mb{g}$ flux binds a single Majorana zero mode because of $(-1)^{N^f_{\uparrow 1}}$.  We can further compute the topological spin of the symmetry defect:
\begin{equation}
	\theta_\mb{g}=e^{\frac{i\pi}{8}}\cdot e^{-\frac{i\pi}{N^2}\cdot \frac{N^2}{8}} = 1,
	\label{}
\end{equation}
which confirms that this is indeed the root phase.

We set up notations for the edge modes. The Majorana edge modes of the $p+ip$ superconductors are denoted by $\gamma_{a}$ for $a=1, \dots, N^2/4$, and the Dirac edge modes of the Chern insulators with opposite chiralities are denoted by $\bar{\psi}_b$ for $b=1, \dots, N^2/8$.

For the generator of the $\Z_{2N}$ subgroup, we consider a bilayer system. The first layer has $C=1$, and the second layer has $C=-1$. The symmetry is simply
\begin{equation}
	U_\mb{g}=e^{\frac{2\pi i}{N}N_1^f},
	\label{}
\end{equation}
where $N_1^f$ is the particle number of fermions in layer $1$.
The topological invariant of this phase:
\begin{equation}
	\theta_\mb{g}^{N} = e^{\frac{i\pi}{N}}.
	\label{}
\end{equation}
We raise $\theta_\mb{g}$ to the $N$-th power to remove dependence on charge attachment.

Next we turn to $G=\Z_{N_1}\times\Z_{N_2}$.
We will not consider those phases protected by $\Z_{N_1}$ or $\Z_{N_2}$ alone. Those that require both $\Z_{N_1}$ and $\Z_{N_2}$ for protection are classified by
\begin{equation}
	\begin{cases}
		\Z_4, & n_1=n_2=1\\
		\Z_{N_{12}}\times\Z_2, & \text{otherwise}
	\end{cases}
	\label{}
\end{equation}
where $N_{12}$ is the greatest common divisor of $N_{1}$ and $N_2$. 

Let us first consider $n_1=n_2=1$. The root phase can be constructed as follows: consider four layers, layer $1$ and $2$ are $p_x+ip_y$ superconductors, layer $3$ and $4$ are $p_x-ip_y$ superconductors. The two $\Z_2$ symmetries are defined as
\begin{equation}
	\mb{g}_1 = (-1)^{N_2+N_4}, \mb{g}_2=(-1)^{N_3+N_4}.
	\label{}
\end{equation}

If either of $n_1,n_2$ is greater than $1$, the generating phase of the $\Z_{N_{12}}$ subgroup is a bosonic one. We will describe below how to construct the generating phase of the $\Z_2$ subgroup. First we take a two-layer construction, where layer 1 is a $p_x+ip_y$ superconductor and layer 2 $p_x-ip_y$. In this system, both generators $\mb{g}_{1,2}$ correspond to $(-1)^{N_2}$. Then we stack two additional fermionic SPT phases, each protected solely by one of the generators. More specifically, for the $\Z_{N_i}$ generator, we need
\begin{enumerate}
	\item $n_i=1$, the $\nu=2$ phase of the $\Z_8$ classification.
	\item $n_i\geq 2$, recall the classification is $\Z_{2N}\times \Z_2$, so we use a tuple $(\nu_1,\nu_2)$ with $\nu_1=0,1,\cdots, 2N_i-1$ and $\nu_2=0,1$ to label the phases. We will need the $(\frac{N_i^2}{8}, 1)$ phase.
\end{enumerate}
This way we realize a phase with $\Theta_{ij}=\frac{\pi N^{ij}}{4}, \Theta_{0ij}=\pi$. We can then stack a bosonic phase to cancel the $\Theta_{ij}$ when $N^{ij}$ is an odd multiple of $4$.

\section{Symmetry properties of $p_x+ip_y$ superconductors}
\label{append:p+ip}

Let us first consider a $p_x+ip_y$ superconductor in continuum. The pairing term reads
\begin{equation}
	\Delta \psi^\dagger (\partial_x+i\partial_y)\psi^\dag + \text{h.c.} 
	\label{}
\end{equation}
Naively the term breaks $\mathrm{SO}(2)$ rotation symmetry, but it can be restored by a gauge transformation $\psi\rightarrow e^{i\alpha/2}\psi$ where $\alpha$ is the rotation angle. As a result, the Hamiltonian is consistent with a $C_M^-$ symmetry.

\begin{figure}[t]
	\centering
	\includegraphics[width=\columnwidth]{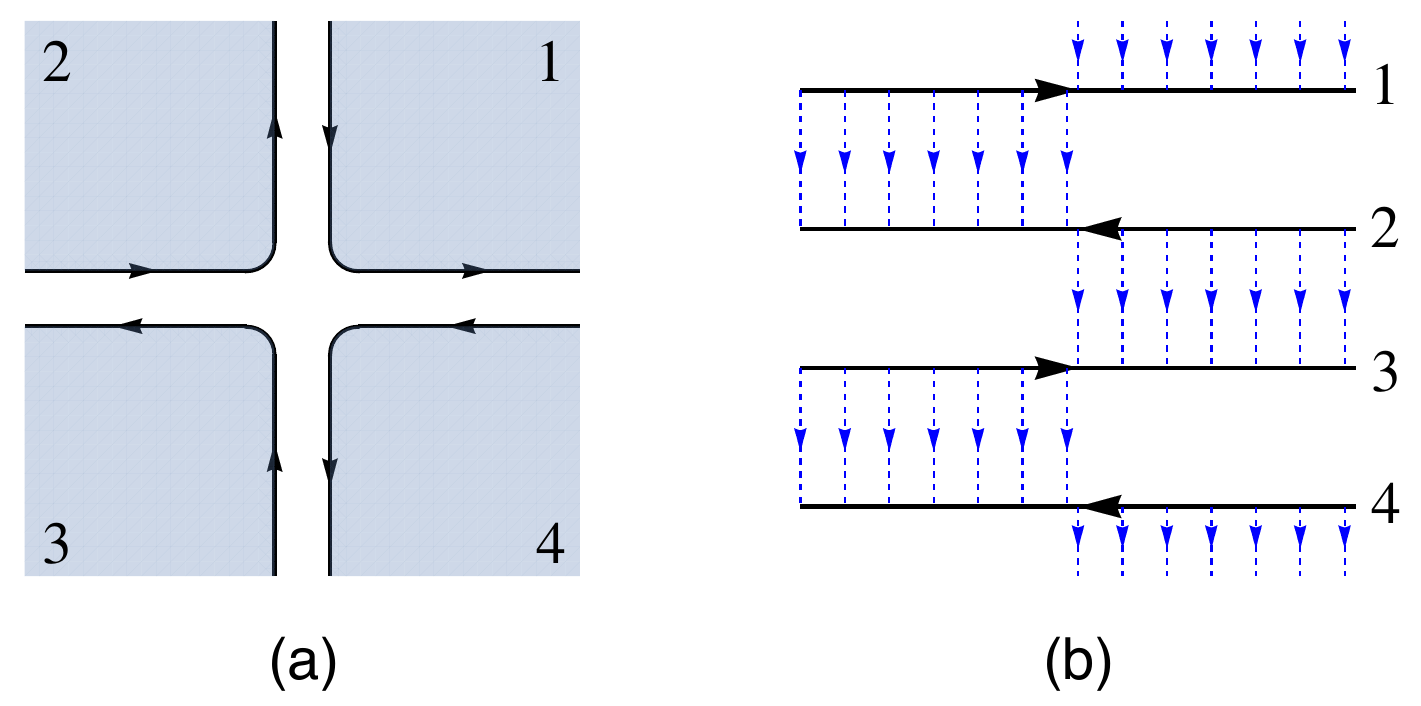}
	\caption{Construction of a $p_x+ip_y$ superconductor with $C_M$ symmetry. (a) One starts from $M$ patches of $p_x+ip_y$ superconductors, related to each other by $C_M$ rotation. To obtain a fully gapped superconductor one couple edge modes from neighboring patches. (b) ``Unfold'' the edge modes to one dimension.}
	\label{fig:pip}
\end{figure}

If we try to enforce the $C_M^+$ symmetry, heuristically it can be done by inserting a superconducting vortex at the origin and therefore a Majorana zero mode is localized there. We now demonstrate this result by patching together $p_x+ip_y$ blocks. We first partition the 2D plane into $M$ regions similar to what is done in the dimension reduction. In each of the $M$ regions we put a $p_x+ip_y$ superconductor. We then couple adjacent edges together to gap out the Majorana edge modes, as illustrated in Fig. \ref{fig:pip}(a). The question is whether there are any low-energy modes left. Clearly if low-energy modes were to exist, they must be localized near the rotation center. We will show below that for $C_M^+$ symmetry, there is exactly one such zero-energy mode.

We can just focus on the low-energy edge modes, and ``unfold'' the $M$ chiral/anti-chiral Majorana fermions to a 1D system, with couplings turned on between neighboring Majorana modes, but only on half lines, see Fig. \ref{fig:pip}(b) for an illustration for $M=4$.  We can write down the following effective Hamiltonian:
\begin{equation}
	\mathcal{H}=\sum_{j=1}^M \Big[\frac{1}{2}(-1)^j \eta_j i\partial_x\eta_j + i\Delta_j(x)\eta_j \eta_{j+1}\Big].
	\label{}
\end{equation}
%Here $\eta_j(x)$ are chiral Majorana fields, and $\Delta_j(x）=m\theta[(-1)^j x]$. We will assume $m>0$. $C_M^+$ symmetry requires $\eta_{M+1}\equiv \eta_1$. 
Here $\eta_j(x)$ are chiral Majorana fields, and $\Delta_j(x)=m\Theta[(-1)^j x]$, where $\Theta$ is the step function. We will assume $m>0$. $C_M^\pm $ symmetry requires $\eta_{M+1}\equiv \pm\eta_1$.

We look for zero-energy bound state:
\begin{equation}
	\xi =  \int_{-\infty}^\infty \di x\, \sum_{n=1}^M f_n(x)\eta_n(x).
	\label{}
\end{equation}
From the equation of motion $[H, \xi]=0$ we obtain the following coupled differential equations:
\begin{equation}
	(-1)^j\frac{df_j}{dx}+\Delta_j(x)f_{j+1}(x)-\Delta_{j-1}(x)f_{j-1}(x)=0.
	\label{}
\end{equation}

An ansatz for a localized solution is 
\begin{equation}
	f_n(x)=f_n[\Theta(x)  e^{-\lambda_+ x} + \Theta(-x)e^{\lambda_- x}].
	\label{}
\end{equation}
Here $\lambda_\pm > 0$ to ensure solutions are normalized.

First for even $j=2k$, we find
\begin{equation}
	\lambda_+ f_{2k} = mf_{2k+1}, \lambda_- f_{2k} = mf_{2k-1}.
	\label{}
\end{equation}
For odd $j=2k-1$, we have
\begin{equation}
	\lambda_+ f_{2k+1} = mf_{2k}, \lambda_- f_{2k-1} = mf_{2k}.
	\label{}
\end{equation}
It immediately follows that $\lambda_\pm = m$, and all $f_j$'s are equal. This is clearly only compatible with $C_M^+$ symmetry.

To summarize we find a zero mode operator
\begin{equation}
	\xi = \sum_{j=1}^M \int_{-\infty}^\infty dx\, e^{-m|x|}\eta_j(x).
	\label{}
\end{equation}
Apparently $\xi$ is rotationally invariant, i.e. $\vr\xi\vr^{-1}=\xi$.

Let us provide an alternative argument for why $p_x+ip_y$ superconductors are compatible with $C_M^-$ symmetry only. We can gauge the fermion parity in a $(p_x+ip_y)$ superconductor to obtain an Ising topological order, with three types of quasiparticles $I, \sigma,\psi$ where $\sigma$ is the fermion parity flux and $\psi$ is the fermion. In the presence of $C_n$ symmetry, anyons in the (now bosonic) Ising topological phase can carry fractional quantum numbers under $C_n$.  Because of the Ising fusion rule $\sigma\times\sigma=I+\psi$, it follows that $\psi$ must have linear $C_M$ quantum number, i.e. $\vr^M=1$ on the $\psi$ quasiparticle. Now we need to relate the fractional quantum numbers on quasiparticles, to symmetry representation on second-quantized operators before gauging. Let us consider a state $\ket{\psi}$ with $M$ of $\psi$ particles, arranged in $C_M$-symmetric positions. Since $M$ is even, such a state can be created physically from vacuum. We can ask what is the $C_M$ quantum number of this state. According to \Ref{QiPRB2015}, we know that the $C_M$ eigenvalue is equal to the $\vr^M$ value on a single $\psi$, which is $+1$ in this case. Now we consider the same state, but in the ``ungauged'' system:
\begin{equation}
	\ket{\psi}=c_\vx^\dag c_{\vr(\vx)}^\dag \cdots c_{\vr^{M-1}(\vx)}^\dag\ket{0}.
	\label{}
\end{equation}
We may need to average over other internal indices. The $\vr$ eigenvalue is given by
\begin{equation}
	\vr\ket{\psi}=U_{\vr^M} (-1)^{M-1}\ket{\psi}=-U_{\vr^M}\ket{\psi}.
	\label{}
\end{equation}
Here $U_{\vr^M}$ is the abbreviation for $\prod_{j=1}^MU_\vr\big(\vr^j(\vx)\big)$.  To match the result in the gauged system, we demand that $U_{\vr^M}=-1$.

\section{More classifications of 2D FSPT phases}
\label{app:2d}

\subsection{$C_M^{-}\times \mathbb{Z}_N$}
\label{sec:2D-CM-minus-2}

For $C_M^{-}\times \mathbb{Z}_N$ symmetry, FSPTs can be divided into (1) those protected by $C_M^{-}$ only, (2) those protected by $\mathbb{Z}_N$ only, and (3) those protected by both $C_M^-$ and $\mathbb{Z}_N$. For FSPTs protected by $C_M^{-}$ only, we have studied them in Sec.~\ref{sec:2D-CM-minus-1}. In principle, we need to check if these FSPTs are compatible with the onsite $\mathbb{Z}_N$. Nevertheless, onsite unitary symmetry (excluding fermion parity $P_f$ or a symmetry that multiplies to $P_f$ ) are always compatible with FSPTs protected by other symmetries --- because there is always the special situation that the unitary symmetry acts trivially on the FSPTs. On the other hand, for FSPTs protected by $\mathbb{Z}_N$ only, they may or may not be compatible with $C_M^-$ symmetry. By definition, rotation $\vr$ acts nontrivial on the Hilbert space. However, through similar analysis as in Appendix \ref{append:p+ip}, one can show that all internal FSPT phases protected by $\Z_N$ are compatible with $C_M^-$ (but not with $C_M^+$). So, they only require protection from $\Z_N$. In this subsection, we will focus on the FSPTs protected jointly by $C_M^-$ and $\mathbb{Z}_N$. Hence, we need to consider (i) possible 0D-block states and (ii) possible 1D-states.

First, 0D-block states correspond to irreducible representations of the group $\mathbb Z_{2M}^f\times\mathbb{Z}_N$. Different $\mathbb{Z}_{2M}^f$ eigenvalues correspond to FSPTs protected by $C_M^-$ only (see Sec.~\ref{sec:2D-CM-minus-1}). Those protected by both $C_M^-$ and $\mathbb{Z}_N$ correspond to the different eigenvalues of $\mathbb{Z}_N$: $1, e^{i2\pi/N}, \dots, e^{i2\pi(N-1)/N}$. However, not all of them correspond to distinct FSPTs. Imagine a product state, in which each site is a state with $\mathbb{Z}_N$ eigenvalue being $e^{i2\pi p/N}$. Then, we can take the 0D block to contain the rotation-related $M$ sites that are closest to the origin. This 0D-block state has a $\mathbb{Z}_N$ eigenvalue $e^{i2\pi  pM/N}$. Properly choosing the value of $p$, we find the that smallest $\mathbb{Z}_N$ eigenvalue of this product state is $e^{i2\pi (M,N)/N}$, where $(M,N)$ is the greatest common divisor of $M$ and $N$.  Accordingly, meaningful $\mathbb{Z}_N$ eigenvalues are
\begin{equation}
1, \ e^{i2\pi/N}, \ \dots,\ e^{i2\pi[(M,N)-1]/N}. \nonumber
\end{equation}
Hence, the 0D-block states protected by both $C_M^-$ and $\mathbb{Z}_N$ form a group $\mathbb{Z}_{(M,N)}$. We remark that these states are essentially bosonic, and our argument above is essentially identical to the discussion in Appendix \ref{sec:bosonic}.

Next, we construct 1D-block states by gluing $M$ semi-infinite line across the origin in a rotation symmetric way. The semi-infinite lines can be either Majorana chains or 1D FSPTs protected by $\mathbb{Z}_N$ symmetry. The case of Majorana chains corresponds to FSPTs protected by $C_M^-$ only, and have been considered in Sec.~\ref{sec:2D-CM-minus-1}. Here, we study the case of 1D FSPTs protected by onsite $\mathbb{Z}_N$ symmetry.

Let us briefly revisit 1D FSPTs protected by $\mathbb{Z}_N$, with the full symmetry group being $\mathbb{Z}_2^f\times\mathbb{Z}_N$ (see a more general review in Appendix \ref{append:review}). The FSPTs are classified by the two cohomology groups,  
\begin{align}
\H^1[\mathbb{Z}_N, \mathbb{Z}_2] & =\mathbb{Z}_{\gcd(2,N)}, \nonumber\\
\H^2[\mathbb{Z}_N, \U] & = \Z_1
\end{align}
Accordingly, only even $N$ allows a nontrivial FSPT, corresponding to the nontrivial element of $\mathbb{Z}_{\gcd(2,N)}$. This FSPT state can be constructed on a 1D lattice with two fermions, $c_i^a$ and $ c_i^b$, on each site $i$. The type-$a$ and type-$b$ fermions each form a Majorana chain. Let $\mb{g}$ be the generator of $\mathbb{Z}_N$. Under the $\mb{g}$ symmetry, 
\begin{equation}
c_i^a \rightarrow c_i^a, \quad  c_i^b \rightarrow - c_i^b
\end{equation}
That is, $\mb{g}$ is the fermion parity $P_f^b$ of the type-$b$ fermions. At the end of the 1D lattice, there are two Majorana zero modes, transforming in the follow way under $\mb{g}$ symmetry
\begin{equation}
\gamma^a \rightarrow \gamma^a, \quad \gamma^b \rightarrow -\gamma^b
\label{eqn:Zn1D}
\end{equation} 
Since the two Majorana fermions has opposite charge under $g$, they cannot be removed by the term $i\gamma^a\gamma^b$. This degeneracy is robust against any $\mathbb{Z}_N$ symmetric perturbation.

Consider $M$ copies of the above 1D FSPTs defined on semi-infinite lines, arranged around the origin in a rotation invariant way (similar to  Fig.~\ref{fig:rotation-connect}). There are $2M$ Majorana zero modes around the origin, $\{\gamma_n^a\}$ and $\{\gamma_n^b\}$. Under rotation $\vr$ and $\mathbb{Z}_N$ symmetry $\mb{g}$, these Majoranas transform as
\begin{align}
	\vr: \gamma_j\rightarrow \gamma_{j+1}, 1\leq j<M; \quad \gamma_M\rightarrow -\gamma_1, 
\end{align}
where we supress the $a/b$ index, and the $\mb{g}$ transformation is already given in Eq. \eqref{eqn:Zn1D}. Then, we can write down the following Hamiltonian to remove all Majoranas:
\begin{equation}
	H=-i\sum_{j=1}^{M/2}(\gamma_j^a\gamma_{j+M/2}^a + \gamma_j^b\gamma_{j+M/2}^b)
	\label{}
\end{equation}
when $M$ is even. When $M$ is odd, we cannot gapped out the Majoranas.

Finally, we need to check if the 1D block states will stack into 0D block states. (We expect that 2D internal FSPTs will not stack into 1D block states, as they only need $\Z_N$ for protection but 1D block states need both rotation and $\Z_N$.) To check this, we stack two copies of 1D block states, such that entanglement in each 1D block can be symmetrically removed. The remaining 0D block state resembles the one in Fig.~\ref{fig:1Dblock-stack}, but now we have two copies $a$ and $b$. Since the fermion parity of each short Majorana chain is odd, it is not hard to see that the 0D block state has an $\Z_N$ eigenvalue $(-1)^{M/2}$. This 0D block state is non-trivial only if $M/2$ is odd. In addition, only when both $N/2$ and $M/2$ are odd, it is not possible to split this state into two other identical 0D block states. Hence, nontrivial extension of $\mathscr{G}_{-1}$ by $\mathscr{G}_{-2}$ occurs only if both $M/2$ and $N/2$ are odd. 

In summary, the classification of $C_M^-\times \Z_N$ 2D FSPT phases are classified by
\begin{equation}
\mathscr{G} = \left\{
\begin{array}{ll}
\Z_{2(M,N)}, & \quad \text{$M/2,N/2$ are odd} \\[3pt]
\Z_{(M,N)}\times \Z_{(2,M,N)} & \quad \text{otherwise}
\end{array}
\right.
\end{equation}
Note that here we only classify those FSPT phases protected jointly by $C_M^-$ and $\Z_N$. This classification agrees with that in Ref.~\onlinecite{WangPRB2017} for $\Z_2^f\times \Z_M\times \Z_N$ internal symmetry.

\subsection{$C_M^{-}\times \mathbb{Z}_{N_1}\times \mathbb{Z}_{N_2}$}
\label{app:2Dcm-zn12}

Let us now consider the internal symmetry group being $\Z_{N_1}\times\Z_{N_2}$. We only consider those FSPT phases that require protection from all $C_M^-$, $\Z_{N_1}$ and $\Z_{N_2}$.   First of all,  0D block states need protection from at most two symmetries among $C_M^-$, $\Z_{N_1}$ and $\Z_{N_2}$. One can find them in Sec.~\ref{sec:2D-CM-minus-1} and Appendix \ref{sec:2D-CM-minus-2}. Next, we claim the the 2D block states (i.e., 2D $\Z_{N_1}\times \Z_{N_2}$ internal FSPT phases) are compatible with the $C_M^-$ symmetry. To see that, we notice that 2D FSPT phases with $\Z_{N_1}\times\Z_{N_2}$ symmetry have been completely classified in \Ref{WangPRB2017}, and the classification is reviewed in Appendix \ref{append:review}. For our purpose, it suffices to know that all states can be obtained by stacking free fermion topological phases (i.e. copies of $p_x\pm ip_y$ superconductors, as well as Chern insulators) and bosonic SPT phases. All of these building blocks are compatible with the $C_M^-$ symmetry. So we conclude that the classification of 2D block states is identical to that of $\Z_{N_1}\times\Z_{N_2}$ FSPT phases. So, they only need protection from  $\Z_{N_1}$ and $\Z_{N_2}$, but not $C_M^-$.

Therefore, we only need to consider 1D block states. To construct 1D block state, we put 1D FSPT phases with internal $\Z_{N_1}\times\Z_{N_2}$ FSPT states on each 1D block.  The 1D internal FSPTs should be protected jointly by $\Z_{N_1}\times \Z_{N_2}$. Accordingly to Appendix \ref{append:review}, such FSPTs are bosonic and are classified by $H^2[\Z_{N_1}\times\Z_{N_2},U(1)]=\Z_{(N_1, N_2)}$. The 1D block state is a collection of $M$ copies of an FSPT phase in the $\Z_{(N_1,N_2)}$ classification, arranged in a rotation invariant way. Taking into accounts the compatibility with $C_M^-$ symmetry, we have the 1D block states to be classified by $\Z_{(M,N_1, N_2)}$. These are all the FSPT phases protected jointly by   $C_M^-$, $\Z_{N_1}$ and $\Z_{N_2}$. All of them are essentially bosonic.

\subsection{{$C_M^{+}\times \mathbb{Z}_N$}}
\label{app:cm+zn}

Again, we are interested in those FSPT phases protected jointly by $C_M^+$ and $\Z_N$. Let us first consider 0D block states. Those protected jointly by $C_M^+$ and $\Z_N$ correspond to 0D blocks with $\Z_N$ eigenvalues:  $1, e^{i2\pi/N}, \dots, e^{i2\pi(N-1)/N}$. Some of them will be trivialized. We claim that there are two ways: (1) Similarly to the $C_M^-\times \Z_N$ case, one may place $\Z_N$ charges around the origin, arranged rotation symmetrically. This is a trivial state but with $\Z_N$  eigenvalues $e^{i2\pi pM/N}$. So, the $\Z_N$ eigenvalues are topologically distinct only modulo $e^{i2\pi M/N}$. (2) Another way to trivialize some of the 0D block states is similar to Fig.~\ref{fig:2DMajoranaChains}. Take two copies of those in Fig.~\ref{fig:2DMajoranaChains}, with the fermions in the two copies transforming under $\Z_N$ as follows
\begin{equation}
c^a\rightarrow -c^a, \quad c^b\rightarrow c^b
\label{app:2Dcm+zn1}
\end{equation} 
where $a,b$ are the copy indices and site indices are neglected. Then, the Majorana chains in Fig.~\ref{fig:2DMajoranaChains} are replaced by 1D FSPT phases in the classification $H^1(\Z_N, \Z_2)$. All arguments there can still go through, when $M, N$ are even. Accordingly, we obtain a 0D block state which is two copies of that in Fig.~\ref{fig:2DMajoranaChains}, and which represents a trivial FSPT state. Note that each copy is a short Majorana chain, which should be odd under fermion parity. According to the transformations \eqref{app:2Dcm+zn1}, the 0D block state has a  $\Z_N$ eigenvalue $-1$. Hence, the $\Z_N$ eigenvalues for the 0D block states should also be modulo $-1$ when $M$ and $N$ are even. Combining the two results, we conclude that 0D block states are classified by
\begin{equation}
\mathscr{G}_{-2} = \left\{
\begin{array}{ll}
\Z_{(M,N/2)}, & \quad \text{$M,N$ even} \\[3pt]
\Z_{(M,N)}, & \quad \text{otherwise}
\end{array}
\right.
\label{eq:g-2cm+}
\end{equation}

Next, let us consider 1D block states. We can assume $N$ is even because otherwise there are no 1D SPT states protected by $\Z_2^f\times \Z_N$ symmetry. To build a 1D block state, we will also need $M$ to be even, since the 1D SPT state is of order two. The end of the SPT state hosts two Majoranas $\gamma^a$ and $\gamma^b$. The generator $\mb{g}$ of $\Z_N$ is implemented by $\gamma^b$ such that $\gamma^a \rightarrow -\gamma^a$ and $\gamma^b\rightarrow \gamma^b$. Taking the $M$ copies together, we have
\begin{equation}
	U_\mb{g}=\prod_{j=1}^M\gamma_j^b.
	\label{eq:cm+zn1}
\end{equation}
The rotation symmetry $\vr$ acts as $\gamma^a_j\rightarrow \gamma^a_{j+1}$, $\gamma^b_j\rightarrow \gamma^b_{j+1}$. We find that
\begin{equation}
	\vr U_\mb{g} \vr^{-1}=-U_\mb{g}.
	\label{eq:cm+zn2}
\end{equation}
In other words, the Majoranas near the rotation center actually form a projective representation of $C_M^+\times \Z_N$. Hence, there is no non-trivial 1D block state.

Finally, we consider 2D $\Z_N$ FSPT phases, and check if they are compatible with $C_M^+$ symmetry. While these phases do not  need protection from $C_M^+$ symmetry, it is still interesting to discuss their compatibility. We take $N=2$ as an example, and it is known that the classification of 2D $\Z_2$ FSPT phases is $\Z_8$. The free fermion realization of the root phase consists of two decoupled layers, one $p_x+ip_y$ superconductor and the other a $p_x-ip_y$ superconductor. The $\Z_2$ symmetry is the fermion parity of one of the layers. This construction is only compatible with $C_M^-$. If we enforce the $C_M^+$ symmetry with even $M$, then we find a pair of Majoranas at the rotation center, $\gamma_1$ and $\gamma_2$, where $\mb{g}$ acts nontrivially on one of them. The two Majorana zero modes form the boundary of a 1D $\Z_2$ FSPT phase.

Another way to see the compatibility is the following: the ground state wave function of the root $\Z_2$ FSPT phase is a superposition of fluctuating $\Z_2$ domain walls decorated by Majorana chains. For such a superposition to be possible, all domain walls must have even fermion parity. Now place a $\Z_2$ domain wall in a $C_M$ symmetric position surrounding the rotation center. The (closed) Majorana chain decorated on the domain wall has periodic/anti-periodic boundary condition if the symmetry is $C_M^+$/$C_M^-$. Only the latter leads to a ground state with even fermion parity.  

The root Majorana phase is still compatible with $C_M^+$ if $M$ is odd. However, $C_M^+$ is not different from $C_M^-$ for odd $M$. One can simply redefine $\vr P_f$ as the new rotation operator, and $C_M^+$ is turned into $C_M^-$. Similar results can be obtained for other even values of $N$. Namely, those ``Majorana'' SPT phases are not compatible with $C_M^+$ symmetry, $M$ being even. For $N=2 \ ({\rm mod}\ 4)$, the $\Z_N$ internal FSPT phases are classified by $\Z_{4N}$, with the root phase being ``Majorana''. For $N=0\ ({\rm mod}\ 4)$, the internal FSPT phases are classified by $\Z_{2N}\times \Z_2$,  with the $\Z_2$ root phase being ``Majorana''.  For  odd $N$, all SPT phases are bosonic and classified by $\Z_N$. Hence,  we derive that 2D block states are classified by
\begin{equation}
	\mathscr{G}_0=\begin{cases}
			\Z_{2N}, & \text{even $M$ and $N$}\\
		\Z_{4N}, & \text{odd $M$, $N=2 \ ({\rm mod}\ 4)$}\\
		\Z_{2N} \times \Z_2, & \text{odd $M$, $N=0 \ ({\rm mod}\ 4)$}\\
		\Z_N, & \text{odd }N
	\end{cases}
	\label{}
\end{equation}
However, all these phases do not need protection from $C_M^+$ symmetry. 

In summary, all FSPT phases that require protection from both $C_M^+$ and $\Z_N$ are 0D block states, given by $\mathscr{G}_{-2}$ in \eqref{eq:g-2cm+}.
If one wants the full classification of $C_M^+\times \Z_N$ FSPT phases, those protected by $C_M^+$ only and those internal FSPT phases in $\mathscr{G}_0$ should also be included. The full classification agrees with that in Ref.~\onlinecite{WangPRB2017} for internal FSPT phases with $\Z_{2M}^f\times \Z_N$ symmetry.

\subsection{{$C_M^{+}\times \Z_{N_1} \times \Z_{N_2}$}}
\label{app:CplusZZ2D}

Again we are interested in those FSPT phases protected jointly by $C_M^+$, $\Z_{N_1}$, and $\Z_{N_2}$.  Similar to the $C_M^-\times\Z_{N_1}\times\Z_{N_2}$ case, there are no 0D block states requiring protection from the full group. 1D block states are also simple, as they are bosonic and classified by $\Z_{(M, N_1, N_2)}$ (see Appendix \ref{sec:bosonic}). What remains are the 2D block states, in particular their compatibility with the $C_M^+$ symmetry.  For simplicity, we will assume both $N_1$ and $N_2$ are even. Also, we take $M$ to be even. For odd $M$, one can simply redefine $P_f\vr$ as the new rotation, and then it becomes $C_M^-$, so that all 2D internal FSPT phases are compatible according to Appendix \ref{app:2Dcm-zn12}. 

To build 2D block states, we put $\Z_2^f \times \Z_{N_1}\times \Z_{N_2}$ internal FSPT phases on each block. While the (free fermion) constructions of root FSPT phases are generally involved, we can focus one of the states with the following physical characterization: on a fermion parity flux, $\Z_{N_1}$ and $\Z_{N_2}$ generators anti-commute. In fact, one can divide all FSPT phases into two families distinguished by having this property or not. Based on the classification in \Ref{WangPRB2017},  we find it sufficient to consider the following four-layer model, where layer $2$ and $3$ are $p_x+ip_y$ superconductors and layer $1$ and $4$ are $p_x-ip_y$ superconductors. The generators of the $\Z_{N_1}$ and $\Z_{N_2}$ symmetries are defined as
\begin{equation}
	\mb{g}_1 = (-1)^{N_2+N_4}, \ \mb{g}_2=(-1)^{N_3+N_4}.
	\label{}
\end{equation}
Other phases can be obtained by adding bosonic SPT phases. Since bosonic SPT phases are always compatible with rotation, so we focus on the above state.

According to Appendix \ref{append:p+ip}, to enforce $C_M^+$ symmetry, it remains a Majorana zero mode $\gamma_i$, with $i=1,2,3,4$, at the rotation center, after we glue the 2D blocks in a layer-by-layer fashion. One can readily see that
\begin{equation}
	\begin{gathered}
		\mb{g}_1: \gamma_{2}\rightarrow -\gamma_{2}, \ \gamma_{4}\rightarrow -\gamma_{4} \\
	\mb{g}_2: \gamma_{3}\rightarrow -\gamma_{3}. \  \gamma_{4}\rightarrow -\gamma_{4}
	\end{gathered}
	\label{}
\end{equation}
We can couple them through an interaction $\gamma_1\gamma_2\gamma_3\gamma_4$ to select the fermion parity even sector (i.e. purely bosonic degrees of freedom), which is two-dimensional. One can easily see that $\mb{g}_1$ and $\mb{g}_2$ anti-commute when acting on this two-fold degenerate space. It may seem like one has to introduce a spin-$1/2$ transforming projectively  under $\Z_{N_1}\times\Z_{N_2}$ at the rotation center to realize the FSPT phase, but this is not always necessary.  Recall that 1D $\Z_{N_1}\times \Z_{N_2}$ bosonic SPT phases are classified by $\H^2[\Z_{N_1}\times\Z_{N_2}, \U]=\Z_{\text{gcd}(N_1,N_2)}$. Given that the projective phase at the rotation center is just $-1$, it must be the order-2 element in $\Z_{\text{gcd}(N_1,N_2)}$. If one is able to split this order-2 element into $M$ copies, i.e. when $2M|\text{gcd}(N_1, N_2)$, then we can introduce $M$ semi-infinite 1D bosonic SPT states joining at the rotation center, each characterized by a projective phase $e^{\frac{\pi i}{M}}$, and the end states can neutralize the projective representation resulting from the Majoranas. In other words, when we glue the 2D blocks, 1D SPT phases are attached along the gluing lines, such that it helps to eventually build a valid 2D block state. Therefore, when $2M|\text{gcd}(N_1, N_2)$, the 2D block state can be realized with the $C_M^+$ symmetry.

In addition, stacking multiple copies of the above 2D block state always gives rise to a 1D block state. We notice that to have $2M|\gcd(N_1, N_2)$ with even $M$, it is required that $\gcd(N_1,N_2)$ must be a multiple of 4. In this case, it is known that the above internal FSPT phase forms a $\Z_2$ classification\cite{WangPRB2017}. Then, stacking two copies of the above 2D block state is a 1D block state.  Moreover, one can show that it is the root 1D block state. Therefore, we have a classification $\Z_{2(M,N_1,N_2)}$ for those FSPTs protected jointly by $C_M^+$, $Z_{N_1}$ and $\Z_{N_2}$ (when $M,N_1,N_2$ are all even).  

This is an example --- the only example we find --- that  $\mathscr{G}_{0}$ is extended non-trivially by $\mathscr{G}_{-1}$. The simplest symmetry to support this phenomenon is $C_2^+\times \Z_4 \times \Z_4$. We point out that the internal counterpart $\Z_4^f \times \Z_4 \times \Z_4$ is also the simplest symmetry that supports \emph{intrinsically interacting } FSPT phases.\cite{WangPRB2017}  It will be interesting to study the connection.

\section{More classifications of 3D FSPT phases}
\label{sec:more3D}

In this appendix, we discuss classifications of 3D FSPT phases for more symmetries of the form $C_M^\pm\times G$. We will discuss 1D and 2D block states. We will not discuss the compatibility of 3D $\mathbb{Z}_2^f\times G$ FSPTs with respect to the rotation symmetry $C_M^\pm$, as we do not know the surface theory of them in general. (In some examples, one may use gapped and topologically ordered surface state to study the compatibility.) However, in Sec.~\ref{sec:sptlsm}, we make a conjecture on the compatibility. In principle, if they are compatible with the rotation symmetry, they may stack into 2D or 1D block states.  Again we will not discuss it in this work. 

For $C_M^\pm \times \Z_N$, we believe that there are no 3D block states (as is indicated by the absence of $C_M^\pm $ rotation SPTs), so our classifications are complete. For  $C_M^\pm \times \Z_{N_1}\times\Z_{N_2}$, we only consider those FSPT phases protected jointly by $C_M^\pm$,  $\Z_{N_1}$, and $\Z_{N_2}$. The complete classification can be inferred by further combining with the classification of $C_M^\pm \times \Z_N$ and the crystalline equivalence principle.

\label{app:more3D}

\subsection{$C_M^-\times\Z_{N_1}\times \Z_{N_2}$}
\label{app:more3D1}

Let $N_i=2^{n_i}$ for simplicity.  First,  we consider 1D block states. Using the notations in Appendix \ref{append:review}, we have $G_b=\Z_M\times\Z_{N_1}\times\Z_{N_2}$ in this case. We are interested in the phases that require protection from both $\Z_{N_1}$ and $\Z_{N_2}$, because otherwise it reduces to the discussion in Sec. \ref{sec:3Dm1d}. Then, according to Appendix \ref{append:review}, the 1D block states are purely bosonic and classified by $\Z_{(N_1, N_2)}$. Taking into accounts the compatibility with $C_M^-$, we have the 1D block states to be classified by $\mathscr{G}_{-2}=\Z_{(N_1,N_2)}/M\Z_{(N_1,N_2)}=\Z_{(M, N_1,N_2)}$.

Next we consider 2D block states, using the root FSPT phases protected by both $\Z_{N_1}$ and $\Z_{N_2}$ (otherwise it should be reduced to the earlier discussions in Sec. \ref{sec:CMmZN}). Review of 2D internal FSPT phases can be found in Appendix \ref{append:review}. 

We discuss the 2D block states in two cases.

\textbf{Case I}. Let us first consider $n_1=n_2=1$. The internal FSPT phases protected by both $\Z_{N_1}$ and $\Z_{N_2}$ are classified by $\Z_4$. The root phase can be constructed as follows: consider four layers, layer $1$ and $4$ are $p_x-ip_y$ superconductors, layer $2$ and $3$ are $p_x+ip_y$ superconductors. The two $\Z_2$ symmetries are defined as
\begin{equation}
	\mb{g}_1 = (-1)^{F_2+F_4}, \ \mb{g}_2=(-1)^{F_3+F_4}.
	\label{appf0}
\end{equation}
where $F_i$ is the fermion number operator in layer $i$. Now we consider $M$ copies of the root phase, related by $C_M^-$ rotations, with the edge being described by Majorana fermions $\bar\gamma_{j1}, \gamma_{j2}, {\gamma}_{j3}, \bar{\gamma}_{j4}$, $j=1, \dots, M$. We notice that because the root phase is $\Z_4$ classified, we must have $m\geq 2$.

Next, we define the following complex fermions
\begin{equation}
	\begin{split}
\psi_{la}&=\sum_{j=1}^{M} \omega^{-lj} \gamma_{j a }, 
\bar{\psi}_{la}=\sum_{j=1}^{M} \omega^{-lj} \bar{\gamma}_{ja}, 
	\end{split}
	\label{}
\end{equation}
where $\omega = e^{i\pi/M}$, $l=1,3,\dots, M-1$, and $a=2,3$  for $\psi_{la}$, $a=1,4$ for $\bar\psi_{la}$. Under the symmetry transformations, we have
\begin{align}
\vr: &\ \psi_{la}\rightarrow\omega^{l}\psi_{la}, \ \bar{\psi}_{la}\rightarrow \omega^l \bar{\psi}_{la}\nonumber\\
\mb{g}_1: & \ \bar\psi_{l1}\rightarrow \bar\psi_{l1}, \ \psi_{l2} \rightarrow -\psi_{l2}  \nonumber\\
& \ \psi_{l3} \rightarrow \psi_{l3}, \ \bar\psi_{l4}\rightarrow -\bar\psi_{l4} \nonumber\\
\mb{g}_2: & \ \bar\psi_{l1}\rightarrow \bar\psi_{l1}, \ \psi_{l2} \rightarrow \psi_{l2}  \nonumber\\
& \ \psi_{l3} \rightarrow -\psi_{l3}, \ \bar\psi_{l4}\rightarrow -\bar\psi_{l4}
\label{appf:symm}
\end{align}
It is worth emphasizing that all transformations are diagonal. The edge modes meet at the rotation axis, and should be gapped out without breaking the symmetries such that we obtain a valid 2D block state. To be able to gap out the edge,  the multi-layer system, obtained by folding the $M$ copies of root states, should be a trivial $\Z_{2M}^f\times\Z_{2}\times\Z_2$ FSPT phase. According to Ref.~\onlinecite{WangPRB2017}, the multi-layer system is a trivial FSPT if the symmetry fluxes have trivial braiding statistics after gauging the symmetries. More precisely, it is required that
\begin{align}
2M\theta_\vr & =N_1\theta_{\mb g_1} = N_2\theta_{\mb g_2}  =[2M, N_1]\theta_{\vr, \mb{g}_1}\nonumber\\
 & = [2M, N_2]\theta_{\vr, \mb{g}_2}=[N_1, N_2]\theta_{\mb{g}_1,\mb{g}_2}  =0
\label{appf:requirement}
\end{align}
where $\theta_{\alpha}= 2\pi h_\alpha$ is the exchange statistics, and $\theta_{\alpha,\beta}$ is the mutual braiding statistics between the fluxes, and $[a,b]$ denotes the least common multiple of $a$ and $b$. All the equations are defined modulo $2\pi$. Strictly speaking, there are additional requirements for the multi-layer FSPT to be trivial in the case that the fluxes are non-Abelian anyons. However, one can show that the transformation \eqref{appf:symm} only leads to Abelian fluxes, so the conditions \eqref{appf:requirement} are enough. 

We need to compute the quantities in \eqref{appf:requirement} according to the symmetry transformations \eqref{appf:symm}. Note that the multi-layer system can be understood as a stack of multiple Chern insulators, so the braiding statistics between fluxes are easy to compute. With some straightforward calculations, we find that  $\theta_{\vr}=\theta_{\mb g_1} = \theta_{\mb g_2}=\theta_{\vr,\mb{g}_1}=\theta_{\vr, \mb{g}_2}=0$, and 
\begin{align}
\theta_{\mb{g}_1, \mb{g}_2} = -\frac{M\pi}{4}
\label{eq:g1g2}
\end{align}
Accordingly, \eqref{appf:requirement} imposes the condition that  $M$ is a multiple of $4$, i.e., $m\ge 2$. This leads to a $\Z_4$ classification. When $M=2$,  we can however use two copies of the root phase, which is bosonic, to build a valid 2D block state. Thus, the classification is $\Z_2$ when $M=2$. For convenience, we write the classification in a closed form $\mathscr{G}_{-1}=\Z_{(M, 2N_1, 2N_2)}$.

\textbf{Case II}. Next we consider other cases that either $n_1\geq 2$ or $n_2\ge 2$. In this case, the 2D FSPT phases with internal $\Z_2^f \times \Z_{N_1} \times \Z_{N_2}$ are classified by $\Z_{(N_1, N_2)}\times \Z_{2}$ (those protected by both $\Z_{N_1}$ and $\Z_{N_2}$). The phases classified by $\Z_{(N_1,N_2)}$ are essentially bosonic. Hence, to build 2D block state, we only require the order of the phase in each block is compatible with $M$-fold rotation. Hence, the 2D block states are classified by $\Z_{(M, N_1, N_2)}$.

The 2D FSPT phases classified by $\Z_2$ are intrinsically fermionic and non-Abelian (in the sense that the fluxes have non-Abelian braiding statistics). The realization of the root state is the same as in Case I, with the transformations given in \eqref{appf:symm}. (To simplify the discussion, we have chosen the root state to be the one given in Ref.~\onlinecite{WangPRB2017} stacked with an order two bosonic SPT.) In this theory, the $\Z_{N_1}$ and $\Z_{N_2}$ are not realized in a faithful way, but this does not affect any of our discussions. Then, the only nontrivial braiding statistics is again $\theta_{\mb{g}_1, \mb{g}_2}$ given in  \eqref{eq:g1g2}. Since $[N_1, N_2]$ is always a multiple of 4,  the condition $[N_1,N_2]\theta_{\mb{g}_1, \mb{g}_2}=0$ only requires $M$ to be even, which we always assume. Accordingly, $\mathscr{G}_{-1} = \Z_{(M, N_1, N_2)} \times \Z_2$.

\textbf{Group structure $\mathscr{G}$.} Finally, we need to check the overall group structure $\mathscr{G}$. This can be analyzed in a similar way as in Sec.~\ref{sec:stackgroup}.  For Case I, the root state in $\mathscr{G}_{-1}$ is either bosonic or stacking into a bosonic state, so no group extension occurs. It is the same for the ``Abelian'' 2D-block states in $\Z_{(M,N_1,N_2)}$ in Case II. For the ``non-Abelian'' root 2D-block state, we consider inserting a $\mb{g}_2$ defect loop as in Fig.~\ref{fig:flux}(a). According to \ref{appf0}, a defect point on each 2D block carries a pair of Majorana zero modes $\gamma_{j,3}$ and $\gamma_{j,4}$, and $\gamma_{j,3}\rightarrow \gamma_{j,3}$,   $\gamma_{j,4}\rightarrow -\gamma_{j,4}$ under $\mb{g_1}$ action. Therefore, we can view the membrane bounded by a $\mb{g}_2$ loop in Fig.~\ref{fig:flux}(a) as a 1D-block states of $C_M^-\times \Z_{N_1}$ symmetry. According to Appendix \ref{sec:2D-CM-minus-2}, this membrane SPT state stacks into 0D-block state only if $M=N_1=2$, which indicates that the 3D FSPT bulk state also stacks into a 1D-block state. Similarly, if one inserts a $\mb{g}_1$ defect loop, one can argue that the 3D FSPT bulk state  stacks into a 1D-block state if $M=N_2=2$.  Therefore, if either $M=N_1=2$ or $M=N_2=2$, $\mathscr{G}_{-1}=\Z_2\times \Z_2$ is extended by $\mathscr{G}_{-2}=\Z_2$, giving $\mathscr{G}=\Z_2\times \Z_4$.

In summary, the classification of FSPT phases protected jointly by  $C_M^-$, $\Z_{N_1}$ and $\Z_{N_2}$ is given by
\begin{equation}
\mathscr{G}=\left\{
\begin{array}{ll}
\Z_2\times \Z_2, & M=N_1=N_2=2;  \\[3pt]
\Z_4\times \Z_2, & N_1=N_2=2, \  M=N_1=2,\\
&  \text{or } M=N_2=2; \\[3pt]
\Z_{(M,N_1,N_2)}^2\times \Z_2, & \text{otherwise}
\end{array}
\right.
\end{equation}
One consistency check is that our classification is expected to agree with FSPT phases with internal $\Z_2^f\times\Z_M\times \Z_{N_1}\times \Z_{N_2}$, where $M, N_1, N_2$ should be on equal footing. It is indeed the case in our classification.

\subsection{$C_M^+\times\Z_N$}
\label{app:more3D2}

Let first consider 1D block states, the problem reduces to classifying $\Z_2^f \times \Z_M\times\Z_N$ internal FSPT phases in 1D. In this section, we are only interested in the SPT phases that are protected by both $\Z_M$ and $\Z_N$. According to Appendix \ref{append:review}, there are bosonic SPT phases classified by $\Z_{(M,N)}$, which we will label as $\nu=0,1,\dots, (M,N)$. However, the one with $\nu = (M,N)/2$ can actually be trivialized. To see it, we note that this 1D FSPT phase hosts zero modes at its endpoints, which are characterized by a projective representation in which the generators $\vr$ and $\mb{g}$ anti-commute. Similarly to the discussion in Sec.~\ref{sec:2dcm+}, one can show that these zero modes can be cancelled by a pure 2D state, if it is attached to surface of the 3D system. The pure 2D state to be attached is the one discussed in Appendix \ref{app:cm+zn}, characterized by Eqs.~\eqref{eq:cm+zn1} and \eqref{eq:cm+zn2}. Hence,  it reduces to a classification of $\mathscr{G}_{-2}=\Z_{(M,N)/2}$ for the 1D block states.

Next, we discuss 2D block states. We separately consider two cases: $n=1$ and $n\ge2$. 

\textbf{Case I}. We first consider the case $n=1$, i.e., $C_M^+\times\Z_2$ symmetry. Each 2D block hosts a $\Z_2^f \times \Z_2$ FSPT phase, which is classified by $\Z_8$. On the edge of each block, there are $\nu$ pairs of counter-propagating Majorana fermions. Then, on the rotation axis, we have Majorana fermions $\gamma_{j,a}$ and $\bar{\gamma}_{j,a}$, where $j=1,\dots,M$ and $a=1,\dots,\nu$. Under rotation $\vr$, they transform as
\begin{equation}
	\gamma_{j,a}\rightarrow \gamma_{j+1,a},\ \bar{\gamma}_{j,a}\rightarrow \bar\gamma_{j+1,a}
	\label{}
\end{equation}
for all indices $j$ and $a$. Under the $\Z_2$ symmetry $\mb{g}$, they transform as
\begin{equation}
	\gamma_{j,a}\rightarrow -\gamma_{j+1,a}, \ 	\bar\gamma_{j,a}\rightarrow \bar \gamma_{j+1,a}, 
	\label{}
\end{equation}
As above, we define
\begin{equation}
	\begin{split}
\psi_{la}&=\frac{1}{\sqrt{M}}\sum_{j} \omega^{-lj} \gamma_{j a }, \ 
\bar{\psi}_{la}=\frac{1}{\sqrt{M}}\sum_{j} \omega^{-lj} \bar{\gamma}_{ja}, 
	\end{split}
	\label{}
\end{equation}
Here $\omega= e^{\frac{2i\pi}{M}}$ and $l$ is an integer. It is not hard to check that under rotation $\vr$,  they transform diagonally $\psi_{la}\rightarrow\omega^{l}\psi_{la}, \ol{\psi}_{la}\rightarrow\omega^{l}\ol{\psi}_{la}$.

There is actually a subtlety here, which is absent for $C_M^-$. One may notices that $\psi_{la}^\dag = \psi_{(M-l),a}$. Therefore, $\psi_{l=0,a}$ and $\psi_{l=M/2,a}$ (as well as $\bar{\psi}_{0a}$ and $\bar{\psi}_{M/2,a}$) are still Majorana fields. They need to be treated separately in the following discussion. On the other hand, $\psi_{la}$ and $\bar\psi_{la}$ with $l=1,\dots, (M/2-1)$ are Dirac fermions. 

To form a valid 2D block state, these edge modes should be gapped out without breaking the symmetries. This is equivalent to check if the multi-layer system, obtained from folding the 2D blocks, is a trivial SPT state of $\Z_2^f\times \Z_M\times \Z_2$. Accordingly to Ref.~\onlinecite{WangPRB2017}, the SPT is trivial if and only if we have
\begin{align}
M\theta_{\vr}= N\theta_{\mb g} = [M,N]\theta_{\vr, \mb{g}} = 0
\label{app:eqcm+zn1}
\end{align}
and
\begin{equation}
\Theta_{P_f,P_f, \vr}=\Theta_{P_f,P_f, \mb{g}}=\Theta_{P_f, \vr, \mb{g}}=0
\label{app:eqcm+zn2}
\end{equation}
The latter quantities of $\Theta_{\alpha,\beta,\gamma}$ are statistical phases associated with a special braiding process: vortex $\alpha$ is first braided around $\beta$, then around $\gamma$, then around $\beta$ again but in an opposite direction, and finally around $\gamma$ in an opposite direction. It is easy to see that these quantities vanish if the vortices are Abelian. In our case, the multi-layer system can be thought of as a stack of Chern insulators and $p_x\pm ip_y$ superconductors. Vortices in Chern insulators are Abelian, so $\Theta_{\alpha,\beta,\gamma}$ always vanish. In a single $p_x\pm ip_y$ superconductor,  it was shown in Ref.~\onlinecite{WangPRB2016} that $\Theta_{P_f,P_f,P_f}=\pi$. This is enough for us to compute the quantities in \eqref{app:eqcm+zn2}.

The calculations are straightforward and very similar to before. We find that
\begin{align}
\theta_{\mb g} &= \nu M\frac{\pi}{8} \nonumber\\
\theta_{\vr, \mb{g}} &=\nu \frac{\pi}{4}\left(\frac{M}{2}-3\right) \nonumber\\
\Theta_{P_f, \vr, \mb{g}} & = \nu\pi
\end{align} 
and  other quantities vanish. (We point out that the non-vanishing contribution to $\theta_{P_f, \vr, \mb{g}}$ results from $\psi_{M/2, a}$ only.) According to \eqref{app:eqcm+zn1} and \eqref{app:eqcm+zn2}, we find the the SPT is trivial if $\nu M=0 \ ({\rm mod} \ 8)$ and $\nu$ is even. Therefore, the $\nu=1$ Majorana phase is never compatible with the rotation symmetry. More specifically, we have
\begin{enumerate}
\item when $m=1$, the root 2D block state corresponds to $\nu=4$, forming a $\Z_2$ classification.

\item when $m\ge 2$, the root 2D block state corresponds to $\nu=2$, forming a $\Z_4$ classification.
\end{enumerate} 
Combining all together, we have $\mathscr{G}_{-1} = \Z_{(M, 4)}$. 

%\begin{equation}
%	\begin{gathered}
%	\psi_{la}\rightarrow -\omega_M^l\psi_{la}, \bar{\psi}_{la}\rightarrow \omega_M^l\bar{\psi}_{la}, 0 < l< M/2\\
%	\psi_{0a}\rightarrow -\psi_{0a}, \psi_{M/2,a}\rightarrow \psi_{M/2,a}\\
%	\bar{\psi}_{0a}\rightarrow \bar{\psi}_{0a}, \bar{\psi}_{M/2,a}\rightarrow -\bar{\psi}_{M/2,a}
%	\end{gathered}
%	\label{}
%\end{equation}
%It is easy to see that the Majorana fermions $\psi_{0a}$ etc do not contribute to the topological spin. It follows that
%\begin{equation}
%	\begin{split}
%		h_{\mb{g}\vr}&= \frac{\nu}{2}\left(\sum_{l=1}^{M/2-1}\left(\frac{l}{M}+\frac{1}{2}\right)^2 - \sum_l\left(\frac{l}{M}%\right)^2\right)\\
%		&=\frac{\nu}{2}\sum_{l=0}^{M/2} \left( \frac{l}{M}+\frac{1}{4} \right)\\
%		%&=\frac{\nu}{4} \left( \frac{1}{M} M\frac{M-1}{4} + \frac{M-1}{4} \right)\\
%		&=\frac{\nu(M-2)}{8}.
%	\end{split}
%	\label{}
%\end{equation}
%The invariant is actually
%\begin{equation}
%	\theta_{\mb{g}\vr}^M=e^{\frac{2\pi i \nu M(M-2)}{8}}=1.
%	\label{}
%\end{equation}

%So the result is identical to the $C_M^-$ case.

\textbf{Case II.} We now consider the case $n\ge 2$. As discussed in Sec.~\ref{sec:2db2Z}, FSPT phases with internal $\Z_N$ symmetry is classified by $\Z_2\times \Z_{2N}$, where the $\Z_2$ root phase is  ``non-Abelian'' and $\Z_{2N}$ root phase is ``Abelian''. We start with the root non-Abelian phase, and use the same notations for edge modes as in Sec. \ref{sec:2db2Z}. While the transformations under $\mb{g}$ are the same as in \ref{eq:ng2-g}, the transformations under $\vr$ are different:
\begin{align}
\gamma_{j,a} \rightarrow \gamma_{j+1,a}, \ \bar\varPsi_{j,b} \rightarrow \bar \varPsi_{j+1,b}
\end{align}
To proceed, we define the complex fermions,
\begin{align}
	\psi_{la} &=\sum_{j=1}^{M} \omega^{-lj} \gamma_{j a}, \nonumber\\
	\bar{\psi}_{pb} &= \frac{1}{\sqrt{M}}\sum_j \omega^{-pj}\bar{\psi}_{j b}.
	\label{}
\end{align}
where $\omega = e^{i2\pi/M}$, and $l=0,1,\dots, M/2$, $p=0,1\dots, (M-1)$. Again, when $l=0$ and $l=M/2$, $\psi_{la}$ are still Majorana fermions. Under the symmetries, we then have
\begin{align}
	\vr:\  & \psi_{la} \rightarrow w^l\psi_{la}, \ \bar\psi_{la} \rightarrow \omega^p\bar\psi_{la} \nonumber\\
	\mb{g}: \ & \psi_{l,1}\rightarrow -\psi_{l,1}, \ \psi_{l,a}\rightarrow \psi_{l,a}, \ a\ge 2 \nonumber\\
	&  \bar{\psi}_{pb}\rightarrow  e^{i2\pi/N}\bar{\psi}_{pb}.
	\label{}
\end{align}

To check if the edge modes can be gapped out in a symmetric way, we calculate the quantities in \eqref{app:eqcm+zn1} and \eqref{app:eqcm+zn2}. In fact, it is enough to check $\Theta_{P_f, \vr, \mb{g}}$, which we show is nontrivial. We find that
\begin{equation}
\Theta_{P_f, \vr, \mb{g}} = \pi
\end{equation}
for any values of $M$ and $N$. This phase is solely contributed by the Majorana fermion $\psi_{\frac{M}{2},1}$. Therefore, the non-Abelian root FPST phase can never be gapped out to form a valid 2D block state.

We now turn to Abelian SPT phases that are classified by $\Z_{2N}$. Again, we use the notations in Sec.~\ref{sec:2db2Z} to describe the edge modes of the 2D blocks.  There are $\nu M$ counter-propagating pairs of Dirac fermions in total, $\varPsi_{ja}$ and $\bar{\varPsi}_{ja}$, which transform under the symmetries as
\begin{align}
	\vr: \ & \varPsi_{ja}\rightarrow \varPsi_{j+1,a}, \ \bar{\varPsi}_{ja}\rightarrow \bar{\varPsi}_{j+1,a}\nonumber\\
	\mb{g}: \ &\varPsi_{ja}\rightarrow e^{i2\pi/N}\varPsi_{ja}, \ \bar{\varPsi}_{ja}\rightarrow \bar{\varPsi}_{ja}.
	\label{app:eqcm+zn3}
\end{align}
As before, we Fourier transform them to $\psi_{pa}$ and $\bar{\psi}_{pa}$ where $p=0,1,\dots, M-1$. Then, we have diagonal symmetry transformations
\begin{equation}
	\begin{gathered}
		\vr: \psi_{pa}\rightarrow \omega^p\psi_{pa}, \bar{\psi}_p\rightarrow \omega^p \bar{\psi}_{pa}.
	\end{gathered}
	\label{}
\end{equation}
where $\omega=e^{i2\pi/M}$. The $\mb{g}$ transformation takes the same form as in \eqref{app:eqcm+zn3}. 

Since all the edge modes are complex fermions, the multi-layer system obtained by folding all the 2D blocks is simply a stack of Chern insulators. Therefore, the non-Abelian invariants in \eqref{app:eqcm+zn2} all vanish. With some calculations, we also find
\begin{align}
\theta_{\vr } &= 0, \nonumber\\
\theta_{\mb{g}} & = \frac{\nu M\pi}{N^2}, \nonumber\\
\theta_{\vr, \mb{g}} & = \frac{\pi \nu (M-1)}{N}
\end{align}
According to \eqref{app:eqcm+zn1}, the multi-layer system is trivial if $\nu M=0 \ ({\rm mod} \ 2N)$ and $[M,N]\nu/N$ is even. More explicitly, we have
\begin{enumerate}
\item when $m>n \ge2$, any choice of $\nu$ is valid, leading to a classification $\mathscr{G}_{-1}=\Z_{2N}$.

\item when $m\le n$,  $\nu$ should be a multiple of $2N/M$, leading to a classification $\mathscr{G}_{-1}=\Z_{M}$.
\end{enumerate}
A single formula $\mathscr{G}_{-1} = \Z_{(M,2N)}$ summarizes the classification.

One can see that the classifications of 2D block states in both cases $n=1$ and $n\ge 2$ are given by $\mathscr{G}_{-1} = \Z_{(M, 2N)}$. 

\textbf{Group structure $\mathscr{G}$}. Since the root state in $\mathscr{G}_{-1}$ is either bosonic or stacking into a bosonic SPT state, no group extension occurs. Accordingly, the overall classification is given by
\begin{equation}
\mathscr{G} = \mathscr{G}_{-2}\times \mathscr{G}_{-1}= \Z_{(M,N)/2} \times \Z_{(M, 2N)}
\end{equation}
where we have inserted $\mathscr{G}_{-2} = \Z_{(M,N)/2}$ and $\mathscr{G}_{-1} = \Z_{(M, 2N)}$. The root phase of $\Z_{(M, 2N)}$ is intrinsically fermionic when $M\ge 2N$, with the simplest example being $C_4^+\times \Z_2$. The internal FSPT counterpart (with $\Z_8^f\times \Z_2$ symmetry) was first discovered in Ref.~\onlinecite{ChengPRX2018}.

\subsection{$C_M^+\times \Z_{N_1}\times \Z_{N_2}$}
\label{app:e3}
In this appendix, we are interested in those FSPT phases protected jointly by $C_M^+$, $\Z_{N_1}$ and  $\Z_{N_2}$. For 1D block states, the relevant internal FSPT phases are the bosonic ones classified by $H^2[\Z_{N_1}\times\Z_{N_2},U(1)]=\Z_{N_{12}}$, where we use $N_{12}$ to denote the greatest common divisor $(N_1,N_2)$ for brevity. Endpoints of the 1D block support projective representations characterized by $\mb{g_1}\mb{g_2}=\theta \mb{g_2}\mb{g_1}$, with the projective phase 
\begin{equation}
\theta=1,e^{i2\pi/N_{12}}, \dots, e^{i2\pi(N_{12}-1)/N_{12}}\nonumber
\end{equation}
However, some values of $\theta$ actually correspond to trivial 1D block states. First, by placing $M$ copies of 1D SPT phases around the rotation axis in a rotation symmetry way, we can modify $\theta$ by $e^{i2\pi M/N_{12}}$. This lead to a result that $\theta$ is topological distinct only modulo $e^{i2\pi M/N_{12}}$. This is discussed in Appendix \ref{sec:bosonic}. Second, as was shown in Appendix   \ref{app:CplusZZ2D}, there exists a purely 2D state that respects  $C_M^+\times \Z_{N_1}\times \Z_{N_2}$ symmetry, but hosts a zero mode at the rotation center, with the symmetries realized projectively $\mb{g_1}\mb{g_2}=- \mb{g_2}\mb{g_1}$. This state can be used to trivialize the surface of 1D block state with the projective phase $\theta=-1$. Accordingly, $\theta$ corresponds to topologically distinct phases only modulo $-1$. Combining the two results, we find that $\theta$ takes topologically distinct values only modulo $e^{i2\pi x/N_{12}}$, with $x = (M, N_{12}/2)$. Therefore, 1D block states are classified by $\mathscr{G}_{-2} = \Z_{(M, N_{12}/2)}$. 

Next, we consider 2D block states. We put those 2D internal FSPT phases protected by $\Z_{N_1}$ and $\Z_{N_2}$ jointly on each block. There are both non-Abelian and Abelian FSPT phases, with the latter classified by $\Z_{N_{12}}$. As mentioned in Appendix \ref{app:more3D1}, depending on $N_1,N_2,$, the non-Abelian ones may or may not stack into the Abelian ones, forming $\Z_{2N_{12}}$ or $\Z_{N_{12}}\times \Z_2$ classification. Below, we show that the non-Abelian FSPTs are not compatible with $C_M^+$ symmetry, regardless of the values of $M,N_1,N_2$. 

To see that, it is enough to consider the edge modes of the root non-Abelian SPT phase, which have already been discussed in Appendix \ref{app:more3D1}. On the rotation axis, we have Majorana fermions $\bar\gamma_{j1}, \gamma_{j2}, \gamma_{j3}, \bar\gamma_{j4}$, with $j=1, \dots, M$. Under rotation $\vr$, we have $\gamma_{ja}\rightarrow \gamma_{j+1,a}$, $\bar\gamma_{ja}\rightarrow \bar\gamma_{j+1,a}$. Under $\mb{g}_1$ and $\mb{g}_2$, the fields transform according to \eqref{appf0}. As always, we define the new fields
\begin{equation}
\psi_{la} = \frac{1}{\sqrt{M}}\sum_{j=1}^M \omega^{-lj}\gamma_{ja}, \quad \bar\psi_{la} = \frac{1}{\sqrt{M}}\sum_{j=1}^M \omega^{-lj}\bar\gamma_{ja}
\end{equation}
where $\omega = e^{i2\pi/M}$, $l=0, \dots, M-1$, and $a=2,3$ for $\psi_{la}$, $a=1,4$ for $\bar\psi_{la}$. We note that $\psi_{la}^\dag = \psi_{(M-l)a}$  and  $\bar\psi_{la}^\dag = \bar\psi_{(M-l)a}$. Accordingly, $\psi_{0a}$ and $\psi_{\frac{M}{2},a}$ are still Majorana fermions. The independent complex fermions correspond to $l=1,\dots, \frac{M}{2}-1$. The symmetry transformations on $\psi_{la}$ and $\bar\psi_{la}$ are exactly in the same form as in \eqref{appf:symm}.   

To form a valid 2D block state, the edge modes should be gapped out without breaking the symmetries. That is, the multi-layer system, obtained by folding all the blocks, should be a trivial SPT phase with $\Z_2^f\times \Z_M\times \Z_{N_1}\times \Z_{N_2}$ internal symmetry. This can be checked by studying the topological invariants introduced in Ref.~\onlinecite{WangPRB2017}. To show it is a non-trivial SPT state, it is actually enough to check the quantity $\Theta_{\vr, \mb{g}_1, \mb{g}_2}$, which we have introduced previously [see the discussion below Eq.~\eqref{app:eqcm+zn2}]. This quantity detects certain non-Abelian braiding statistics among the vortices $\vr$, $\mb{g}_1$ and $\mb{g}_2$. So, all the complex fermions, which correspond to Chern insulators, do not contribute to $\Theta_{\vr, \mb{g}_1, \mb{g}_2}$. The only nontrivial contribution comes from $\bar\psi_{\frac{M}{2}, 4}$,  giving rise to $\Theta_{\vr, \mb{g}_1, \mb{g}_2}=\pi$. Therefore, the root non-Abelian internal FSPT state cannot be used to build a valid 2D block state.

Accordingly, we are left with the Abelian SPT phases, which are purely bosonic. Then, to be compatible with $M$-fold rotation, it is easy to see that the classification is reduced from $\Z_{N_{12}}$ to $\Z_{(M, N_{12})}$. Combining 1D and 2D block states, all of which are bosonic,  we find that the SPT phases protected jointly by $C_M^+$, $\Z_{N_1}$ and $\Z_{N_2}$ have a classification $\Z_{(M, N_{12}/2)}\times \Z_{(M, N_{12})}$.

\bibliography{TI.bib}

\begin{thebibliography}{74}%
\makeatletter
\providecommand \@ifxundefined [1]{%
 \@ifx{#1\undefined}
}%
\providecommand \@ifnum [1]{%
 \ifnum #1\expandafter \@firstoftwo
 \else \expandafter \@secondoftwo
 \fi
}%
\providecommand \@ifx [1]{%
 \ifx #1\expandafter \@firstoftwo
 \else \expandafter \@secondoftwo
 \fi
}%
\providecommand \natexlab [1]{#1}%
\providecommand \enquote  [1]{``#1''}%
\providecommand \bibnamefont  [1]{#1}%
\providecommand \bibfnamefont [1]{#1}%
\providecommand \citenamefont [1]{#1}%
\providecommand \href@noop [0]{\@secondoftwo}%
\providecommand \href [0]{\begingroup \@sanitize@url \@href}%
\providecommand \@href[1]{\@@startlink{#1}\@@href}%
\providecommand \@@href[1]{\endgroup#1\@@endlink}%
\providecommand \@sanitize@url [0]{\catcode `\\12\catcode `\$12\catcode
  `\&12\catcode `\#12\catcode `\^12\catcode `\_12\catcode `\%12\relax}%
\providecommand \@@startlink[1]{}%
\providecommand \@@endlink[0]{}%
\providecommand \url  [0]{\begingroup\@sanitize@url \@url }%
\providecommand \@url [1]{\endgroup\@href {#1}{\urlprefix }}%
\providecommand \urlprefix  [0]{URL }%
\providecommand \Eprint [0]{\href }%
\providecommand \doibase [0]{https://doi.org/}%
\providecommand \selectlanguage [0]{\@gobble}%
\providecommand \bibinfo  [0]{\@secondoftwo}%
\providecommand \bibfield  [0]{\@secondoftwo}%
\providecommand \translation [1]{[#1]}%
\providecommand \BibitemOpen [0]{}%
\providecommand \bibitemStop [0]{}%
\providecommand \bibitemNoStop [0]{.\EOS\space}%
\providecommand \EOS [0]{\spacefactor3000\relax}%
\providecommand \BibitemShut  [1]{\csname bibitem#1\endcsname}%
\let\auto@bib@innerbib\@empty
%</preamble>
\bibitem [{\citenamefont {Fu}(2011)}]{FuPRL2011}%
  \BibitemOpen
  \bibfield  {author} {\bibinfo {author} {\bibfnamefont {L.}~\bibnamefont
  {Fu}},\ }\bibfield  {title} {\bibinfo {title} {Topological crystalline
  insulators},\ }\href {https://doi.org/10.1103/PhysRevLett.106.106802}
  {\bibfield  {journal} {\bibinfo  {journal} {Phys. Rev. Lett.}\ }\textbf
  {\bibinfo {volume} {106}},\ \bibinfo {pages} {106802} (\bibinfo {year}
  {2011})}\BibitemShut {NoStop}%
\bibitem [{\citenamefont {{Slager}}\ \emph {et~al.}(2013)\citenamefont
  {{Slager}}, \citenamefont {{Mesaros}}, \citenamefont {{Juri{\v c}i{\'c}}},\
  and\ \citenamefont {{Zaanen}}}]{SlagerNP2012}%
  \BibitemOpen
  \bibfield  {author} {\bibinfo {author} {\bibfnamefont {R.-J.}\ \bibnamefont
  {{Slager}}}, \bibinfo {author} {\bibfnamefont {A.}~\bibnamefont {{Mesaros}}},
  \bibinfo {author} {\bibfnamefont {V.}~\bibnamefont {{Juri{\v c}i{\'c}}}},\
  and\ \bibinfo {author} {\bibfnamefont {J.}~\bibnamefont {{Zaanen}}},\
  }\bibfield  {title} {\bibinfo {title} {{The space group classification of
  topological band-insulators}},\ }\href@noop {} {\bibfield  {journal}
  {\bibinfo  {journal} {Nature Physics}\ }\textbf {\bibinfo {volume} {9}}
  (\bibinfo {year} {2013})},\ \Eprint {https://arxiv.org/abs/1209.2610}
  {arXiv:1209.2610 [cond-mat.mes-hall]} \BibitemShut {NoStop}%
\bibitem [{\citenamefont {Teo}\ and\ \citenamefont
  {Hughes}(2013)}]{TeoPRL2013}%
  \BibitemOpen
  \bibfield  {author} {\bibinfo {author} {\bibfnamefont {J.~C.~Y.}\
  \bibnamefont {Teo}}\ and\ \bibinfo {author} {\bibfnamefont {T.~L.}\
  \bibnamefont {Hughes}},\ }\bibfield  {title} {\bibinfo {title} {Existence of
  majorana-fermion bound states on disclinations and the classification of
  topological crystalline superconductors in two dimensions},\ }\href
  {https://doi.org/10.1103/PhysRevLett.111.047006} {\bibfield  {journal}
  {\bibinfo  {journal} {Phys. Rev. Lett.}\ }\textbf {\bibinfo {volume} {111}},\
  \bibinfo {pages} {047006} (\bibinfo {year} {2013})}\BibitemShut {NoStop}%
\bibitem [{\citenamefont {Shiozaki}\ and\ \citenamefont
  {Sato}(2014)}]{ShiozakiPRB2014}%
  \BibitemOpen
  \bibfield  {author} {\bibinfo {author} {\bibfnamefont {K.}~\bibnamefont
  {Shiozaki}}\ and\ \bibinfo {author} {\bibfnamefont {M.}~\bibnamefont
  {Sato}},\ }\bibfield  {title} {\bibinfo {title} {Topology of crystalline
  insulators and superconductors},\ }\href
  {https://doi.org/10.1103/PhysRevB.90.165114} {\bibfield  {journal} {\bibinfo
  {journal} {Phys. Rev. B}\ }\textbf {\bibinfo {volume} {90}},\ \bibinfo
  {pages} {165114} (\bibinfo {year} {2014})}\BibitemShut {NoStop}%
\bibitem [{\citenamefont {Shiozaki}\ \emph
  {et~al.}(2017{\natexlab{a}})\citenamefont {Shiozaki}, \citenamefont {Sato},\
  and\ \citenamefont {Gomi}}]{ShiozakiPRB2017}%
  \BibitemOpen
  \bibfield  {author} {\bibinfo {author} {\bibfnamefont {K.}~\bibnamefont
  {Shiozaki}}, \bibinfo {author} {\bibfnamefont {M.}~\bibnamefont {Sato}},\
  and\ \bibinfo {author} {\bibfnamefont {K.}~\bibnamefont {Gomi}},\ }\bibfield
  {title} {\bibinfo {title} {Topological crystalline materials: General
  formulation, module structure, and wallpaper groups},\ }\href
  {https://doi.org/10.1103/PhysRevB.95.235425} {\bibfield  {journal} {\bibinfo
  {journal} {Phys. Rev. B}\ }\textbf {\bibinfo {volume} {95}},\ \bibinfo
  {pages} {235425} (\bibinfo {year} {2017}{\natexlab{a}})}\BibitemShut
  {NoStop}%
\bibitem [{\citenamefont {Song}\ \emph
  {et~al.}(2017{\natexlab{a}})\citenamefont {Song}, \citenamefont {Huang},
  \citenamefont {Fu},\ and\ \citenamefont {Hermele}}]{SongPRX2017}%
  \BibitemOpen
  \bibfield  {author} {\bibinfo {author} {\bibfnamefont {H.}~\bibnamefont
  {Song}}, \bibinfo {author} {\bibfnamefont {S.-J.}\ \bibnamefont {Huang}},
  \bibinfo {author} {\bibfnamefont {L.}~\bibnamefont {Fu}},\ and\ \bibinfo
  {author} {\bibfnamefont {M.}~\bibnamefont {Hermele}},\ }\bibfield  {title}
  {\bibinfo {title} {Topological phases protected by point group symmetry},\
  }\href {https://doi.org/10.1103/PhysRevX.7.011020} {\bibfield  {journal}
  {\bibinfo  {journal} {Phys. Rev. X}\ }\textbf {\bibinfo {volume} {7}},\
  \bibinfo {pages} {011020} (\bibinfo {year} {2017}{\natexlab{a}})}\BibitemShut
  {NoStop}%
\bibitem [{\citenamefont {Kruthoff}\ \emph {et~al.}(2017)\citenamefont
  {Kruthoff}, \citenamefont {de~Boer}, \citenamefont {van Wezel}, \citenamefont
  {Kane},\ and\ \citenamefont {Slager}}]{KruthoffPRX2017}%
  \BibitemOpen
  \bibfield  {author} {\bibinfo {author} {\bibfnamefont {J.}~\bibnamefont
  {Kruthoff}}, \bibinfo {author} {\bibfnamefont {J.}~\bibnamefont {de~Boer}},
  \bibinfo {author} {\bibfnamefont {J.}~\bibnamefont {van Wezel}}, \bibinfo
  {author} {\bibfnamefont {C.~L.}\ \bibnamefont {Kane}},\ and\ \bibinfo
  {author} {\bibfnamefont {R.-J.}\ \bibnamefont {Slager}},\ }\bibfield  {title}
  {\bibinfo {title} {Topological classification of crystalline insulators
  through band structure combinatorics},\ }\href
  {https://doi.org/10.1103/PhysRevX.7.041069} {\bibfield  {journal} {\bibinfo
  {journal} {Phys. Rev. X}\ }\textbf {\bibinfo {volume} {7}},\ \bibinfo {pages}
  {041069} (\bibinfo {year} {2017})}\BibitemShut {NoStop}%
\bibitem [{\citenamefont {{Bradlyn}}\ \emph {et~al.}(2017)\citenamefont
  {{Bradlyn}}, \citenamefont {{Elcoro}}, \citenamefont {{Cano}}, \citenamefont
  {{Vergniory}}, \citenamefont {{Wang}}, \citenamefont {{Felser}},
  \citenamefont {{Aroyo}},\ and\ \citenamefont
  {{Bernevig}}}]{BradlynNature2017}%
  \BibitemOpen
  \bibfield  {author} {\bibinfo {author} {\bibfnamefont {B.}~\bibnamefont
  {{Bradlyn}}}, \bibinfo {author} {\bibfnamefont {L.}~\bibnamefont {{Elcoro}}},
  \bibinfo {author} {\bibfnamefont {J.}~\bibnamefont {{Cano}}}, \bibinfo
  {author} {\bibfnamefont {M.~G.}\ \bibnamefont {{Vergniory}}}, \bibinfo
  {author} {\bibfnamefont {Z.}~\bibnamefont {{Wang}}}, \bibinfo {author}
  {\bibfnamefont {C.}~\bibnamefont {{Felser}}}, \bibinfo {author}
  {\bibfnamefont {M.~I.}\ \bibnamefont {{Aroyo}}},\ and\ \bibinfo {author}
  {\bibfnamefont {B.~A.}\ \bibnamefont {{Bernevig}}},\ }\bibfield  {title}
  {\bibinfo {title} {{Topological quantum chemistry}},\ }\href
  {https://doi.org/10.1038/nature23268} {\bibfield  {journal} {\bibinfo
  {journal} {Nature}\ }\textbf {\bibinfo {volume} {547}},\ \bibinfo {pages}
  {298} (\bibinfo {year} {2017})},\ \Eprint {https://arxiv.org/abs/1703.02050}
  {arXiv:1703.02050 [cond-mat.mes-hall]} \BibitemShut {NoStop}%
\bibitem [{\citenamefont {{Fang}}\ and\ \citenamefont
  {{Fu}}(2017)}]{FangArxiv2017}%
  \BibitemOpen
  \bibfield  {author} {\bibinfo {author} {\bibfnamefont {C.}~\bibnamefont
  {{Fang}}}\ and\ \bibinfo {author} {\bibfnamefont {L.}~\bibnamefont {{Fu}}},\
  }\bibfield  {title} {\bibinfo {title} {{Rotation Anomaly and Topological
  Crystalline Insulators}},\ }\href@noop {} {\bibfield  {journal} {\bibinfo
  {journal} {ArXiv e-prints}\ } (\bibinfo {year} {2017})},\ \Eprint
  {https://arxiv.org/abs/1709.01929} {arXiv:1709.01929 [cond-mat.mes-hall]}
  \BibitemShut {NoStop}%
\bibitem [{\citenamefont {Po}\ \emph {et~al.}(2017{\natexlab{a}})\citenamefont
  {Po}, \citenamefont {Vishwanath},\ and\ \citenamefont {Watanabe}}]{PoNC2018}%
  \BibitemOpen
  \bibfield  {author} {\bibinfo {author} {\bibfnamefont {H.~C.}\ \bibnamefont
  {Po}}, \bibinfo {author} {\bibfnamefont {A.}~\bibnamefont {Vishwanath}},\
  and\ \bibinfo {author} {\bibfnamefont {H.}~\bibnamefont {Watanabe}},\
  }\href@noop {} {\bibfield  {journal} {\bibinfo  {journal} {Nature Commun.}\
  }\textbf {\bibinfo {volume} {8}},\ \bibinfo {pages} {50} (\bibinfo {year}
  {2017}{\natexlab{a}})}\BibitemShut {NoStop}%
\bibitem [{\citenamefont {Song}\ \emph {et~al.}(2018)\citenamefont {Song},
  \citenamefont {Zhang}, \citenamefont {Fang},\ and\ \citenamefont
  {Fang}}]{SongNC2018}%
  \BibitemOpen
  \bibfield  {author} {\bibinfo {author} {\bibfnamefont {Z.}~\bibnamefont
  {Song}}, \bibinfo {author} {\bibfnamefont {T.}~\bibnamefont {Zhang}},
  \bibinfo {author} {\bibfnamefont {Z.}~\bibnamefont {Fang}},\ and\ \bibinfo
  {author} {\bibfnamefont {C.}~\bibnamefont {Fang}},\ }\href@noop {} {\bibfield
   {journal} {\bibinfo  {journal} {Nature Commun.}\ }\textbf {\bibinfo {volume}
  {9}},\ \bibinfo {pages} {3530} (\bibinfo {year} {2018})}\BibitemShut
  {NoStop}%
\bibitem [{\citenamefont {Khalaf}\ \emph {et~al.}(2018)\citenamefont {Khalaf},
  \citenamefont {Po}, \citenamefont {Vishwanath},\ and\ \citenamefont
  {Watanabe}}]{KhalafPRX2018}%
  \BibitemOpen
  \bibfield  {author} {\bibinfo {author} {\bibfnamefont {E.}~\bibnamefont
  {Khalaf}}, \bibinfo {author} {\bibfnamefont {H.~C.}\ \bibnamefont {Po}},
  \bibinfo {author} {\bibfnamefont {A.}~\bibnamefont {Vishwanath}},\ and\
  \bibinfo {author} {\bibfnamefont {H.}~\bibnamefont {Watanabe}},\ }\bibfield
  {title} {\bibinfo {title} {Symmetry indicators and anomalous surface states
  of topological crystalline insulators},\ }\href
  {https://doi.org/10.1103/PhysRevX.8.031070} {\bibfield  {journal} {\bibinfo
  {journal} {Phys. Rev. X}\ }\textbf {\bibinfo {volume} {8}},\ \bibinfo {pages}
  {031070} (\bibinfo {year} {2018})}\BibitemShut {NoStop}%
\bibitem [{\citenamefont {{Han}}\ \emph {et~al.}(2018)\citenamefont {{Han}},
  \citenamefont {{Wang}},\ and\ \citenamefont {{Ye}}}]{HanArxiv2018}%
  \BibitemOpen
  \bibfield  {author} {\bibinfo {author} {\bibfnamefont {B.}~\bibnamefont
  {{Han}}}, \bibinfo {author} {\bibfnamefont {H.}~\bibnamefont {{Wang}}},\ and\
  \bibinfo {author} {\bibfnamefont {P.}~\bibnamefont {{Ye}}},\ }\bibfield
  {title} {\bibinfo {title} {{Symmetry-protected topological phases with both
  spatial and internal symmetries}},\ }\href@noop {} {\bibfield  {journal}
  {\bibinfo  {journal} {ArXiv e-prints}\ } (\bibinfo {year} {2018})},\ \Eprint
  {https://arxiv.org/abs/1807.10844} {arXiv:1807.10844 [cond-mat.str-el]}
  \BibitemShut {NoStop}%
\bibitem [{\citenamefont {{Zhang}}\ \emph {et~al.}(2018)\citenamefont
  {{Zhang}}, \citenamefont {{Jiang}}, \citenamefont {{Song}}, \citenamefont
  {{Huang}}, \citenamefont {{He}}, \citenamefont {{Fang}}, \citenamefont
  {{Weng}},\ and\ \citenamefont {{Fang}}}]{ZhangCatalogue2018}%
  \BibitemOpen
  \bibfield  {author} {\bibinfo {author} {\bibfnamefont {T.}~\bibnamefont
  {{Zhang}}}, \bibinfo {author} {\bibfnamefont {Y.}~\bibnamefont {{Jiang}}},
  \bibinfo {author} {\bibfnamefont {Z.}~\bibnamefont {{Song}}}, \bibinfo
  {author} {\bibfnamefont {H.}~\bibnamefont {{Huang}}}, \bibinfo {author}
  {\bibfnamefont {Y.}~\bibnamefont {{He}}}, \bibinfo {author} {\bibfnamefont
  {Z.}~\bibnamefont {{Fang}}}, \bibinfo {author} {\bibfnamefont
  {H.}~\bibnamefont {{Weng}}},\ and\ \bibinfo {author} {\bibfnamefont
  {C.}~\bibnamefont {{Fang}}},\ }\bibfield  {title} {\bibinfo {title}
  {{Catalogue of Topological Electronic Materials}},\ }\href@noop {} {\bibfield
   {journal} {\bibinfo  {journal} {ArXiv e-prints}\ } (\bibinfo {year}
  {2018})},\ \Eprint {https://arxiv.org/abs/1807.08756} {arXiv:1807.08756
  [cond-mat.mtrl-sci]} \BibitemShut {NoStop}%
\bibitem [{\citenamefont {{Vergniory}}\ \emph {et~al.}(2018)\citenamefont
  {{Vergniory}}, \citenamefont {{Elcoro}}, \citenamefont {{Felser}},
  \citenamefont {{Bernevig}},\ and\ \citenamefont {{Wang}}}]{Vergniory2018}%
  \BibitemOpen
  \bibfield  {author} {\bibinfo {author} {\bibfnamefont {M.~G.}\ \bibnamefont
  {{Vergniory}}}, \bibinfo {author} {\bibfnamefont {L.}~\bibnamefont
  {{Elcoro}}}, \bibinfo {author} {\bibfnamefont {C.}~\bibnamefont {{Felser}}},
  \bibinfo {author} {\bibfnamefont {B.~A.}\ \bibnamefont {{Bernevig}}},\ and\
  \bibinfo {author} {\bibfnamefont {Z.}~\bibnamefont {{Wang}}},\ }\bibfield
  {title} {\bibinfo {title} {{The (High Quality) Topological Materials In The
  World}},\ }\href@noop {} {\bibfield  {journal} {\bibinfo  {journal} {ArXiv
  e-prints}\ } (\bibinfo {year} {2018})},\ \Eprint
  {https://arxiv.org/abs/1807.10271} {arXiv:1807.10271 [cond-mat.mtrl-sci]}
  \BibitemShut {NoStop}%
\bibitem [{\citenamefont {{Tang}}\ \emph {et~al.}(2018)\citenamefont {{Tang}},
  \citenamefont {{Po}}, \citenamefont {{Vishwanath}},\ and\ \citenamefont
  {{Wan}}}]{Tang2018}%
  \BibitemOpen
  \bibfield  {author} {\bibinfo {author} {\bibfnamefont {F.}~\bibnamefont
  {{Tang}}}, \bibinfo {author} {\bibfnamefont {H.~C.}\ \bibnamefont {{Po}}},
  \bibinfo {author} {\bibfnamefont {A.}~\bibnamefont {{Vishwanath}}},\ and\
  \bibinfo {author} {\bibfnamefont {X.}~\bibnamefont {{Wan}}},\ }\bibfield
  {title} {\bibinfo {title} {{Towards ideal topological materials:
  Comprehensive database searches using symmetry indicators}},\ }\href@noop {}
  {\bibfield  {journal} {\bibinfo  {journal} {ArXiv e-prints}\ } (\bibinfo
  {year} {2018})},\ \Eprint {https://arxiv.org/abs/1807.09744}
  {arXiv:1807.09744 [cond-mat.mes-hall]} \BibitemShut {NoStop}%
\bibitem [{\citenamefont {Hermele}\ and\ \citenamefont
  {Chen}(2016)}]{hermele2016}%
  \BibitemOpen
  \bibfield  {author} {\bibinfo {author} {\bibfnamefont {M.}~\bibnamefont
  {Hermele}}\ and\ \bibinfo {author} {\bibfnamefont {X.}~\bibnamefont {Chen}},\
  }\bibfield  {title} {\bibinfo {title} {Flux-fusion anomaly test and bosonic
  topological crystalline insulators},\ }\href
  {https://doi.org/10.1103/PhysRevX.6.041006} {\bibfield  {journal} {\bibinfo
  {journal} {Phys. Rev. X}\ }\textbf {\bibinfo {volume} {6}},\ \bibinfo {pages}
  {041006} (\bibinfo {year} {2016})},\ \Eprint
  {https://arxiv.org/abs/arXiv:1508.00573} {arXiv:1508.00573} \BibitemShut
  {NoStop}%
\bibitem [{\citenamefont {Huang}\ \emph {et~al.}(2017)\citenamefont {Huang},
  \citenamefont {Song}, \citenamefont {Huang},\ and\ \citenamefont
  {Hermele}}]{HuangPRB2017}%
  \BibitemOpen
  \bibfield  {author} {\bibinfo {author} {\bibfnamefont {S.-J.}\ \bibnamefont
  {Huang}}, \bibinfo {author} {\bibfnamefont {H.}~\bibnamefont {Song}},
  \bibinfo {author} {\bibfnamefont {Y.-P.}\ \bibnamefont {Huang}},\ and\
  \bibinfo {author} {\bibfnamefont {M.}~\bibnamefont {Hermele}},\ }\bibfield
  {title} {\bibinfo {title} {Building crystalline topological phases from
  lower-dimensional states},\ }\href
  {https://doi.org/10.1103/PhysRevB.96.205106} {\bibfield  {journal} {\bibinfo
  {journal} {Phys. Rev. B}\ }\textbf {\bibinfo {volume} {96}},\ \bibinfo
  {pages} {205106} (\bibinfo {year} {2017})}\BibitemShut {NoStop}%
\bibitem [{\citenamefont {{Rasmussen}}\ and\ \citenamefont
  {{Lu}}(2018)}]{RasmussenArxiv2018}%
  \BibitemOpen
  \bibfield  {author} {\bibinfo {author} {\bibfnamefont {A.}~\bibnamefont
  {{Rasmussen}}}\ and\ \bibinfo {author} {\bibfnamefont {Y.-M.}\ \bibnamefont
  {{Lu}}},\ }\bibfield  {title} {\bibinfo {title} {{Classification and
  construction of higher-order symmetry protected topological phases of
  interacting bosons}},\ }\href@noop {} {\bibfield  {journal} {\bibinfo
  {journal} {ArXiv e-prints}\ } (\bibinfo {year} {2018})},\ \Eprint
  {https://arxiv.org/abs/1809.07325} {arXiv:1809.07325 [cond-mat.str-el]}
  \BibitemShut {NoStop}%
\bibitem [{\citenamefont {'t~Hooft}(1980)}]{tHooftbook}%
  \BibitemOpen
  \bibfield  {author} {\bibinfo {author} {\bibfnamefont {G.}~\bibnamefont
  {'t~Hooft}},\ }\href@noop {} {\emph {\bibinfo {title} {Recent Developments in
  Gauge Theories}}}\ (\bibinfo  {publisher} {Plenum, New York},\ \bibinfo
  {year} {1980})\ p.\ \bibinfo {pages} {135}\BibitemShut {NoStop}%
\bibitem [{\citenamefont {Kapustin}\ and\ \citenamefont
  {Thorngren}(2014)}]{KapustinPRL2014}%
  \BibitemOpen
  \bibfield  {author} {\bibinfo {author} {\bibfnamefont {A.}~\bibnamefont
  {Kapustin}}\ and\ \bibinfo {author} {\bibfnamefont {R.}~\bibnamefont
  {Thorngren}},\ }\bibfield  {title} {\bibinfo {title} {Anomalous discrete
  symmetries in three dimensions and group cohomology},\ }\href
  {https://doi.org/10.1103/PhysRevLett.112.231602} {\bibfield  {journal}
  {\bibinfo  {journal} {Phys. Rev. Lett.}\ }\textbf {\bibinfo {volume} {112}},\
  \bibinfo {pages} {231602} (\bibinfo {year} {2014})}\BibitemShut {NoStop}%
\bibitem [{\citenamefont {Witten}(2016)}]{WittenRMP2016}%
  \BibitemOpen
  \bibfield  {author} {\bibinfo {author} {\bibfnamefont {E.}~\bibnamefont
  {Witten}},\ }\bibfield  {title} {\bibinfo {title} {Fermion path integrals and
  topological phases},\ }\href {https://doi.org/10.1103/RevModPhys.88.035001}
  {\bibfield  {journal} {\bibinfo  {journal} {Rev. Mod. Phys.}\ }\textbf
  {\bibinfo {volume} {88}},\ \bibinfo {pages} {035001} (\bibinfo {year}
  {2016})}\BibitemShut {NoStop}%
\bibitem [{\citenamefont {Vishwanath}\ and\ \citenamefont
  {Senthil}(2013)}]{vishwanath2013}%
  \BibitemOpen
  \bibfield  {author} {\bibinfo {author} {\bibfnamefont {A.}~\bibnamefont
  {Vishwanath}}\ and\ \bibinfo {author} {\bibfnamefont {T.}~\bibnamefont
  {Senthil}},\ }\bibfield  {title} {\bibinfo {title} {Physics of
  three-dimensional bosonic topological insulators: Surface-deconfined
  criticality and quantized magnetoelectric effect},\ }\href
  {https://doi.org/10.1103/PhysRevX.3.011016} {\bibfield  {journal} {\bibinfo
  {journal} {Phys. Rev. X}\ }\textbf {\bibinfo {volume} {3}},\ \bibinfo {pages}
  {011016} (\bibinfo {year} {2013})}\BibitemShut {NoStop}%
\bibitem [{\citenamefont {Lieb}\ \emph {et~al.}(1961)\citenamefont {Lieb},
  \citenamefont {Schultz},\ and\ \citenamefont {Mattis}}]{LSM}%
  \BibitemOpen
  \bibfield  {author} {\bibinfo {author} {\bibfnamefont {E.}~\bibnamefont
  {Lieb}}, \bibinfo {author} {\bibfnamefont {T.}~\bibnamefont {Schultz}},\ and\
  \bibinfo {author} {\bibfnamefont {D.}~\bibnamefont {Mattis}},\ }\bibfield
  {title} {\bibinfo {title} {Two soluble models of an antiferromagnetic
  chain},\ }\href@noop {} {\bibfield  {journal} {\bibinfo  {journal} {Annals of
  Physics}\ }\textbf {\bibinfo {volume} {16}},\ \bibinfo {pages} {407}
  (\bibinfo {year} {1961})}\BibitemShut {NoStop}%
\bibitem [{\citenamefont {Oshikawa}(2000)}]{OshikawaLSM}%
  \BibitemOpen
  \bibfield  {author} {\bibinfo {author} {\bibfnamefont {M.}~\bibnamefont
  {Oshikawa}},\ }\bibfield  {title} {\bibinfo {title} {Topological approach to
  luttinger's theorem and the fermi surface of a kondo lattice},\ }\href@noop
  {} {\bibfield  {journal} {\bibinfo  {journal} {Phys. Rev. Lett.}\ }\textbf
  {\bibinfo {volume} {84}},\ \bibinfo {pages} {1535} (\bibinfo {year}
  {2000})},\ \Eprint {https://arxiv.org/abs/cond-mat/0002392}
  {cond-mat/0002392} \BibitemShut {NoStop}%
\bibitem [{\citenamefont {Hastings}(2004)}]{HastingsLSM}%
  \BibitemOpen
  \bibfield  {author} {\bibinfo {author} {\bibfnamefont {M.~B.}\ \bibnamefont
  {Hastings}},\ }\bibfield  {title} {\bibinfo {title} {Lieb-schultz-mattis in
  higher dimensions},\ }\href@noop {} {\bibfield  {journal} {\bibinfo
  {journal} {Phys. Rev. B}\ }\textbf {\bibinfo {volume} {69}},\ \bibinfo
  {pages} {104431} (\bibinfo {year} {2004})},\ \Eprint
  {https://arxiv.org/abs/cond-mat/0305505} {cond-mat/0305505} \BibitemShut
  {NoStop}%
\bibitem [{\citenamefont {Cheng}\ \emph {et~al.}(2016)\citenamefont {Cheng},
  \citenamefont {Zaletel}, \citenamefont {Barkeshli}, \citenamefont
  {Vishwanath},\ and\ \citenamefont {Bonderson}}]{ChengPRX2016}%
  \BibitemOpen
  \bibfield  {author} {\bibinfo {author} {\bibfnamefont {M.}~\bibnamefont
  {Cheng}}, \bibinfo {author} {\bibfnamefont {M.}~\bibnamefont {Zaletel}},
  \bibinfo {author} {\bibfnamefont {M.}~\bibnamefont {Barkeshli}}, \bibinfo
  {author} {\bibfnamefont {A.}~\bibnamefont {Vishwanath}},\ and\ \bibinfo
  {author} {\bibfnamefont {P.}~\bibnamefont {Bonderson}},\ }\bibfield  {title}
  {\bibinfo {title} {Translational symmetry and microscopic constraints on
  symmetry-enriched topological phases: A view from the surface},\ }\href@noop
  {} {\bibfield  {journal} {\bibinfo  {journal} {Phys. Rev. X}\ }\textbf
  {\bibinfo {volume} {6}},\ \bibinfo {pages} {041068} (\bibinfo {year}
  {2016})}\BibitemShut {NoStop}%
\bibitem [{\citenamefont {Jian}\ \emph {et~al.}(2018)\citenamefont {Jian},
  \citenamefont {Bi},\ and\ \citenamefont {Xu}}]{JianPRB2018}%
  \BibitemOpen
  \bibfield  {author} {\bibinfo {author} {\bibfnamefont {C.-M.}\ \bibnamefont
  {Jian}}, \bibinfo {author} {\bibfnamefont {Z.}~\bibnamefont {Bi}},\ and\
  \bibinfo {author} {\bibfnamefont {C.}~\bibnamefont {Xu}},\ }\bibfield
  {title} {\bibinfo {title} {Lieb-schultz-mattis theorem and its
  generalizations from the perspective of the symmetry-protected topological
  phase},\ }\href {https://doi.org/10.1103/PhysRevB.97.054412} {\bibfield
  {journal} {\bibinfo  {journal} {Phys. Rev. B}\ }\textbf {\bibinfo {volume}
  {97}},\ \bibinfo {pages} {054412} (\bibinfo {year} {2018})}\BibitemShut
  {NoStop}%
\bibitem [{\citenamefont {Po}\ \emph {et~al.}(2017{\natexlab{b}})\citenamefont
  {Po}, \citenamefont {Watanabe}, \citenamefont {Jian},\ and\ \citenamefont
  {Zaletel}}]{PoPRL2017}%
  \BibitemOpen
  \bibfield  {author} {\bibinfo {author} {\bibfnamefont {H.~C.}\ \bibnamefont
  {Po}}, \bibinfo {author} {\bibfnamefont {H.}~\bibnamefont {Watanabe}},
  \bibinfo {author} {\bibfnamefont {C.-M.}\ \bibnamefont {Jian}},\ and\
  \bibinfo {author} {\bibfnamefont {M.~P.}\ \bibnamefont {Zaletel}},\
  }\bibfield  {title} {\bibinfo {title} {Lattice homotopy constraints on phases
  of quantum magnets},\ }\href {https://doi.org/10.1103/PhysRevLett.119.127202}
  {\bibfield  {journal} {\bibinfo  {journal} {Phys. Rev. Lett.}\ }\textbf
  {\bibinfo {volume} {119}},\ \bibinfo {pages} {127202} (\bibinfo {year}
  {2017}{\natexlab{b}})}\BibitemShut {NoStop}%
\bibitem [{\citenamefont {{Cheng}}(2018)}]{ChengArxiv2018}%
  \BibitemOpen
  \bibfield  {author} {\bibinfo {author} {\bibfnamefont {M.}~\bibnamefont
  {{Cheng}}},\ }\bibfield  {title} {\bibinfo {title} {{Fermionic
  Lieb-Schultz-Mattis Theorems and Weak Symmetry-Protected Phases}},\
  }\href@noop {} {\bibfield  {journal} {\bibinfo  {journal} {ArXiv e-prints}\ }
  (\bibinfo {year} {2018})},\ \Eprint {https://arxiv.org/abs/1804.10122}
  {arXiv:1804.10122 [cond-mat.str-el]} \BibitemShut {NoStop}%
\bibitem [{\citenamefont {Yang}\ \emph {et~al.}(2018)\citenamefont {Yang},
  \citenamefont {Jiang}, \citenamefont {Vishwanath},\ and\ \citenamefont
  {Ran}}]{YangPRB2018}%
  \BibitemOpen
  \bibfield  {author} {\bibinfo {author} {\bibfnamefont {X.}~\bibnamefont
  {Yang}}, \bibinfo {author} {\bibfnamefont {S.}~\bibnamefont {Jiang}},
  \bibinfo {author} {\bibfnamefont {A.}~\bibnamefont {Vishwanath}},\ and\
  \bibinfo {author} {\bibfnamefont {Y.}~\bibnamefont {Ran}},\ }\bibfield
  {title} {\bibinfo {title} {Dyonic lieb-schultz-mattis theorem and symmetry
  protected topological phases in decorated dimer models},\ }\href
  {https://doi.org/10.1103/PhysRevB.98.125120} {\bibfield  {journal} {\bibinfo
  {journal} {Phys. Rev. B}\ }\textbf {\bibinfo {volume} {98}},\ \bibinfo
  {pages} {125120} (\bibinfo {year} {2018})}\BibitemShut {NoStop}%
\bibitem [{\citenamefont {{Lu}}(2017)}]{LuLSMSPT}%
  \BibitemOpen
  \bibfield  {author} {\bibinfo {author} {\bibfnamefont {Y.-M.}\ \bibnamefont
  {{Lu}}},\ }\bibfield  {title} {\bibinfo {title} {{Lieb-Schultz-Mattis
  theorems for symmetry protected topological phases}},\ }\href@noop {}
  {\bibfield  {journal} {\bibinfo  {journal} {ArXiv e-prints}\ } (\bibinfo
  {year} {2017})},\ \Eprint {https://arxiv.org/abs/1705.04691}
  {arXiv:1705.04691 [cond-mat.str-el]} \BibitemShut {NoStop}%
\bibitem [{\citenamefont {Benalcazar}\ \emph
  {et~al.}(2017{\natexlab{a}})\citenamefont {Benalcazar}, \citenamefont
  {Bernevig},\ and\ \citenamefont {Hughes}}]{BenalcazarScience}%
  \BibitemOpen
  \bibfield  {author} {\bibinfo {author} {\bibfnamefont {W.~A.}\ \bibnamefont
  {Benalcazar}}, \bibinfo {author} {\bibfnamefont {B.~A.}\ \bibnamefont
  {Bernevig}},\ and\ \bibinfo {author} {\bibfnamefont {T.~L.}\ \bibnamefont
  {Hughes}},\ }\href@noop {} {\bibfield  {journal} {\bibinfo  {journal}
  {Science}\ }\textbf {\bibinfo {volume} {357}},\ \bibinfo {pages} {61}
  (\bibinfo {year} {2017}{\natexlab{a}})}\BibitemShut {NoStop}%
\bibitem [{\citenamefont {{Schindler}}\ \emph {et~al.}(2018)\citenamefont
  {{Schindler}}, \citenamefont {{Cook}}, \citenamefont {{Vergniory}},
  \citenamefont {{Wang}}, \citenamefont {{Parkin}}, \citenamefont
  {{Bernevig}},\ and\ \citenamefont {{Neupert}}}]{SchindlerSciAdv2018}%
  \BibitemOpen
  \bibfield  {author} {\bibinfo {author} {\bibfnamefont {F.}~\bibnamefont
  {{Schindler}}}, \bibinfo {author} {\bibfnamefont {A.~M.}\ \bibnamefont
  {{Cook}}}, \bibinfo {author} {\bibfnamefont {M.~G.}\ \bibnamefont
  {{Vergniory}}}, \bibinfo {author} {\bibfnamefont {Z.}~\bibnamefont {{Wang}}},
  \bibinfo {author} {\bibfnamefont {S.~S.~P.}\ \bibnamefont {{Parkin}}},
  \bibinfo {author} {\bibfnamefont {B.~A.}\ \bibnamefont {{Bernevig}}},\ and\
  \bibinfo {author} {\bibfnamefont {T.}~\bibnamefont {{Neupert}}},\ }\bibfield
  {title} {\bibinfo {title} {{Higher-order topological insulators}},\ }\href
  {https://doi.org/10.1126/sciadv.aat0346} {\bibfield  {journal} {\bibinfo
  {journal} {Science Advances}\ }\textbf {\bibinfo {volume} {4}},\ \bibinfo
  {pages} {eaat0346} (\bibinfo {year} {2018})},\ \Eprint
  {https://arxiv.org/abs/1708.03636} {arXiv:1708.03636 [cond-mat.mes-hall]}
  \BibitemShut {NoStop}%
\bibitem [{\citenamefont {Benalcazar}\ \emph
  {et~al.}(2017{\natexlab{b}})\citenamefont {Benalcazar}, \citenamefont
  {Bernevig},\ and\ \citenamefont {Hughes}}]{BenalcazarPRB2017}%
  \BibitemOpen
  \bibfield  {author} {\bibinfo {author} {\bibfnamefont {W.~A.}\ \bibnamefont
  {Benalcazar}}, \bibinfo {author} {\bibfnamefont {B.~A.}\ \bibnamefont
  {Bernevig}},\ and\ \bibinfo {author} {\bibfnamefont {T.~L.}\ \bibnamefont
  {Hughes}},\ }\bibfield  {title} {\bibinfo {title} {Electric multipole
  moments, topological multipole moment pumping, and chiral hinge states in
  crystalline insulators},\ }\href {https://doi.org/10.1103/PhysRevB.96.245115}
  {\bibfield  {journal} {\bibinfo  {journal} {Phys. Rev. B}\ }\textbf {\bibinfo
  {volume} {96}},\ \bibinfo {pages} {245115} (\bibinfo {year}
  {2017}{\natexlab{b}})}\BibitemShut {NoStop}%
\bibitem [{\citenamefont {Song}\ \emph
  {et~al.}(2017{\natexlab{b}})\citenamefont {Song}, \citenamefont {Fang},\ and\
  \citenamefont {Fang}}]{SongPRL2017}%
  \BibitemOpen
  \bibfield  {author} {\bibinfo {author} {\bibfnamefont {Z.}~\bibnamefont
  {Song}}, \bibinfo {author} {\bibfnamefont {Z.}~\bibnamefont {Fang}},\ and\
  \bibinfo {author} {\bibfnamefont {C.}~\bibnamefont {Fang}},\ }\bibfield
  {title} {\bibinfo {title} {$(d\ensuremath{-}2)$-dimensional edge states of
  rotation symmetry protected topological states},\ }\href
  {https://doi.org/10.1103/PhysRevLett.119.246402} {\bibfield  {journal}
  {\bibinfo  {journal} {Phys. Rev. Lett.}\ }\textbf {\bibinfo {volume} {119}},\
  \bibinfo {pages} {246402} (\bibinfo {year} {2017}{\natexlab{b}})}\BibitemShut
  {NoStop}%
\bibitem [{\citenamefont {{Wang}}\ \emph {et~al.}(2018)\citenamefont {{Wang}},
  \citenamefont {{Lin}},\ and\ \citenamefont {{Hughes}}}]{YuxuanWang2018}%
  \BibitemOpen
  \bibfield  {author} {\bibinfo {author} {\bibfnamefont {Y.}~\bibnamefont
  {{Wang}}}, \bibinfo {author} {\bibfnamefont {M.}~\bibnamefont {{Lin}}},\ and\
  \bibinfo {author} {\bibfnamefont {T.~L.}\ \bibnamefont {{Hughes}}},\
  }\bibfield  {title} {\bibinfo {title} {{Weak-Pairing Higher Order Topological
  Superconductors}},\ }\href@noop {} {\bibfield  {journal} {\bibinfo  {journal}
  {ArXiv e-prints}\ } (\bibinfo {year} {2018})},\ \Eprint
  {https://arxiv.org/abs/1804.01531} {arXiv:1804.01531} \BibitemShut {NoStop}%
\bibitem [{\citenamefont {Langbehn}\ \emph {et~al.}(2017)\citenamefont
  {Langbehn}, \citenamefont {Peng}, \citenamefont {Trifunovic}, \citenamefont
  {von Oppen},\ and\ \citenamefont {Brouwer}}]{JosiasPRL2017}%
  \BibitemOpen
  \bibfield  {author} {\bibinfo {author} {\bibfnamefont {J.}~\bibnamefont
  {Langbehn}}, \bibinfo {author} {\bibfnamefont {Y.}~\bibnamefont {Peng}},
  \bibinfo {author} {\bibfnamefont {L.}~\bibnamefont {Trifunovic}}, \bibinfo
  {author} {\bibfnamefont {F.}~\bibnamefont {von Oppen}},\ and\ \bibinfo
  {author} {\bibfnamefont {P.~W.}\ \bibnamefont {Brouwer}},\ }\bibfield
  {title} {\bibinfo {title} {Reflection-symmetric second-order topological
  insulators and superconductors},\ }\href
  {https://doi.org/10.1103/PhysRevLett.119.246401} {\bibfield  {journal}
  {\bibinfo  {journal} {Phys. Rev. Lett.}\ }\textbf {\bibinfo {volume} {119}},\
  \bibinfo {pages} {246401} (\bibinfo {year} {2017})}\BibitemShut {NoStop}%
\bibitem [{\citenamefont {{Matsugatani}}\ and\ \citenamefont
  {{Watanabe}}(2018)}]{Matsugatani2018}%
  \BibitemOpen
  \bibfield  {author} {\bibinfo {author} {\bibfnamefont {A.}~\bibnamefont
  {{Matsugatani}}}\ and\ \bibinfo {author} {\bibfnamefont {H.}~\bibnamefont
  {{Watanabe}}},\ }\bibfield  {title} {\bibinfo {title} {{Connecting
  higher-order topological insulators to lower-dimensional topological
  insulators}},\ }\href@noop {} {\bibfield  {journal} {\bibinfo  {journal}
  {ArXiv e-prints}\ } (\bibinfo {year} {2018})},\ \Eprint
  {https://arxiv.org/abs/1804.02794} {arXiv:1804.02794 [cond-mat.str-el]}
  \BibitemShut {NoStop}%
\bibitem [{\citenamefont {Khalaf}(2018)}]{KhalafPRB2018}%
  \BibitemOpen
  \bibfield  {author} {\bibinfo {author} {\bibfnamefont {E.}~\bibnamefont
  {Khalaf}},\ }\bibfield  {title} {\bibinfo {title} {Higher-order topological
  insulators and superconductors protected by inversion symmetry},\ }\href
  {https://doi.org/10.1103/PhysRevB.97.205136} {\bibfield  {journal} {\bibinfo
  {journal} {Phys. Rev. B}\ }\textbf {\bibinfo {volume} {97}},\ \bibinfo
  {pages} {205136} (\bibinfo {year} {2018})}\BibitemShut {NoStop}%
\bibitem [{\citenamefont {{You}}\ \emph {et~al.}(2018)\citenamefont {{You}},
  \citenamefont {{Devakul}}, \citenamefont {{Burnell}},\ and\ \citenamefont
  {{Neupert}}}]{YouHO2018}%
  \BibitemOpen
  \bibfield  {author} {\bibinfo {author} {\bibfnamefont {Y.}~\bibnamefont
  {{You}}}, \bibinfo {author} {\bibfnamefont {T.}~\bibnamefont {{Devakul}}},
  \bibinfo {author} {\bibfnamefont {F.~J.}\ \bibnamefont {{Burnell}}},\ and\
  \bibinfo {author} {\bibfnamefont {T.}~\bibnamefont {{Neupert}}},\ }\bibfield
  {title} {\bibinfo {title} {{Higher order symmetry-protected topological
  states for interacting bosons and fermions}},\ }\href@noop {} {\bibfield
  {journal} {\bibinfo  {journal} {ArXiv e-prints}\ } (\bibinfo {year}
  {2018})},\ \Eprint {https://arxiv.org/abs/1807.09788} {arXiv:1807.09788}
  \BibitemShut {NoStop}%
\bibitem [{\citenamefont {{Dubinkin}}\ and\ \citenamefont
  {{Hughes}}(2018)}]{DubinkinHO2018}%
  \BibitemOpen
  \bibfield  {author} {\bibinfo {author} {\bibfnamefont {O.}~\bibnamefont
  {{Dubinkin}}}\ and\ \bibinfo {author} {\bibfnamefont {T.~L.}\ \bibnamefont
  {{Hughes}}},\ }\bibfield  {title} {\bibinfo {title} {{Higher Order Bosonic
  Topological Phases in Spin Models}},\ }\href@noop {} {\bibfield  {journal}
  {\bibinfo  {journal} {ArXiv e-prints}\ } (\bibinfo {year} {2018})},\ \Eprint
  {https://arxiv.org/abs/1807.09781} {arXiv:1807.09781} \BibitemShut {NoStop}%
\bibitem [{\citenamefont {Wang}\ and\ \citenamefont
  {Senthil}(2014)}]{ChongWangPRB2014}%
  \BibitemOpen
  \bibfield  {author} {\bibinfo {author} {\bibfnamefont {C.}~\bibnamefont
  {Wang}}\ and\ \bibinfo {author} {\bibfnamefont {T.}~\bibnamefont {Senthil}},\
  }\bibfield  {title} {\bibinfo {title} {Interacting fermionic topological
  insulators/superconductors in three dimensions},\ }\href
  {https://doi.org/10.1103/PhysRevB.89.195124} {\bibfield  {journal} {\bibinfo
  {journal} {Phys. Rev. B}\ }\textbf {\bibinfo {volume} {89}},\ \bibinfo
  {pages} {195124} (\bibinfo {year} {2014})}\BibitemShut {NoStop}%
\bibitem [{\citenamefont {Freed}\ and\ \citenamefont
  {Hopkins}(2016)}]{freed2016}%
  \BibitemOpen
  \bibfield  {author} {\bibinfo {author} {\bibfnamefont {D.~S.}\ \bibnamefont
  {Freed}}\ and\ \bibinfo {author} {\bibfnamefont {M.~J.}\ \bibnamefont
  {Hopkins}},\ }\Eprint {https://arxiv.org/abs/arXiv:1604.06527}
  {arXiv:1604.06527}  (\bibinfo {year} {2016})\BibitemShut {NoStop}%
\bibitem [{\citenamefont {Gu}\ and\ \citenamefont {Wen}(2014)}]{GuPRB2014}%
  \BibitemOpen
  \bibfield  {author} {\bibinfo {author} {\bibfnamefont {Z.-C.}\ \bibnamefont
  {Gu}}\ and\ \bibinfo {author} {\bibfnamefont {X.-G.}\ \bibnamefont {Wen}},\
  }\bibfield  {title} {\bibinfo {title} {Symmetry-protected topological orders
  for interacting fermions: Fermionic topological nonlinear
  $\ensuremath{\sigma}$ models and a special group supercohomology theory},\
  }\href@noop {} {\bibfield  {journal} {\bibinfo  {journal} {Phys. Rev. B}\
  }\textbf {\bibinfo {volume} {90}},\ \bibinfo {pages} {115141} (\bibinfo
  {year} {2014})}\BibitemShut {NoStop}%
\bibitem [{\citenamefont {{Kapustin}}\ \emph {et~al.}(2015)\citenamefont
  {{Kapustin}}, \citenamefont {{Thorngren}}, \citenamefont {{Turzillo}},\ and\
  \citenamefont {{Wang}}}]{Kapustin2015b}%
  \BibitemOpen
  \bibfield  {author} {\bibinfo {author} {\bibfnamefont {A.}~\bibnamefont
  {{Kapustin}}}, \bibinfo {author} {\bibfnamefont {R.}~\bibnamefont
  {{Thorngren}}}, \bibinfo {author} {\bibfnamefont {A.}~\bibnamefont
  {{Turzillo}}},\ and\ \bibinfo {author} {\bibfnamefont {Z.}~\bibnamefont
  {{Wang}}},\ }\bibfield  {title} {\bibinfo {title} {{Fermionic symmetry
  protected topological phases and cobordisms}},\ }\href
  {https://doi.org/10.1007/JHEP12(2015)052} {\bibfield  {journal} {\bibinfo
  {journal} {Journal of High Energy Physics}\ }\textbf {\bibinfo {volume}
  {2015}},\ \bibinfo {eid} {52} (\bibinfo {year} {2015})}\BibitemShut {NoStop}%
\bibitem [{\citenamefont {{Kapustin}}\ and\ \citenamefont
  {{Thorngren}}(2017)}]{KapustinThorngren}%
  \BibitemOpen
  \bibfield  {author} {\bibinfo {author} {\bibfnamefont {A.}~\bibnamefont
  {{Kapustin}}}\ and\ \bibinfo {author} {\bibfnamefont {R.}~\bibnamefont
  {{Thorngren}}},\ }\bibfield  {title} {\bibinfo {title} {{Fermionic SPT phases
  in higher dimensions and bosonization}},\ }\href@noop {} {\bibfield
  {journal} {\bibinfo  {journal} {Journal of High Energy Physics}\ } (\bibinfo
  {year} {2017})},\ \Eprint {https://arxiv.org/abs/1701.08264}
  {arXiv:1701.08264 [cond-mat.str-el]} \BibitemShut {NoStop}%
\bibitem [{\citenamefont {Wang}\ and\ \citenamefont {Gu}(2018)}]{WangPRX2018}%
  \BibitemOpen
  \bibfield  {author} {\bibinfo {author} {\bibfnamefont {Q.-R.}\ \bibnamefont
  {Wang}}\ and\ \bibinfo {author} {\bibfnamefont {Z.-C.}\ \bibnamefont {Gu}},\
  }\bibfield  {title} {\bibinfo {title} {Towards a complete classification of
  symmetry-protected topological phases for interacting fermions in three
  dimensions and a general group supercohomology theory},\ }\href
  {https://doi.org/10.1103/PhysRevX.8.011055} {\bibfield  {journal} {\bibinfo
  {journal} {Phys. Rev. X}\ }\textbf {\bibinfo {volume} {8}},\ \bibinfo {pages}
  {011055} (\bibinfo {year} {2018})}\BibitemShut {NoStop}%
\bibitem [{\citenamefont {Thorngren}\ and\ \citenamefont
  {Else}(2018)}]{ThorngrenPRX2018}%
  \BibitemOpen
  \bibfield  {author} {\bibinfo {author} {\bibfnamefont {R.}~\bibnamefont
  {Thorngren}}\ and\ \bibinfo {author} {\bibfnamefont {D.~V.}\ \bibnamefont
  {Else}},\ }\bibfield  {title} {\bibinfo {title} {Gauging spatial symmetries
  and the classification of topological crystalline phases},\ }\href
  {https://doi.org/10.1103/PhysRevX.8.011040} {\bibfield  {journal} {\bibinfo
  {journal} {Phys. Rev. X}\ }\textbf {\bibinfo {volume} {8}},\ \bibinfo {pages}
  {011040} (\bibinfo {year} {2018})}\BibitemShut {NoStop}%
\bibitem [{\citenamefont {Freed}\ and\ \citenamefont
  {Hopkins}(2021)}]{freed2021reflection}%
  \BibitemOpen
  \bibfield  {author} {\bibinfo {author} {\bibfnamefont {D.~S.}\ \bibnamefont
  {Freed}}\ and\ \bibinfo {author} {\bibfnamefont {M.~J.}\ \bibnamefont
  {Hopkins}},\ }\bibfield  {title} {\bibinfo {title} {Reflection positivity and
  invertible topological phases},\ }\href@noop {} {\bibfield  {journal}
  {\bibinfo  {journal} {Geometry \& Topology}\ }\textbf {\bibinfo {volume}
  {25}},\ \bibinfo {pages} {1165} (\bibinfo {year} {2021})}\BibitemShut
  {NoStop}%
\bibitem [{Note1()}]{Note1}%
  \BibitemOpen
  \bibinfo {note} {Certain types of anomalies involving crystalline symmetries
  can be implemented by allowing symmetries to act \protect \emph
  {projectively}, instead of \protect \emph {linearly}, such that the anomalous
  system can be realized in the same dimension without a bulk. Not every
  anomaly can be realized in this way. The conditions making this to happen are
  refereed to as ``LSM-type conditions'' (see e.g. Ref.~\cite
  {ChengPRX2016}).}\BibitemShut {Stop}%
\bibitem [{\citenamefont {Shiozaki}\ \emph
  {et~al.}(2017{\natexlab{b}})\citenamefont {Shiozaki}, \citenamefont
  {Shapourian},\ and\ \citenamefont {Ryu}}]{ShiozakiPRB2017b}%
  \BibitemOpen
  \bibfield  {author} {\bibinfo {author} {\bibfnamefont {K.}~\bibnamefont
  {Shiozaki}}, \bibinfo {author} {\bibfnamefont {H.}~\bibnamefont
  {Shapourian}},\ and\ \bibinfo {author} {\bibfnamefont {S.}~\bibnamefont
  {Ryu}},\ }\bibfield  {title} {\bibinfo {title} {Many-body topological
  invariants in fermionic symmetry-protected topological phases: Cases of point
  group symmetries},\ }\href {https://doi.org/10.1103/PhysRevB.95.205139}
  {\bibfield  {journal} {\bibinfo  {journal} {Phys. Rev. B}\ }\textbf {\bibinfo
  {volume} {95}},\ \bibinfo {pages} {205139} (\bibinfo {year}
  {2017}{\natexlab{b}})}\BibitemShut {NoStop}%
\bibitem [{\citenamefont {Shapourian}\ \emph {et~al.}(2017)\citenamefont
  {Shapourian}, \citenamefont {Shiozaki},\ and\ \citenamefont
  {Ryu}}]{ShapourianPRL2017}%
  \BibitemOpen
  \bibfield  {author} {\bibinfo {author} {\bibfnamefont {H.}~\bibnamefont
  {Shapourian}}, \bibinfo {author} {\bibfnamefont {K.}~\bibnamefont
  {Shiozaki}},\ and\ \bibinfo {author} {\bibfnamefont {S.}~\bibnamefont
  {Ryu}},\ }\bibfield  {title} {\bibinfo {title} {Many-body topological
  invariants for fermionic symmetry-protected topological phases},\ }\href
  {https://doi.org/10.1103/PhysRevLett.118.216402} {\bibfield  {journal}
  {\bibinfo  {journal} {Phys. Rev. Lett.}\ }\textbf {\bibinfo {volume} {118}},\
  \bibinfo {pages} {216402} (\bibinfo {year} {2017})}\BibitemShut {NoStop}%
\bibitem [{\citenamefont {Fidkowski}\ and\ \citenamefont
  {Kitaev}(2011)}]{Fidkowski2011}%
  \BibitemOpen
  \bibfield  {author} {\bibinfo {author} {\bibfnamefont {L.}~\bibnamefont
  {Fidkowski}}\ and\ \bibinfo {author} {\bibfnamefont {A.}~\bibnamefont
  {Kitaev}},\ }\bibfield  {title} {\bibinfo {title} {Topological phases of
  fermions in one dimension},\ }\href
  {https://doi.org/10.1103/PhysRevB.83.075103} {\bibfield  {journal} {\bibinfo
  {journal} {Phys. Rev. B}\ }\textbf {\bibinfo {volume} {83}},\ \bibinfo
  {pages} {075103} (\bibinfo {year} {2011})}\BibitemShut {NoStop}%
\bibitem [{\citenamefont {{Ludwig}}(2016)}]{Ludwig_review}%
  \BibitemOpen
  \bibfield  {author} {\bibinfo {author} {\bibfnamefont {A.~W.~W.}\
  \bibnamefont {{Ludwig}}},\ }\bibfield  {title} {\bibinfo {title}
  {{Topological phases: classification of topological insulators and
  superconductors of non-interacting fermions, and beyond}},\ }\href
  {https://doi.org/10.1088/0031-8949/2015/T168/014001} {\bibfield  {journal}
  {\bibinfo  {journal} {Physica Scripta Volume T}\ }\textbf {\bibinfo {volume}
  {168}},\ \bibinfo {pages} {014001} (\bibinfo {year} {2016})},\ \Eprint
  {https://arxiv.org/abs/1512.08882} {arXiv:1512.08882 [cond-mat.mes-hall]}
  \BibitemShut {NoStop}%
\bibitem [{\citenamefont {{Chen}}\ \emph {et~al.}(2018)\citenamefont {{Chen}},
  \citenamefont {{Kapustin}}, \citenamefont {{Turzillo}},\ and\ \citenamefont
  {{You}}}]{ChenArxiv2018}%
  \BibitemOpen
  \bibfield  {author} {\bibinfo {author} {\bibfnamefont {Y.-A.}\ \bibnamefont
  {{Chen}}}, \bibinfo {author} {\bibfnamefont {A.}~\bibnamefont {{Kapustin}}},
  \bibinfo {author} {\bibfnamefont {A.}~\bibnamefont {{Turzillo}}},\ and\
  \bibinfo {author} {\bibfnamefont {M.}~\bibnamefont {{You}}},\ }\bibfield
  {title} {\bibinfo {title} {{Free and Interacting Short-Range Entangled Phases
  of Fermions: Beyond the Ten-Fold Way}},\ }\href@noop {} {\bibfield  {journal}
  {\bibinfo  {journal} {ArXiv e-prints}\ } (\bibinfo {year} {2018})},\ \Eprint
  {https://arxiv.org/abs/1809.04958} {arXiv:1809.04958 [cond-mat.str-el]}
  \BibitemShut {NoStop}%
\bibitem [{\citenamefont {Wang}\ and\ \citenamefont
  {Levin}(2015)}]{WangPRB2015}%
  \BibitemOpen
  \bibfield  {author} {\bibinfo {author} {\bibfnamefont {C.}~\bibnamefont
  {Wang}}\ and\ \bibinfo {author} {\bibfnamefont {M.}~\bibnamefont {Levin}},\
  }\bibfield  {title} {\bibinfo {title} {Topological invariants for gauge
  theories and symmetry-protected topological phases},\ }\href
  {https://doi.org/10.1103/PhysRevB.91.165119} {\bibfield  {journal} {\bibinfo
  {journal} {Phys. Rev. B}\ }\textbf {\bibinfo {volume} {91}},\ \bibinfo
  {pages} {165119} (\bibinfo {year} {2015})}\BibitemShut {NoStop}%
\bibitem [{\citenamefont {Tantivasadakarn}\ and\ \citenamefont
  {Vishwanath}(2018)}]{TantivasadakarnPRB2018}%
  \BibitemOpen
  \bibfield  {author} {\bibinfo {author} {\bibfnamefont {N.}~\bibnamefont
  {Tantivasadakarn}}\ and\ \bibinfo {author} {\bibfnamefont {A.}~\bibnamefont
  {Vishwanath}},\ }\bibfield  {title} {\bibinfo {title} {Full commuting
  projector hamiltonians of interacting symmetry-protected topological phases
  of fermions},\ }\href {https://doi.org/10.1103/PhysRevB.98.165104} {\bibfield
   {journal} {\bibinfo  {journal} {Phys. Rev. B}\ }\textbf {\bibinfo {volume}
  {98}},\ \bibinfo {pages} {165104} (\bibinfo {year} {2018})}\BibitemShut
  {NoStop}%
\bibitem [{\citenamefont {Wang}\ \emph {et~al.}(2017)\citenamefont {Wang},
  \citenamefont {Lin},\ and\ \citenamefont {Gu}}]{WangPRB2017}%
  \BibitemOpen
  \bibfield  {author} {\bibinfo {author} {\bibfnamefont {C.}~\bibnamefont
  {Wang}}, \bibinfo {author} {\bibfnamefont {C.-H.}\ \bibnamefont {Lin}},\ and\
  \bibinfo {author} {\bibfnamefont {Z.-C.}\ \bibnamefont {Gu}},\ }\bibfield
  {title} {\bibinfo {title} {Interacting fermionic symmetry-protected
  topological phases in two dimensions},\ }\href
  {https://doi.org/10.1103/PhysRevB.95.195147} {\bibfield  {journal} {\bibinfo
  {journal} {Phys. Rev. B}\ }\textbf {\bibinfo {volume} {95}},\ \bibinfo
  {pages} {195147} (\bibinfo {year} {2017})}\BibitemShut {NoStop}%
\bibitem [{\citenamefont {Cheng}\ \emph
  {et~al.}(2018{\natexlab{a}})\citenamefont {Cheng}, \citenamefont {Bi},
  \citenamefont {You},\ and\ \citenamefont {Gu}}]{ChengPRB2018}%
  \BibitemOpen
  \bibfield  {author} {\bibinfo {author} {\bibfnamefont {M.}~\bibnamefont
  {Cheng}}, \bibinfo {author} {\bibfnamefont {Z.}~\bibnamefont {Bi}}, \bibinfo
  {author} {\bibfnamefont {Y.-Z.}\ \bibnamefont {You}},\ and\ \bibinfo {author}
  {\bibfnamefont {Z.-C.}\ \bibnamefont {Gu}},\ }\bibfield  {title} {\bibinfo
  {title} {Classification of symmetry-protected phases for interacting fermions
  in two dimensions},\ }\href {https://doi.org/10.1103/PhysRevB.97.205109}
  {\bibfield  {journal} {\bibinfo  {journal} {Phys. Rev. B}\ }\textbf {\bibinfo
  {volume} {97}},\ \bibinfo {pages} {205109} (\bibinfo {year}
  {2018}{\natexlab{a}})}\BibitemShut {NoStop}%
\bibitem [{\citenamefont {Cheng}\ \emph
  {et~al.}(2018{\natexlab{b}})\citenamefont {Cheng}, \citenamefont
  {Tantivasadakarn},\ and\ \citenamefont {Wang}}]{ChengPRX2018}%
  \BibitemOpen
  \bibfield  {author} {\bibinfo {author} {\bibfnamefont {M.}~\bibnamefont
  {Cheng}}, \bibinfo {author} {\bibfnamefont {N.}~\bibnamefont
  {Tantivasadakarn}},\ and\ \bibinfo {author} {\bibfnamefont {C.}~\bibnamefont
  {Wang}},\ }\bibfield  {title} {\bibinfo {title} {Loop braiding statistics and
  interacting fermionic symmetry-protected topological phases in three
  dimensions},\ }\href {https://doi.org/10.1103/PhysRevX.8.011054} {\bibfield
  {journal} {\bibinfo  {journal} {Phys. Rev. X}\ }\textbf {\bibinfo {volume}
  {8}},\ \bibinfo {pages} {011054} (\bibinfo {year}
  {2018}{\natexlab{b}})}\BibitemShut {NoStop}%
\bibitem [{\citenamefont {Zhou}\ \emph {et~al.}(2021)\citenamefont {Zhou},
  \citenamefont {Wang}, \citenamefont {Wang},\ and\ \citenamefont
  {Gu}}]{Zhou2021}%
  \BibitemOpen
  \bibfield  {author} {\bibinfo {author} {\bibfnamefont {J.-R.}\ \bibnamefont
  {Zhou}}, \bibinfo {author} {\bibfnamefont {Q.-R.}\ \bibnamefont {Wang}},
  \bibinfo {author} {\bibfnamefont {C.}~\bibnamefont {Wang}},\ and\ \bibinfo
  {author} {\bibfnamefont {Z.-C.}\ \bibnamefont {Gu}},\ }\bibfield  {title}
  {\bibinfo {title} {{Non-Abelian three-loop braiding statistics for 3D
  fermionic topological phases}},\ }\href@noop {} {\bibfield  {journal}
  {\bibinfo  {journal} {Nat. Commun.}\ }\textbf {\bibinfo {volume} {12}},\
  \bibinfo {pages} {2191} (\bibinfo {year} {2021})}\BibitemShut {NoStop}%
\bibitem [{Note2()}]{Note2}%
  \BibitemOpen
  \bibinfo {note} {This argument is due to D. Freed}\BibitemShut {NoStop}%
\bibitem [{\citenamefont {{Song}}\ \emph {et~al.}(2018)\citenamefont {{Song}},
  \citenamefont {{Fang}},\ and\ \citenamefont {{Qi}}}]{SongArxiv2018}%
  \BibitemOpen
  \bibfield  {author} {\bibinfo {author} {\bibfnamefont {Z.}~\bibnamefont
  {{Song}}}, \bibinfo {author} {\bibfnamefont {C.}~\bibnamefont {{Fang}}},\
  and\ \bibinfo {author} {\bibfnamefont {Y.}~\bibnamefont {{Qi}}},\ }\bibfield
  {title} {\bibinfo {title} {{Real-space recipes for general topological
  crystalline states}},\ }\href@noop {} {\bibfield  {journal} {\bibinfo
  {journal} {ArXiv e-prints}\ } (\bibinfo {year} {2018})},\ \Eprint
  {https://arxiv.org/abs/1810.11013} {arXiv:1810.11013 [cond-mat.str-el]}
  \BibitemShut {NoStop}%
\bibitem [{\citenamefont {Barkeshli}\ \emph {et~al.}()\citenamefont
  {Barkeshli}, \citenamefont {Bonderson}, \citenamefont {Cheng},\ and\
  \citenamefont {Wang}}]{SET}%
  \BibitemOpen
  \bibfield  {author} {\bibinfo {author} {\bibfnamefont {M.}~\bibnamefont
  {Barkeshli}}, \bibinfo {author} {\bibfnamefont {P.}~\bibnamefont
  {Bonderson}}, \bibinfo {author} {\bibfnamefont {M.}~\bibnamefont {Cheng}},\
  and\ \bibinfo {author} {\bibfnamefont {Z.}~\bibnamefont {Wang}},\ }\bibfield
  {title} {\bibinfo {title} {Symmetry, defects, and gauging of topological
  phases},\ }\Eprint {https://arxiv.org/abs/arXiv:1410.4540} {arXiv:1410.4540}
  \BibitemShut {NoStop}%
\bibitem [{\citenamefont {Tarantino}\ \emph {et~al.}(2016)\citenamefont
  {Tarantino}, \citenamefont {Lindner},\ and\ \citenamefont
  {Fidkowski}}]{Tarantino_SET}%
  \BibitemOpen
  \bibfield  {author} {\bibinfo {author} {\bibfnamefont {N.}~\bibnamefont
  {Tarantino}}, \bibinfo {author} {\bibfnamefont {N.}~\bibnamefont {Lindner}},\
  and\ \bibinfo {author} {\bibfnamefont {L.}~\bibnamefont {Fidkowski}},\
  }\href@noop {} {\bibfield  {journal} {\bibinfo  {journal} {New J. Phys.}\
  }\textbf {\bibinfo {volume} {18}},\ \bibinfo {pages} {035006} (\bibinfo
  {year} {2016})},\ \Eprint {https://arxiv.org/abs/arXiv:1506.06754}
  {arXiv:1506.06754} \BibitemShut {NoStop}%
\bibitem [{\citenamefont {{Shiozaki}}\ \emph {et~al.}(2018)\citenamefont
  {{Shiozaki}}, \citenamefont {{Zhaoxi Xiong}},\ and\ \citenamefont
  {{Gomi}}}]{ShiozakiArxiv2018}%
  \BibitemOpen
  \bibfield  {author} {\bibinfo {author} {\bibfnamefont {K.}~\bibnamefont
  {{Shiozaki}}}, \bibinfo {author} {\bibfnamefont {C.}~\bibnamefont {{Zhaoxi
  Xiong}}},\ and\ \bibinfo {author} {\bibfnamefont {K.}~\bibnamefont
  {{Gomi}}},\ }\bibfield  {title} {\bibinfo {title} {{Generalized homology and
  Atiyah-Hirzebruch spectral sequence in crystalline symmetry protected
  topological phenomena}},\ }\href@noop {} {\bibfield  {journal} {\bibinfo
  {journal} {ArXiv e-prints}\ } (\bibinfo {year} {2018})},\ \Eprint
  {https://arxiv.org/abs/1810.00801} {arXiv:1810.00801 [cond-mat.str-el]}
  \BibitemShut {NoStop}%
\bibitem [{\citenamefont {{Else}}\ and\ \citenamefont
  {{Thorngren}}(2018)}]{ElseArxiv2018}%
  \BibitemOpen
  \bibfield  {author} {\bibinfo {author} {\bibfnamefont {D.~V.}\ \bibnamefont
  {{Else}}}\ and\ \bibinfo {author} {\bibfnamefont {R.}~\bibnamefont
  {{Thorngren}}},\ }\bibfield  {title} {\bibinfo {title} {{Crystalline
  topological phases as defect networks}},\ }\href@noop {} {\bibfield
  {journal} {\bibinfo  {journal} {ArXiv e-prints}\ } (\bibinfo {year}
  {2018})},\ \Eprint {https://arxiv.org/abs/1810.10539} {arXiv:1810.10539
  [cond-mat.str-el]} \BibitemShut {NoStop}%
\bibitem [{\citenamefont {Rasmussen}\ and\ \citenamefont
  {Lu}(2018)}]{RasmussenArxiv2018b}%
  \BibitemOpen
  \bibfield  {author} {\bibinfo {author} {\bibfnamefont {A.}~\bibnamefont
  {Rasmussen}}\ and\ \bibinfo {author} {\bibfnamefont {Y.-M.}\ \bibnamefont
  {Lu}},\ }\href@noop {} {\bibfield  {journal} {\bibinfo  {journal} {ArXiv
  e-prints}\ } (\bibinfo {year} {2018})},\ \Eprint
  {https://arxiv.org/abs/1810.12317} {arXiv:1810.12317} \BibitemShut {NoStop}%
\bibitem [{\citenamefont {Wen}(2015)}]{WenPRB2015}%
  \BibitemOpen
  \bibfield  {author} {\bibinfo {author} {\bibfnamefont {X.-G.}\ \bibnamefont
  {Wen}},\ }\bibfield  {title} {\bibinfo {title} {Construction of bosonic
  symmetry-protected-trivial states and their topological invariants via
  $g\ifmmode\times\else\texttimes\fi{}so(\ensuremath{\infty})$ nonlinear
  $\ensuremath{\sigma}$ models},\ }\href
  {https://doi.org/10.1103/PhysRevB.91.205101} {\bibfield  {journal} {\bibinfo
  {journal} {Phys. Rev. B}\ }\textbf {\bibinfo {volume} {91}},\ \bibinfo
  {pages} {205101} (\bibinfo {year} {2015})}\BibitemShut {NoStop}%
\bibitem [{\citenamefont {Kapustin}\ \emph {et~al.}(2018)\citenamefont
  {Kapustin}, \citenamefont {Turzillo},\ and\ \citenamefont
  {You}}]{KapustinPRB2018}%
  \BibitemOpen
  \bibfield  {author} {\bibinfo {author} {\bibfnamefont {A.}~\bibnamefont
  {Kapustin}}, \bibinfo {author} {\bibfnamefont {A.}~\bibnamefont {Turzillo}},\
  and\ \bibinfo {author} {\bibfnamefont {M.}~\bibnamefont {You}},\ }\bibfield
  {title} {\bibinfo {title} {Spin topological field theory and fermionic matrix
  product states},\ }\href {https://doi.org/10.1103/PhysRevB.98.125101}
  {\bibfield  {journal} {\bibinfo  {journal} {Phys. Rev. B}\ }\textbf {\bibinfo
  {volume} {98}},\ \bibinfo {pages} {125101} (\bibinfo {year}
  {2018})}\BibitemShut {NoStop}%
\bibitem [{\citenamefont {{Turzillo}}\ and\ \citenamefont
  {{You}}(2017)}]{TurzilloYou}%
  \BibitemOpen
  \bibfield  {author} {\bibinfo {author} {\bibfnamefont {A.}~\bibnamefont
  {{Turzillo}}}\ and\ \bibinfo {author} {\bibfnamefont {M.}~\bibnamefont
  {{You}}},\ }\bibfield  {title} {\bibinfo {title} {{Fermionic Matrix Product
  States and One-Dimensional Short-Range Entangled Phases with Anti-Unitary
  Symmetries}},\ }\href@noop {} {\bibfield  {journal} {\bibinfo  {journal}
  {ArXiv e-prints}\ } (\bibinfo {year} {2017})},\ \Eprint
  {https://arxiv.org/abs/1710.00140} {arXiv:1710.00140 [cond-mat.str-el]}
  \BibitemShut {NoStop}%
\bibitem [{\citenamefont {Qi}\ and\ \citenamefont {Fu}(2015)}]{QiPRB2015}%
  \BibitemOpen
  \bibfield  {author} {\bibinfo {author} {\bibfnamefont {Y.}~\bibnamefont
  {Qi}}\ and\ \bibinfo {author} {\bibfnamefont {L.}~\bibnamefont {Fu}},\
  }\bibfield  {title} {\bibinfo {title} {Anomalous crystal symmetry
  fractionalization on the surface of topological crystalline insulators},\
  }\href {https://doi.org/10.1103/PhysRevLett.115.236801} {\bibfield  {journal}
  {\bibinfo  {journal} {Phys. Rev. Lett.}\ }\textbf {\bibinfo {volume} {115}},\
  \bibinfo {pages} {236801} (\bibinfo {year} {2015})}\BibitemShut {NoStop}%
\bibitem [{\citenamefont {Wang}(2016)}]{WangPRB2016}%
  \BibitemOpen
  \bibfield  {author} {\bibinfo {author} {\bibfnamefont {C.}~\bibnamefont
  {Wang}},\ }\bibfield  {title} {\bibinfo {title} {Braiding statistics and
  classification of two-dimensional charge-$2m$ superconductors},\ }\href
  {https://doi.org/10.1103/PhysRevB.94.085130} {\bibfield  {journal} {\bibinfo
  {journal} {Phys. Rev. B}\ }\textbf {\bibinfo {volume} {94}},\ \bibinfo
  {pages} {085130} (\bibinfo {year} {2016})}\BibitemShut {NoStop}%
\end{thebibliography}%

\end{document}